\documentclass[review,12pt]{elsarticle}
\usepackage[margin=1in,footskip=0.25in]{geometry}
\usepackage{times}
\usepackage{amssymb}
\usepackage{graphicx}
\usepackage{amsmath,amsthm,mathtools}
\usepackage[version=4]{mhchem}
\usepackage{siunitx}
\usepackage{float,subcaption}
\usepackage{longtable,tabularx}
\usepackage{booktabs}
\usepackage{multirow}
\usepackage{xcolor}
\usepackage{hyperref}
\hypersetup{colorlinks, breaklinks, citecolor=blue, linkcolor=blue, urlcolor=blue}
\usepackage[nameinlink,noabbrev,sort&compress]{cleveref}
\Crefname{equation}{Eq.}{Eqs.}
\Crefname{figure}{Fig.}{Figs.}
\crefname{figure}{Figure}{Figures}

\newtheorem{lemma}{Lemma}

\newtheorem{problem}{Problem}
\newtheorem{remark}{Remark}
\newtheorem{assumption}{Assumption}

\newcommand{\diag}{\mathrm{diag}}

\biboptions{numbers,sort&compress}
\makeatletter
\def\ps@pprintTitle{%
	\let\@oddhead\@empty
	\let\@evenhead\@empty
	\let\@oddfoot\@empty
	\let\@evenfoot\@empty
}
\makeatother
\begin{document}
\begin{frontmatter}
\title{Trajectory Tracking for Uncrewed Surface Vessels with Input Saturation and Dynamic Motion Constraints} 
\author[rmv]{Ram Milan Kumar Verma\corref{cor1}}
\cortext[cor1]{Corresponding author}
\ead{rmverma@aero.iitb.ac.in}
\author[rmv]{Shashi Ranjan Kumar}
\ead{srk@aero.iitb.ac.in}
\author[rmv]{Hemendra Arya}
\ead{arya@aero.iitb.ac.in}
\affiliation[rmv]{organization={Indian Institute of Technology Bombay},
            addressline={Powai}, 
            city={Mumbai},
            postcode={400076}, 
            state={Maharashtra},
            country={India}}
\begin{abstract}
This work addresses the problem of constrained motion control of the uncrewed surface vessels. 
The constraints are imposed on states/inputs of the vehicles due to the physical limitations, mission requirements, and safety considerations.
We develop a nonlinear feedback controller utilizing log-type Barrier Lyapunov Functions to enforce static and dynamic motion constraints.
The proposed scheme uniquely addresses asymmetric constraints on position and heading alongside symmetric constraints on surge, sway, and yaw rates. Additionally, a smooth input saturation model is incorporated in the design to guarantee stability even under actuator bounds, which, if unaccounted for, can lead to severe performance degradation and poor tracking. Rigorous Lyapunov stability analysis shows that the closed-loop system remains stable and that all state variables remain within their prescribed bounds at all times, provided the initial conditions also lie within those bounds. Numerical simulations demonstrate the effectiveness of the proposed strategies for surface vessels without violating the motion and actuator constraints. 
\end{abstract}
\begin{keyword}
Uncrewed surface vessel \sep Asymmetric input saturation \sep Trajectory tracking\sep Marine vehicles \sep Motion constraints.
\end{keyword}

\end{frontmatter}
\section{Introduction}\label{sec:introduction}
In recent years, the uncrewed surface vessels (USVs) have become an indispensable asset in the maritime domain. USVs are more suitable for performing various critical tasks. Their applications range from scientific exploration to defense and industrial monitoring. 
USVs are being used as a primary tool for oceanography and meteorology, collecting real-time data on bathymetry, salinity, and climate change across remote regions. The USVs can be conveniently deployed to perform repetitive or monotonous tasks in high-risk environments. The ability to operate autonomously on the surface of water without a human crew further adds to its utility for solving maritime-related problems. The offshore energy sector utilizes USVs for offshore infrastructure inspections, wind farm monitoring, and sea mining. Furthermore, the USVs are used as support ships for other underwater vehicles, which may involve coordination among multiple vehicles to achieve the mission objectives. 

Despite numerous favorable use cases, adopting USVs for such applications presents significant challenges. The major issues requiring attention are actuator saturation, uncertainty in modeling system dynamics, and vulnerability to environmental disturbances arising from winds, ocean waves, and currents \cite{zhao2014,Neto2021_paramEstim,yang2014_dyamics}. An additional challenge is to enforce motion constraints imposed due to safety or mission challenges, such as navigating in narrow waterways. While numerous works in literature addressed the trajectory tracking problem under parametric uncertainties and environmental disturbances  \cite{zhao2014,Neto2021_paramEstim,yang2014_dyamics}, what remains often ignored is the consideration of control input bounds in the controller design itself. Real-world actuators have physical bounds on the force/torque they can produce, and the speed at which they can move. Ignoring these restrictions may drastically affect the tracking performance. The problem becomes even more challenging when motion constraints must be enforced with bounded control authority, which is precisely what this work addresses. Furthermore, unlike aerial/ground vehicles, the dynamics of the USVs are influenced by significant Coriolis, added mass, and damping components, which further adds to the complexity.

The USVs are equipped with thrusters and control fins capable of changing speed and heading to follow a prespecified trajectory  \cite{10.1016/j.arcontrol.2022.07.001}. However, due to space or physical limitations, the vehicle's actuation system can only render bounded control authority. The actuator bounds are typically known, and active consideration in the design can enhance performance. The other challenge comes from mission objectives and a safety perspective. Often, vehicles need to pass through physically constrained spaces, such as ponds, canals, rivers, and lakes. Further, if multiple vehicles are in the vicinity, there will be a requirement to design suitable control laws that comply with the imposed hard constraints to prevent possible collisions among different vehicles. It is extremely crucial for USV to adhere to the motion constraints while simultaneously performing its goals. Motion constraints can be both static and dynamic. For missions being performed in a pond, the USV should not hit the boundaries. Since the boundaries are fixed in time, it is a case of static constraints. The rivers or canals often take a curvy form, which can be thought of as dynamic constraints on the USV states as it navigates in the waterways. The nature of constraints can also be symmetric or asymmetric. 

Numerous trajectory tracking approaches, including artificial potential field-based, moving horizon optimal control, reference governor, and model predictive control-based methods, were employed to respect the imposed constraints \cite{zhao2014,sun2018robust}. For nonlinear systems, barrier Lyapunov functions (BLFs) \cite{ngo2005,kpt2009} and control barrier functions (CBFs) \cite{10.1016/j.arcontrol.2024.100945} are popular choices for developing safety-critical controllers. The authors in \cite{10.1109/JOE.2024.3423869} employed relaxed CBF to follow a trajectory while avoiding a moving target. While our goal is to address motion constraints imposed by boundaries, we explored BLFs. Different BLFs used in literature were log-type BLF \cite{zhao2017}, tan-type BLF \cite{QIN2020}, universal BLF \cite{10.1016/j.oceaneng.2025.122344}, and integral BLF \cite{kpt2012_iBLF,wang2019}, etc. The function value of BLF becomes unbounded when its argument approaches some pre-specified values. This property is leveraged to enforce constraints while designing stable controllers. 
The authors in \cite{kpt2009} proposed a control design technique for a general class of single-input, single-output (SISO) nonlinear systems in strict feedback form while imposing asymmetric static constraints on the output states. Building upon this, to deal with time-varying constraints, the authors in \cite{kpt2009_time_varying} proposed an asymmetric time-varying BLF approach for addressing output constraints of time-varying nature.
We also leverage this approach into our design. However, the dynamics of the USV are characterized as a multiple-input and multiple-output (MIMO) system. 
In \cite{zhao2017}, the authors present a BLF-based controller design using the Moore-Penrose inverse. However, they employ only symmetric constraints on the USV states, which is a special case of the asymmetric constraints case. Additionally, they considered quadratic and log-type BLFs to constrain the velocity states of the USV, assuming the control input was available for the BLF-based term and designed using the quadratic term, which is inappropriate, as the control input is what is being designed. 
 
In \cite{QIN2020}, the authors adopted tan-type BLF to address the error constraints in the presence of input saturation. They approximated the input saturation using smooth hyperbolic tangent functions. In \cite{li2021}, the author first transformed the system dynamics into two subsystem dynamics. Then, by using a tan-type BLF, they demonstrated the trajectory tracking while avoiding the constraints. However, they only handle static constraints, and it is worth noting that using a tan-type BLF makes the controller more complex. The authors in \cite{10.1016/j.oceaneng.2025.122344} employed a universal BLF to address asymmetric dynamic constraints and proposed an auxiliary dynamic system to handle input saturation. However, they only entertain symmetric bounds of the actuators. In real-world applications, the capabilities of actuators in {positive and negative} directions may differ. This can be the case due to economic decisions about selecting the actuators or due to wear and tear. In fact, most of the existing works \cite{10.1016/j.oceaneng.2025.121720,10.1016/j.oceaneng.2025.122472,10.1016/j.oceaneng.2024.118626,10.1016/j.oceaneng.2024.116879,10.1016/j.oceaneng.2025.122344,10.1016/j.oceaneng.2025.121021,10.2514/6.2025-1903} address the case of symmetric input saturation. In \cite{10.1016/j.oceaneng.2024.118207}, the authors present a high-order barrier function-based controller to handle time-varying constraints. But they handle only symmetric dynamic constraints, which is a special case of asymmetric constraints. Also, they did not account for actuator saturation in the design.  

As evident from the above discussion, many works often overlook input saturation in the design, instead enforcing saturation bounds at the maximum or minimum values when the control input demand exceeds the bounds \cite{10.1016/j.oceaneng.2024.118207,10.1109/JOE.2025.3529062,10.1016/j.oceaneng.2024.117622}. Other commonly employed approaches involve approximating the input saturation using functions such as the hyperbolic tangent, sigmoid \cite{10.1016/j.oceaneng.2024.118626}, and the Nussbaum function \cite{zheng2018}. However, these only employ the symmetric control input bounds. Also, these approximations of the input saturation nonlinearity result in a non-affine form in the model input, which complicates the controller design. 

In light of the aforementioned challenges and discussions, the key contributions of this work are outlined below: 
\begin{itemize}
    \item A BLF-based nonlinear feedback controller is proposed for USVs that enables it to achieve accurate trajectory tracking while strictly adhering to the static and dynamic state constraints. Unlike in \cite{zhao2017}, this work considers asymmetric state constraints on the USV's position and heading, which is more general and close to practice. For example, the USV might need to navigate through narrow, curvy rivers, which requires adherence to time-varying constraints for safe travel.
    \item The proposed work also accounts for the actuator saturation in the controller design by augmenting a smooth input saturation model with the USV dynamics. Unlike applying a saturation block as in \cite{10.1016/j.oceaneng.2024.118207,10.1109/JOE.2025.3529062,10.1016/j.oceaneng.2024.117622}, the proposed design ensures that the control demand never exceeds the bounds while guaranteeing the stability of the overall system. At the same time, the control demand profile becomes smoother, leading to safer operation and longer actuator life. 
    \item In comparison to \cite{QIN2020,10.1016/j.oceaneng.2025.122344}, this work accounts for asymmetric actuator saturation constraints in the design, which is more general and closer to a representative of real scenarios, as the actuators' characteristics may be different in forward and reverse directions. 
    \item Owing to the nonlinear framework used for deriving control inputs, it remains applicable even for situations where the initial deviations of states are significantly large, and the control methods based on linearization may not be adequate. It is also ensured that all system states always remain bounded during the vehicle's motion, provided the initial conditions lie within the bounds. 
\end{itemize} 
The performance of proposed design are validated using numerical simulations against various engagement scenarios, and results are found to be promising. 

\section{Problem Formulation}\label{sec: Problem}
This section first details about the kinematics and dynamics of the USVs and then subsequently formulate the main problem addressed in this paper. Following which a brief discussion on BLFs and input saturation model is also included. As the USVs operate on the water's surface, it is common practice to model the dynamics of the USV as 3 degrees of freedom (3-DOF), namely surge, sway, and yaw. The ignored DOFs are heave, roll, and pitch dynamics. The USVs are made metacentrically stable with small amplitudes of roll, pitch angles, and their rates. Therefore, for all practical purposes, the roll and pitch dynamics can be safely neglected. Similarly, the dynamics in the heave direction can also be neglected as the vessel is floating on the water surface with zero mean.  
\begin{figure}[ht]
    \centering
    \begin{subfigure}[b]{0.4\textwidth}
        \centering
        \includegraphics[width=\textwidth]{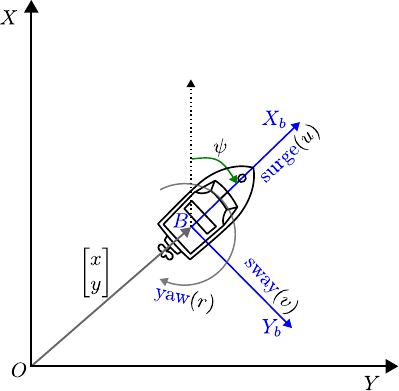}
        \caption{Planar motion of the USV.}
        \label{fig:a_usv_frame}
    \end{subfigure}
    \hfill
    \begin{subfigure}[b]{0.4\textwidth}
        \centering
        \includegraphics[width=\textwidth]{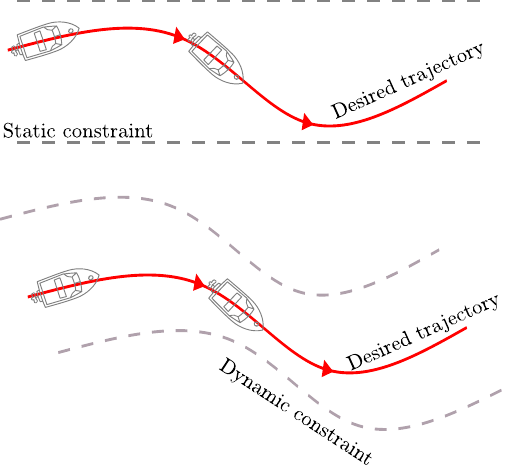}
        \caption{Desired trajectory with static and dynamic constraints.}
        \label{fig:usv_with_pos_constraint}
    \end{subfigure}
    \caption{Planar motion of the USV under static and dynamic motion constraints.}
    \label{fig:main}
\end{figure}

To describe the equation of motion of the USVs, we use the Earth-fixed frame $(OXY)$ and body-fixed frame $(BX_{b}Y_{b})$. As illustrated in \Cref{fig:a_usv_frame}, the points $O$ and $B$ denote the origin of the respective coordinate frames. The axes $OX$ and $OY$ are pointing towards the North and the East, respectively. The vector $BX_b$ is pointing from the aft to the fore and $BY_b$ is towards the starboard side. The position of the USV in the Earth-fixed frame is denoted with $(x,\,y)$ and the heading (measured from North; see \Cref{fig:a_usv_frame}) is denoted with $\psi$. Together, they are represented as vector $\pmb{\eta} = [x~~ y~~ \psi]^\top$. Similarly, the vector $\pmb{\nu} = [u~~ v~~ r ]^\top$ denotes the velocity vector components of the USV in the body-fixed frame. The variables $u$, $v$, and $r$ denote the linear velocity in the surge, the linear velocity in sway, and the yaw rate of the USV, respectively.
The matrix $\pmb{J}$, which relates $\dot{\pmb{\eta}}$ and $\pmb{\nu}$, is given by
\begin{align}
    \pmb{J}(\psi) = \begin{bmatrix}
        \cos\psi & -\sin \psi &0\\
        \sin\psi & \cos\psi & 0 \\
        0 & 0 & 1
    \end{bmatrix}.
\end{align}
The 3-DOF non-linear equation of motion of the USV is expressed as \cite{fossen2021handbook}
\begin{align}\label{eqn: kinematics}
    \dot{\pmb{\eta}}  &= \pmb{J}(\psi) \pmb{\nu},
\\
\label{eqn: dynamics}
    \dot{\pmb{\nu}} &= \pmb{M}^{-1}\left[\pmb{\tau} - \pmb{C}(\pmb{\nu})\pmb{\nu} - \pmb{D}(\pmb{\nu})\pmb{\nu}  + \pmb{b}\right],
\end{align}
where $\pmb{\tau} = [\tau_1~~\tau_2~~ \tau_3]^\top=[\tau_u~~ \tau_v~~ \tau_r]^\top$ is the control input in the body-fixed frame, $\pmb{M}\in \mathbb{R}^{3\times 3}$ is a positive definite mass-inertia matrix, $\pmb{C}\in \mathbb{R}^{3\times 3}$ is the Coriolis matrix, $\pmb{D}\in \mathbb{R}^{3\times 3}$ is the damping matrix, and $\pmb{b}\in \mathbb{R}^{3}$ is the disturbance vector. Note that $\pmb{M}$ and $\pmb{C}$ contains the added components due to the acceleration of the USV. The expressions for these matrices can be written as 
\begin{align}\label{eqn: M}
 \pmb{M} &= 
\begin{bmatrix}
    m_{11} &0 &0\\
    0 &m_{22} &m_{23} \\
    0  &m_{32} &m_{33}
\end{bmatrix}, \quad 
 \pmb{C}(\pmb{\nu}) = 
\begin{bmatrix}
    0 &0 &-m_{22}v-m_{23}r\\
    0 &0 &m_{11}u \\
    m_{22}v+m_{23}r  &-m_{11}u &0
\end{bmatrix},\\
\label{eqn: D}
 \pmb{D}(\pmb{\nu}) &= 
\begin{bmatrix}
    d_{11}(u) &0 &0\\
    0 &d_{22}(v,r) &d_{23}(v,r) \\
    0  &d_{32}(v,r) &d_{33}(v,r)
\end{bmatrix}. 
\end{align}
The expressions for the matrix elements used in \Cref{eqn: M} and \Cref{eqn: D} are given by
\begin{subequations}
  \begin{align}
    m_{11} &= \text{m} - X_{\dot{u}},
    m_{22} = \text{m} -  Y_{\dot{v}},
    m_{23} = \text{m}x_g -  Y_{\dot{r}},
    m_{32} = \text{m}x_g -  N_{\dot{v}},\\
    m_{33} &= I_z -  N_{\dot{r}},
    d_{11}(u) = -X_u - X_{|u|u}|u|,\\
    d_{22}(v,r) &= -Y_v - Y_{|v|v}|v| - Y_{|r|v}|r|, \quad
    d_{23}(v,r) = -Y_r - Y_{|v|r}|v| - Y_{|r|r}|r|,\\
    d_{32}(v,r) &= -N_v -N_{|v|v}|v| - N_{|r|v}|r|, \quad
    d_{33}(v,r) = -N_r - N_{|v|r}|v| - N_{|r|r}|r|,
\end{align}  
\end{subequations}
where $m$ is the mass of the USV, $x_g$ is the distance of the geometric center of the USV from the center of gravity (CG), and $I_z$  is the yaw mass moment of inertia of the USV. The other symbols are related to hydrodynamic derivatives and have the usual meanings, as per the SNAME convention \cite{SNAME1950nomenclature}. 

Let the desired trajectory be represented by $\pmb{\eta}_d(t) =[x_d(t)~~  y_d(t)~~  \psi_d(t)]^\top$ and the desired surge, sway, and yaw rates are denoted by $\pmb{\nu}_d(t) = [u_d(t)~~ v_d(t)~~ r_d(t)]^\top = \pmb{J}^\top(\psi_d)\dot{\pmb{\eta}}_d(t)$. 

\begin{assumption}\label{ass:static}
    The desired trajectory and its time derivatives are bounded. That is, for any $k_{ci}>0~~ (i\in \{x,y,\psi\})$, there exist positive constants $\bar{\lambda}_i$, $\underline{\lambda}_i$ , $\Lambda_{0i}$, $\Lambda_{1i}$, $k_{\nu i}$ and they satisfy the relation $\max \{\underline{\lambda}_i,~\bar{\lambda}_i\} \leq \Lambda_{0i} < k_{ci}$ such that $-\underline{\lambda}_i \leq \eta_{di} \leq \bar{\lambda}_i$, and $|\nu_{di}| <k_{\nu i}$. The time derivatives of the desired trajectory also satisfy $\dot{\eta}_{di}\leq \Lambda_{1i}$ and $\ddot{\eta}_{di}\leq \Lambda_{1i}$.
\end{assumption}
\begin{assumption}\label{ass:dyn}
 The USV is fully actuated, and the system dynamics are known. Also, the external disturbances are absent. 
\end{assumption}
\begin{remark}
    Note that $||\pmb{J}(\psi)||$ and $||\pmb{M}^{-1}||$ are bounded and the system dynamics are in a strict feedback form.
\end{remark}
The assumptions about fully actuated is required in order to control all the three DOFs independently. Also, even if the system dynamics are not known accurately then different techniques (see \cite{kpt2009}) can be used to estimate the model parameters. Taking note of these assumptions, we now state the problem statement addressed in this work. 

\begin{problem}[Static constraints] \label{prob:static}
Consider the USV dynamics governed by \Cref{eqn: kinematics} and ~\Cref{eqn: dynamics} under the Assumptions \ref{ass:static} and \ref{ass:dyn}. The objective is to design a nonlinear feedback controller to steer the USV to track a specified smooth trajectory without violating the state constraints (static) while also ensuring that the control input demand never exceeds their bounds. At the same time, all closed-loop signals should remain bounded.  
Mathematically, the objective is to design control input $\pmb{\tau}$ such that $(\pmb{\eta}(t)-\pmb{\eta}_d(t)) \rightarrow \pmb{0}$ as $t\rightarrow \infty$ while ensuring $\eta_i(t) - \eta_{di}(t) \in (-k_{ai},k_{bi})$, $|\nu_i|<k_{\nu i}$, and $|\tau_{i}|<\tau_{i,\text{M}}$ for all $t\geq 0$ and $i\in\{x,y,\psi\}$, provided the initial conditions lie within the bounds, that is $\eta_i(0)\in(-k_{ci}, k_{ci})$. In this case, $k_{ai} = k_{ci}-\underline{\lambda}_i$, $k_{bi} = k_{ci}- \bar{\lambda}_i$, and $\tau_{i,\text{M}}$ is the actuator bound.
\end{problem}
\begin{assumption}\label{ass:tv}
    The desired trajectory is bounded by functions of time. That is, there exist functions $\underline{\lambda}_i:\mathbb{R}_+ \rightarrow \mathbb{R}_+$ and $\bar{\lambda}_i : \mathbb{R}_+ \rightarrow \mathbb{R}_+$ satisfying $\bar{\lambda}_i(t) < \bar{k}_{ci}(t)$ and $\underline{\lambda}_i(t) > \underline{k}_{ci}(t)$ for all $t\geq 0$ such that $-\underline{\lambda}_i(t) \leq \eta_{di}(t) \leq \bar{\lambda}_i(t)$. The time derivatives of the desired trajectory also satisfy $\dot{\eta}_{di}(t)\leq \Lambda_{1i}$ and $\ddot{\eta}_{di}(t)\leq \Lambda_{1i}$ $\forall ~t \geq 0$.
\end{assumption}
\begin{problem}[Dynamic constraints] \label{prob:tv}
    Given the USV dynamics as in \Cref{eqn: kinematics} and ~\Cref{eqn: dynamics} under the Assumptions \ref{ass:dyn} and \ref{ass:tv}, the aim is to make the USV track a specified smooth trajectory while adhering to asymmetric time-varying constraints on position and heading and static constraints on surge, sway and yaw rates while accounting for availability of bounded control authority. Mathematically, the aim is to  design control input $\pmb{\tau}$ such that $(\pmb{\eta}(t)-\pmb{\eta}_d(t)) \rightarrow \pmb{0}$ as $t\rightarrow \infty$ while ensuring $\eta_i(t) - \eta_{di}(t) \in (-k_{ai}(t),k_{bi}(t))$, and $|\tau_{i}|<\tau_{i,\text{M}}$ for all $t\geq 0$ and $i\in\{x,y,\psi\}$, provided the initial conditions lie within the bounds, that is $\eta_i(0)\in(-k_{ci}(t), k_{ci}(t))$. In this case, $k_{ai}(t) = \eta_{di}(t) - \underline{k}_{ci}(t)$, $k_{bi}(t) = \bar{k}_{ci}(t)-\eta_{di}(t)$ and $\tau_{i,\text{M}}$ is the actuator bound.
\end{problem}

Preceding the main results, we briefly present the preliminaries that aids to the design of controller later. 

\begin{lemma}\cite{kpt2009}\label{lem1}
 For any positive constants $k_{a_1}, k_{b_1}$,
 let $Z_1 := \left\{z_1 \in \mathbb{R}: -k_{a_1}<z_1 <k_{b_1}\right\}\subset \mathbb{R}$ and $\mathbb{N} := \mathbb{R}^l \times \mathbb{Z}_1 \subset \mathbb{R}^{l+1}$ be open sets. Consider the system $\dot \eta= h(t,\eta),$ where $\eta := [w,~~z_1]^\top\in \mathbb{N}$, and $h : \mathbb{R}_+ \times \mathbb{N} \to \mathbb{R}^{l+1}$ is piece-wise continuous in $t$ and locally Lipschitz in $\eta$, uniformly in $t$, on $\mathbb{R}_+ \times \mathbb{N}$. Suppose that there exist functions $U: \mathbb{R}^l \to \mathbb{R}_+$ and $\mathcal{V}_1: Z_1 \to \mathbb{R}_+$, continuously differentiable and positive definite in their respective domains, such that $V_1(z_1)\to\infty~\text{as}~ z_1\to -k_{a_1}~\text{or}~z_1\to k_{b_1},$ and $\gamma_1(||w||)\le \mathcal{U}(w)\le \gamma_2(||w||),$ where $\gamma_1$ and $\gamma_2$ are class $K_{\infty}$ functions. Let $\mathcal{V}(\eta) := \mathcal{V}_1(z_1) + \mathcal{U}(w)$, and $z_1(0)$ belong to the set $z_1 \in (-k_{a_1}, k_{b_1})$. If the inequality holds: $\dot{\mathcal{V}} = -\dfrac{\partial\mathcal{V}}{\partial \eta}h\le 0,$ then $z_1(t)$ remains in the open set $z_1 \in (-k_{a_1}, k_{b_1})~\forall\,\,t \,\in [0,\infty)$.
\end{lemma}

\begin{lemma}\cite{ren2010}\label{lem2}
For all $|\zeta|<1$ and any positive integer $p$, the following inequality holds  
\begin{align}
\ln\left(\dfrac{1}{1-\zeta^{2p}}\right) <\dfrac{\zeta^{2p}}{1-\zeta^{2p}}.
\end{align}
\end{lemma}

\subsection{Input saturation model with asymmetric bounds}
Real-world systems often face actuator saturation. In many cases, their capabilities differ in forward and reverse directions, such as the forward and reverse thrust of a USV. While \cite{10.1049/iet-cta.2016.1097} uses a symmetric model with uniform bounds, our previous work \cite{verma2025trajectory} extends this approach to account for asymmetric saturation as follows:
\begin{equation}\label{eqn:actuator}
    \dot{\pmb{\zeta}} = \pmb{Q}(\pmb{I} - \pmb{G}_M) \pmb{\tau}_c - \pmb{\rho} \pmb{\zeta} + (\pmb{I}-\pmb{Q})(\pmb{I} - \pmb{G}_m) \pmb{\tau}_c, \quad \pmb{\zeta}(0) = \pmb{0}, \quad \pmb{\tau} = \pmb{\zeta},
\end{equation}
where  $\pmb{\tau}_{c}$, $\pmb{\tau}$, and $\pmb{\zeta} \in \mathbb{R}^3$ are the input, the output, and the state of the actuator saturation model, respectively. The term $\pmb{\rho} = \diag [\rho_1~~ \rho_2 ~~ \rho_3] $ where $\rho_1, \rho_2, \rho_3 > 0 $ are the design constants. The matrices $\pmb{I},\pmb{G}_M, \pmb{G}_m \in \mathbb{R}^{3 \times 3} $, where $\pmb{I}$ is the identity matrix, $\pmb{G}_M=\diag [ g_{1M} ~~ g_{2M} ~~ g_{3M}]$ where the terms $g_{iM} = \left(\dfrac{\tau_i}{\tau_{iM}}\right)^n$  $\forall~i=\{1,2,3\}$, and $\pmb{G}_m=\diag [g_{1m} ~~ g_{2m} ~~ g_{3m}]$ where the terms $g_{im} = \left(\dfrac{\tau_i}{\tau_{im}}\right)^n$  $\forall~i=\{1,2,3\}$ and $n\geq 2$ being an even integer. Here, $\tau_{iM}$ and $\tau_{im}$ are the actuator limits in forward and reverse directions. The matrix $\pmb{Q} = \diag \begin{bmatrix}
    q_1 & q_2& q_3
\end{bmatrix}$ helps in enforcing the asymmetric actuator saturation bounds, where 
\begin{equation}\label{eqn:q_zeta}
    q_i (\zeta_i) = \begin{cases}
    1, & \text{if $\zeta_i$} > 0\\
    0, & \text{$\zeta_i$} \leq 0
\end{cases}.
\end{equation}
The actuator saturation bounds in positive and negative directions are $\tau_{iM}$ and $\tau_{im}$, respectively. The input saturation model in \Cref{eqn:actuator} ensures that $\pmb{\tau}$ satisfies the actuator bounds while $\pmb{\tau}_c$ can exceed the actuator bounds but is bounded by some finite positive value (see \cite{verma2025trajectory} for the detailed proof). 

\subsection{Barrier Lyapunov function}
BLFs are used to design nonlinear controllers while accounting for constraints on output and state. The function value grows to infinity as its argument nears some pre-specified values. In literature, the most commonly used BLFs are either log-type or tan-type. The log-type and tan-type asymmetric BLFs can be respectively expressed as 
\begin{align*}
    \mathcal{V} &= \dfrac{q(e)}{2} \ln \left[\dfrac{k_b^2}{k_b^2-e^2}\right] + \dfrac{1-q(e)}{2} \ln \left[\dfrac{k_a^2}{k_a^2-e^2}\right], \\
     \mathcal{V} &= q(e)\dfrac{k_a^2}{\pi} \tan \left[\dfrac{\pi e^2}{2k_a^2}\right] + (1-q(e))\dfrac{k_b^2}{\pi} \tan \left[\dfrac{\pi e^2}{2k_b^2}\right],
\end{align*}
where the term $q(e)$ that takes discrete values is defined as 
\begin{align*}
q(e)= 
\begin{cases}
   1,&  e> 0\\
    0,  & e \leq 0
\end{cases}.    
\end{align*}   
With the problems defined and preliminaries in place, we now discuss our approach and proposed methodology to tackle the problem.

\section{Main Results}\label{sec:main results}
In this section, we present the proposed control strategies to tackle the objectives mentioned in Problems \ref{prob:static} and \ref{prob:tv}. From a safety and control perspective, it is important to account for static and dynamic constraints while developing control strategies. For example, a USV navigating in a river should not hit the banks, which becomes a case of dynamic constraint as the river takes turns. Furthermore, a USV performing its mission in a pond becomes a case of static constraint. To enforce static as well as dynamic constraints on the USV states, we employ a BLF-based control strategy. Additionally, it is equally crucial to develop strategies aware of the bounds on the actuators. In the following subsections, we tackle these challenges systematically. First, we present the control strategy for the motion of the USV under static constraint with bounded control authority. Following this, we divide the dynamic constraint case into two parts: without employing the input saturation model and with the input saturation model.

\subsection{Control under static state constraints and input saturation} \label{sec:st_ip}
Let the tracking error in position and heading be denoted by 
\begin{equation}
\pmb{e_1} = \pmb{\eta} - \pmb{\eta_d} = [
    e_{11}~~ e_{12}~~ e_{13}]^\top 
= [x-x_d~~ ~ y-y_d~~ ~ \psi-\psi_d]^\top,  \label{eqn:e1}   
\end{equation}
 and the errors in the surge, sway, and yaw rates as 
 \begin{align}
 \pmb{e_2} = \pmb{\nu} - \pmb{\nu_d} = [e_{21}~ ~e_{22}~ ~e_{23}]^\top = [u-u_d~ ~ v-v_d~  ~r-r_d]^\top.
 \end{align} 
To constrain $\pmb{e}_1$ into an asymmetric set $(-k_{ai},k_{bi})$, consider the Lyapunov function candidate given by
\begin{align}
  \mathcal{V}_1 =& \dfrac{1}{2} \sum_{i=1}^{3} \left[ q(e_{1i}) \ln\left(\dfrac{k_{bi}^2}{k_{bi}^2 - e_{1i}^2}\right) + \left(1-q(e_{1i})\right) \ln\left(\dfrac{k_{ai}^2}{k_{ai}^2 - e_{1i}^2}\right) \right]. \label{eqn:ablf_st}
\end{align}
On differentiating the chosen asymmetric BLF candidate from \Cref{eqn:ablf_st} with respect to time, one can obtain
\begin{equation}\label{eqn: v1dot}
    \dot{\mathcal{V}}_1 = \sum_{i=1}^{3} \left [q(e_{1i})\dfrac{e_{1i} \dot{e}_{1i}}{k_{bi}^2-e_{1i}^2} + (1-q(e_{1i}))\dfrac{e_{1i} \dot{e}_{1i}}{k_{ai}^2-e_{1i}^2} \right].
\end{equation}
Differentiating \Cref{eqn:e1} with respect to time and using \Cref{eqn: kinematics} yields $\dot{\pmb{e}}_1$ as
\begin{equation}\label{eqn:e1_dot}
    \dot{\pmb{e}}_1 = \pmb{J}(\pmb{\eta}) \pmb{\nu} - \dot{\pmb{\eta}}_d.
\end{equation}
In order to design the controller using the backstepping technique, let us define a tracking error $\pmb{z}:= \pmb{\nu} -\pmb{\alpha} = [z_{11}~~z_{12}~~z_{13}]^\top$, where \(\pmb{\alpha}\) is the stabilizing function to be designed. Using this, \Cref{eqn:e1_dot} can be rewritten as 
\begin{equation}\label{eqn:e1dot1}
    \dot{\pmb{e}}_1 = J(\pmb{\eta}) (\pmb{z} + \pmb{\alpha}) - \dot{\pmb{\eta}}_d.
\end{equation}
On writing $\dot{\pmb{e}}_1$ from  \Cref{eqn:e1dot1} in component-wise form, we get
\begin{equation}
    \dot{{e}}_{1i} = J_{i}(\pmb{\eta}) (\pmb{z} + \pmb{\alpha}) - \dot{{\eta}}_{di},\label{eqn:e1doti}
\end{equation}
where $\dot{{e}}_{1i}$, $J_{i}(\pmb{\eta})$, and $\dot{{\eta}}_{di}$ represent the $i^{\rm th}$ component of $\dot{\pmb{e}}_1$, the $i^{\rm th}$ row of matrix $J_{i}(\pmb{\eta})$ and $i^{\rm th}$ component of $\dot{\pmb{\eta}}_d$, respectively. On substituting \Cref{eqn:e1doti} into \Cref{eqn: v1dot}, we get
\begin{equation}
    \dot{\mathcal{V}}_1 = \sum_{i=1}^{3} \left[q(e_{1i})\dfrac{e_{1i} \left(J_{i}(\pmb{\eta}) (\pmb{z} + \pmb{\alpha}) - \dot{{\eta}}_{di}\right)}{k_{bi}^2-e_{1i}^2} 
    + (1-q(e_{1i}))\dfrac{e_{1i} \left(J_{i}(\pmb{\eta}) (\pmb{z} + \pmb{\alpha}) - \dot{{\eta}}_{di}\right)}{k_{ai}^2-e_{1i}^2} \right].\label{eqn:V1dot_1}
\end{equation}
Now, by designing the stabilizing function, $\pmb{\alpha}$, as 
\begin{equation}\label{eqn: alpha}
    \pmb{\alpha} = \pmb{J}(\pmb{\eta})^\top \left( \dot{\pmb{\eta}}_d - 
                                \begin{bmatrix}
        k_{11}e_{11}\left\{q(e_{11})(k_{b1}^2-e_{11}^2) + (1-q(e_{11}))(k_{a1}^2-e_{11}^2) \right\}\\
        k_{12}e_{12}\left\{q(e_{12})(k_{b2}^2-e_{12}^2) + (1-q(e_{12}))(k_{a2}^2-e_{12}^2)\right\}\\
        k_{13}e_{13}\left\{q(e_{13})(k_{b3}^2-e_{13}^2) + (1-q(e_{13}))(k_{a3}^2-e_{13}^2)\right\}
                                \end{bmatrix}
                        \right),
\end{equation} 
and on substituting this expression of $\pmb{\alpha}$ from \Cref{eqn: alpha} into \Cref{eqn:V1dot_1} results in
\begin{align}
    \dot{\mathcal{V}}_1 = -\sum_{i=1}^{3} k_{1i}e_{1i}^2 + 
    \sum_{i=1}^{3} \left[ 
    \dfrac{q(e_{1i})}{k_{ai}^2-e_{1i}^2}e_{1i}J_i(\pmb{\eta})\pmb{z} + 
    \dfrac{(1-q(e_{1i}))}{k_{bi}^2-e_{1i}^2}e_{1i}J_i(\pmb{\eta})\pmb{z}
    \right], \label{eqn:V1dot_2}
\end{align}
where $J_i(\pmb{\eta})$ represents the $i^{\rm th}$ row of the $\pmb{J}$ matrix. 
To constrain the error $\pmb{e}_2$ within the bound on individual components $(|e_{2i}|<k_{2i})$, consider another Lyapunov function candidate as  
\begin{equation}
    \mathcal{V}_2 = \mathcal{V}_1 + \dfrac{1}{2}\sum_{i=1}^{3}\ln \dfrac{k_{2i}^2}{k_{2i}^2-z_{1i}^2}
    ,
\end{equation}
where $k_{2i}> 0$ for $i=\{1,2,3\}$. On differentiating $\mathcal{V}_2$ with respect to time and using \Cref{eqn:V1dot_2}, we get
\begin{align}
    \nonumber \dot{\mathcal{V}}_2 &= \dot{\mathcal{V}}_1 + \sum_{i=1}^{3}\dfrac{z_{1i} \dot{z}_{1i}}{k_{2i}^2-z_{1i}^2} 
    \\
    &= -\sum_{i=1}^{3} k_{1i}e_{1i}^2 + 
    \sum_{i=1}^{3} \left[ 
    \dfrac{q(e_{1i})}{k_{bi}^2-e_{1i}^2}e_{1i}J_i(\pmb{\eta})\pmb{z} + 
    \dfrac{(1-q(e_{1i}))}{k_{ai}^2-e_{1i}^2}e_{1i}J_i(\pmb{\eta})\pmb{z}
    \right] + \sum_{i=1}^{3}\dfrac{z_{1i} \dot{z}_{1i}}{k_{2i}^2-z_{1i}^2}. \label{eqn:V2_dot_ip}
\end{align}
Now, let us define another intermediate variable $\pmb{z}_2 = \pmb{\tau} - \pmb{\alpha}_2$, where $\pmb{\alpha}_2$ is stabilizing function to be designed next. On differentiating variable $\pmb{z}$ with respect to time and using \Cref{eqn: dynamics}, one can obtain
\begin{align}
    \dot{\pmb{z}} = \dot{\pmb{\nu}} - \dot{\pmb{\alpha}}  &= \pmb{M}^{-1}\left[\pmb{\tau} - \pmb{C}(\pmb{\nu})\pmb{\nu} - \pmb{D}(\pmb{\nu})\pmb{\nu} \right] - \dot{\pmb{\alpha}} =\pmb{M}^{-1}\left[\pmb{z}_2 + \pmb{\alpha}_2 - \pmb{C}(\pmb{\nu})\pmb{\nu} - \pmb{D}(\pmb{\nu})\pmb{\nu} \right] - \dot{\pmb{\alpha}}.
    \label{eqn:z_dot_ip}
\end{align}
The component of $\dot{\pmb{z}}$ can be written as 
\begin{align}
    \dot{z}_{1i} =\pmb{M}^{-1}_i\left[\pmb{z}_2 + \pmb{\alpha}_2 - \pmb{C}(\pmb{\nu})\pmb{\nu} - \pmb{D}(\pmb{\nu})\pmb{\nu} \right] - \dot{{\alpha}}_{1i}, \label{eqn:z1i_dot_ip}
\end{align}
where $\pmb{M}^{-1}_i$ represents the $i^\text{th}$ row of $\pmb{M}^{-1}$ matrix.
Substituting \Cref{eqn:z1i_dot_ip} into \Cref{eqn:V2_dot_ip} yields
\begin{align}
     \dot{\mathcal{V}}_2
    =& -\sum_{i=1}^{3} k_{1i}e_{1i}^2 + 
    \sum_{i=1}^{3} \left[ 
    \dfrac{q(e_{1i})}{k_{bi}^2-e_{1i}^2}e_{1i}J_i(\pmb{\eta})\pmb{z} + 
    \dfrac{(1-q(e_{1i}))}{k_{ai}^2-e_{1i}^2}e_{1i}J_i(\pmb{\eta})\pmb{z}
    \right] \\
    &+ \sum_{i=1}^{3}\dfrac{z_{1i} \left(\pmb{M}^{-1}_i\left[\pmb{z}_2 + \pmb{\alpha}_2 - \pmb{C}(\pmb{\nu})\pmb{\nu} - \pmb{D}(\pmb{\nu})\pmb{\nu} \right] - \dot{{\alpha}}_{1i}\right)}{k_{2i}^2-z_{1i}^2}\label{eqn:V2_dot_ip1}
\end{align}
The last term of \Cref{eqn:V2_dot_ip1} can be rewritten into matrix-vector form as 
\begin{align}
    &\sum_{i=1}^{3}\dfrac{z_{1i} \left(\pmb{M}^{-1}_i\left[\pmb{z}_2 + \pmb{\alpha}_2 - \pmb{C}(\pmb{\nu})\pmb{\nu} - \pmb{D}(\pmb{\nu})\pmb{\nu} \right] - \dot{{\alpha}}_{1i}\right)}{k_{2i}^2-z_{1i}^2} \\
    &\quad = \pmb{z}^\top \begin{bmatrix}
        \dfrac{1}{k_{21}^2-z_{11}^2} & 0 & 0 \\
        0 & \dfrac{1}{k_{22}^2-z_{12}^2} & 0 \\
        0 & 0 & \dfrac{1}{k_{23}^2-z_{13}^2} 
    \end{bmatrix} \left(\pmb{M}^{-1}\left[\pmb{z}_2 + \pmb{\alpha}_2 - \pmb{C}(\pmb{\nu})\pmb{\nu} - \pmb{D}(\pmb{\nu})\pmb{\nu} \right] - \dot{\pmb{\alpha}}\right).
\end{align}
Now, we choose the stabilizing function $\pmb{\alpha}_2$ as 
\begin{align}
    \nonumber \pmb{\alpha}_2 =&~ \pmb{C}(\pmb{\nu})\pmb{\nu} + \pmb{D}(\pmb{\nu})\pmb{\nu} +\pmb{M}\dot{\pmb{\alpha}}  - \pmb{M} 
    \begin{bmatrix}
        k_{21}z_{11}\left(k_{21}^2-z_{11}^2\right) \\
         k_{22}z_{12}\left(k_{22}^2-z_{12}^2\right) \\
          k_{23}z_{13}\left(k_{23}^2-z_{13}^2\right)
    \end{bmatrix} \\
    &~ - \pmb{M} \begin{bmatrix}
        {k_{21}^2-z_{11}^2} & 0 & 0 \\
        0 & {k_{22}^2-z_{12}^2} & 0 \\
        0 & 0 & {k_{23}^2-z_{13}^2} 
    \end{bmatrix} \left(\sum_{i=1}^{3} \left[ 
    \dfrac{q(e_{1i})}{k_{bi}^2-e_{1i}^2}e_{1i}\pmb{J}_i^\top(\pmb{\eta}) + 
    \dfrac{(1-q(e_{1i}))}{k_{ai}^2-e_{1i}^2}e_{1i}\pmb{J}_i^\top(\pmb{\eta})
    \right]\right), \label{eqn:alpha2_ip}
\end{align}
where $\pmb{J}_i^\top(\pmb{\eta})\in \mathbb{R}^3$ is a column vector denoting the transpose of the $i^{\text{th}}$ row of the $\pmb{J}(\pmb{\eta})$ matrix. The parameters $k_{2i}>0~\forall~i\in\{1,2,3\}$ denote the symmetric bounds on the body rates, that is, $u,\,v,\,$ and $r$, respectively. On substituting the expression of $\pmb{\alpha}_2$ from \Cref{eqn:alpha2_ip} into the Lyapunov function derivative $\dot{\mathcal{V}}_2$ from \Cref{eqn:V2_dot_ip1}, we get
\begin{align}
    \dot{\mathcal{V}}_2 = -\sum_{i=1}^{3} k_{1i}e_{1i}^2 - \pmb{z}^\top \pmb{K}_2 \pmb{z} + \pmb{z}^\top \begin{bmatrix}
        \dfrac{1}{k_{21}^2-z_{11}^2} & 0 & 0 \\
        0 & \dfrac{1}{k_{22}^2-z_{12}^2} & 0 \\
        0 & 0 & \dfrac{1}{k_{23}^2-z_{13}^2} 
    \end{bmatrix}\pmb{M}^{-1}\pmb{z}_2,
\end{align}
where $\pmb{K}_2 = \diag[k_{21}~~k_{22}~~k_{23}]$.
Next, to cancel out the remaining cross-term, consider another Lyapunov function candidate, $\mathcal{V}_3$ given by
\begin{align}\label{eqn:V3_ip}
    \mathcal{V}_3 = \mathcal{V}_2 + \dfrac{1}{2}\pmb{z}_2^\top \pmb{z}_2.
\end{align}
On taking the time differentiation of \Cref{eqn:V3_ip}, we get
\begin{align}
    \dot{\mathcal{V}}_3 = \dot{\mathcal{V}}_2 + \pmb{z}_2^\top \dot{\pmb{z}}_2. \label{eqn:V3dot_ip}
\end{align}
Differentiating $\pmb{z}_2$ with respect to time and using \Cref{eqn:actuator}, we get 
\begin{align}
     \dot{\pmb{z}}_2 = \dot{\pmb{\tau}} -\dot{\pmb{\alpha}}_2  = \pmb{Q}(\pmb{I} - \pmb{G}_M) \pmb{\tau}_c - \pmb{\rho} \pmb{\zeta} + (\pmb{I}-\pmb{Q})(\pmb{I} - \pmb{G}_m) \pmb{\tau}_c -\dot{\pmb{\alpha}}_2. \label{eqn:z2_dot_ip}
\end{align}
Choosing the control input $\pmb{\tau}_c$ as 
\begin{align}
    \nonumber \pmb{\tau}_c &= \left[\pmb{Q}(\pmb{I} - \pmb{G}_M) + (\pmb{I}-\pmb{Q})(\pmb{I} - \pmb{G}_m) \right]^{-1}\left(\pmb{\rho \tau} + \dot{\pmb{\alpha}}_2\right) \\
    & \quad - \pmb{K}_3\pmb{z}_2 - \pmb{M}^{-\top}\begin{bmatrix}
        \dfrac{1}{k_{21}^2-z_{11}^2} & 0 & 0 \\
        0 & \dfrac{1}{k_{22}^2-z_{12}^2} & 0 \\
        0 & 0 & \dfrac{1}{k_{23}^2-z_{13}^2} 
    \end{bmatrix}\pmb{z}, \label{eqn:tauc_ip}
\end{align}
and using \Cref{eqn:tauc_ip,eqn:z2_dot_ip,eqn:V3dot_ip}, the time derivative of the Lyapunov function candidate reduces to 
\begin{align}
 \dot{\mathcal{V}}_3 & = -\sum_{i=1}^{3} k_{1i}e_{1i}^{2} - \pmb{z}^\top K_2 \pmb{z} - \pmb{z}_2^\top K_3 \pmb{z}_2.
\end{align}
Since the gains $k_{1i}, k_{2i}$, and $k_{3i}$ are positive real numbers, the square of a real number is already positive. Hence, $\dot{\mathcal{V}}_3$ is negative semidefinite, that is, $\dot{\mathcal{V}}_3 \leq 0$.
Therefore, according to the Lemma \ref{lem1}, the errors $e_{1i}(t)$ remains in the open set $e_{1i} \in (-k_{ai}, k_{bi})~\forall\,\,t \,\in [0,\infty)$, provided that $e_{1i}(0) \in (-k_{ai}, k_{bi})$ for $i \in \{1,2,3\}$.  Thus by virtue of the definition, $-k_{ci}<\eta_{i}<k_{ci}$ and also 
$-k_{2i}<e_{2i}<k_{2i}$.

Next, we tackle a more challenging problem of motion control under dynamic state constraints.

\subsection{Control under asymmetric dynamic state constraints}\label{sec:tv}

When the USV navigates in a long canal or a river, its boundaries may have a varying nature. Also, due to USV traffic, the nature of constraints of the states of the USV becomes time-varying. This problem is very critical to handle for safe operation in waterways. In this section, we present a strategy to elegantly handle the time-varying state constraints in the controller design. Let each component of the desired trajectory be bounded by two time-varying functions with $\underline{k}_{c_i}(t)$ and $\bar{k}_{c_i}(t)$ for $i \in \{1,2,3\}$. Therefore, the time-varying barriers can now be defined as 
\begin{align}
    k_{ai}(t) := \eta_{di}(t) - \underline{k}_{ci}(t) \quad  k_{bi}(t) := \bar{k}_{ci}(t) - \eta_{di}(t).
\end{align}
We employ an asymmetric barrier Lyapunov function to address the asymmetric constraints on the position and heading of the USV. Towards this, we consider the Lyapunov function candidate in terms of the error variable $\pmb{e}_1$ as 
\begin{align}
    \mathcal{V}_1 =\sum_{i=1}^3\left[ \dfrac{q(e_{1i})}{2p} \ln \dfrac{k_{bi}^{2p}(t)}{k_{bi}^{2p}(t) - e_{1i}^{2p}} + \dfrac{1 - q(e_{1i})}{2p} \ln \dfrac{k_{ai}^{2p}(t)}{k_{ai}^{2p}(t) - e_{1i}^{2p}}\right], \label{eqn:V1_tv}
\end{align}
where $p$ is a positive integer and satisfies $2p$ greater than equal to the order of the system (that is, $2p\geq 2$). The time-varying constraints on states are difficult to handle directly. This necessitates changing the error variables to transform the problem into one with constant bounds. Let the transformed error variables be denoted as 
\begin{align}
    \varepsilon_{a_i} = \dfrac{e_{1i}}{k_{ai}}, \quad \varepsilon_{b_i} = \dfrac{e_{1i}}{k_{bi}}, \qquad \varepsilon_i = q \varepsilon_{b_i} + (1-q)\varepsilon_{a_i} \forall i \in \{1,2,3\}. \label{eqn:var_change}
\end{align} 
Now, \Cref{eqn:V1_tv} can be rewritten into a form which is not an explicit function of time:
\begin{align*}
    \mathcal{V}_1 =\sum_{i=1}^3 \dfrac{1}{2p} \ln \dfrac{1}{1 - \varepsilon_{i}^{2p}}. 
\end{align*}
On differentiating $\mathcal{V}_1$ with respect to time using the chain rule, one can obtain
\begin{align}
    \nonumber \dot{\mathcal{V}}_1 &= \sum_{i=1}^3 \left[ \dfrac{q \varepsilon_{b_i}^{2p-1}}{(1 - \varepsilon_{b_i}^{2p})} \left( \dfrac{k_{b_i}\dot{{e}}_{1i} - e_{1i}\dot{k}_{b_i}}{k_{b_i}^2} \right) + 
    \dfrac{(1-q) \varepsilon_{a_i}^{2p-1}}{k_{a_i}(1 - \varepsilon_{a_i}^{2p})} \left( \dfrac{k_{a_i}\dot{{e}}_{1i} - e_{1i}\dot{k}_{a_i}}{k_{a_i}^2} \right)
    \right]
    \\
    &=\sum_{i=1}^3 \left[ \dfrac{q \varepsilon_{b_i}^{2p-1}}{k_{b_i}(1 - \varepsilon_{b_i}^{2p})} \left( \dot{{e}}_{1i} - e_{1i} \dfrac{\dot{k}_{b_i}}{k_{b_i}} \right) + 
    \dfrac{(1-q) \varepsilon_{a_i}^{2p-1}}{k_{a_i}(1 - \varepsilon_{a_i}^{2p})} \left( \dot{{e}}_{1i} - e_{1i} \dfrac{\dot{k}_{a_i}}{k_{a_i}} \right)
    \right] \label{eqn:V1_dot_tv1}
\end{align}
To stabilize the dynamics of $\pmb{e}_1$, we now design $\pmb{\alpha}$. On substituting $\dot{{e}}_{1i}$ from \Cref{eqn:e1doti} into \Cref{eqn:V1_dot_tv1}, we get
\begin{align}
    \dot{\mathcal{V}}_1 &=\sum_{i=1}^3 \left[ \dfrac{q \varepsilon_{b_i}^{2p-1}}{k_{b_i}(1 - \varepsilon_{b_i}^{2p})} \left( J_{i}(\pmb{\eta}) (\pmb{z} + \pmb{\alpha}) - \dot{{\eta}}_{di} - e_{1i} \dfrac{\dot{k}_{b_i}}{k_{b_i}} \right) \right. \nonumber \\
    & \qquad \quad \left. + 
    \dfrac{(1-q) \varepsilon_{a_i}^{2p-1}}{k_{a_i}(1 - \varepsilon_{a_i}^{2p})} \left( J_{i}(\pmb{\eta}) (\pmb{z} + \pmb{\alpha}) - \dot{{\eta}}_{di} - e_{1i} \dfrac{\dot{k}_{a_i}}{k_{a_i}} \right)
    \right] .\label{eqn:V1_dot_tv}
\end{align}
Using variables defined as in \Cref{eqn:var_change}, one can write \Cref{eqn:V1_dot_tv} as 
\begin{align}
    \dot{\mathcal{V}}_1 &=\sum_{i=1}^3 \left[ \dfrac{q \varepsilon_{b_i}^{2p}}{e_{1i}(1 - \varepsilon_{b_i}^{2p})} \left( J_{i}(\pmb{\eta}) (\pmb{z} + \pmb{\alpha}) - \dot{{\eta}}_{di} - e_{1i} \dfrac{\dot{k}_{b_i}}{k_{b_i}} \right) \right. \nonumber \\
    & \qquad \quad \left. + 
    \dfrac{(1-q) \varepsilon_{a_i}^{2p}}{e_{1i}(1 - \varepsilon_{a_i}^{2p})} \left( J_{i}(\pmb{\eta}) (\pmb{z} + \pmb{\alpha}) - \dot{{\eta}}_{di} - e_{1i} \dfrac{\dot{k}_{a_i}}{k_{a_i}} \right)
    \right] \label{eqn:V1_dot_tv2}
\end{align}
Now, we select the stabilizing function, $\pmb{\alpha}$, as given by 
\begin{equation}\label{eqn:alpha_tv}
    \pmb{\alpha} = \pmb{J}(\pmb{\eta})^\top \left[ \dot{\pmb{\eta}}_d - \left(\pmb{K}_1 +\bar{\pmb{K}}(t)\right)\pmb{e}_1 \right],
\end{equation} 
where the gains are defined as, $\pmb{K}_1 := \diag [k_{11}~~k_{12}~~k_{13} ]$ and $\bar{\pmb{K}}(t):=\diag [\bar{k}_{11}(t)~~\bar{k}_{12}(t)~~\bar{k}_{13}(t)] $. The elements of the gain matrix, $k_{1i}>0~(i\in\{1,2,3\})$, are positive constants, while the time-varying gain is given by
\begin{equation}
    \bar{k}_{1i}(t) = \sqrt{\left(\dfrac{\dot{k}_{a_i}}{k_{a_i}}\right)^2 + \left(\dfrac{\dot{k}_{b_i}}{k_{b_i}}\right)^2 + a_i} \label{eqn:k1i_tv}
\end{equation}
for any positive constants $a_i$ ($i \in \{1,2,3\}$). 
Using \Cref{eqn:alpha_tv} into \Cref{eqn:V1_dot_tv2}, the derivative of Lyapunov function candidate, $\dot{\mathcal{V}}_1$, is given by 
\begin{align}
    \dot{\mathcal{V}}_1  &= -\sum_{i=1}^3 \left[ \dfrac{q \varepsilon_{i}^{2p}}{(1 - \varepsilon_{i}^{2p})}\left( k_{1i} + \bar{k}_{1i} + q \dfrac{\dot{k}_{b_i}}{k_{b_i}} + (1 - q) \dfrac{\dot{k}_{a_i}}{k_{a_i}}\right) \right] \nonumber \\
    & \quad + \sum_{i=1}^{3} \left[ 
    \dfrac{q(e_{1i})}{k_{ai}^2-e_{1i}^2}e_{1i}^{2p-1}J_i(\pmb{\eta})\pmb{z} + 
    \dfrac{(1-q(e_{1i}))}{k_{bi}^2-e_{1i}^2}e_{1i}^{2p-1}J_i(\pmb{\eta})\pmb{z}
    \right]. \label{eqn:V1_dot_tv4}
\end{align}
From \Cref{eqn:k1i_tv}, it can be inferred that
\begin{equation}
    \bar{k}_{1i} + q \dfrac{\dot{k}_{b_i}}{k_{b_i}} + (1 - q) \dfrac{\dot{k}_{a_i}}{k_{a_i}} \geq 0.
\end{equation}
Also, $\dfrac{k_{1i} \varepsilon_i^{2p}}{1 - \varepsilon_i^{2p}} >0$ as $k_{1i}>0$. Using the fact that $c \leq -(a+b) \implies c \leq -a$ or $c \leq -b$ for $c \in \mathbb{R}, a \in \mathbb{R}_+, b \in \mathbb{R}_+$, we can write  
\begin{align}
    \nonumber \dot{\mathcal{V}}_1
    &\leq -\sum_{i=1}^{3}\dfrac{k_{1i} \varepsilon_i^{2p}}{1 - \varepsilon_i^{2p}} +
    \sum_{i=1}^{3} \left[ 
    \dfrac{q(e_{1i})}{k_{ai}^2-e_{1i}^2}e_{1i}^{2p-1}J_i(\pmb{\eta})\pmb{z} + 
    \dfrac{(1-q(e_{1i}))}{k_{bi}^2-e_{1i}^2}e_{1i}^{2p-1}J_i(\pmb{\eta})\pmb{z}
    \right]\\
    &\leq -\sum_{i=1}^{3}\dfrac{k_{1i} \varepsilon_i^{2p}}{1 - \varepsilon_i^{2p}} + \sum_{i=1}^{3} \left[  \mu_{1i}e_{1i}^{2p-1}J_i(\pmb{\eta})\pmb{z} \right],\label{eqn:V1dot_tv_ineq}
\end{align}
 where $\mu_{1i} := \dfrac{q(e_{1i})}{k_{b_i}^{2p} - e_{1i}^{2p}} + \dfrac{1-q(e_{1_i})}{k_{a_i}^{2p} - e_{1_i}^{2p}}$.
Next, for the purpose of being consistent with the earlier notations, we take the error variable as $\pmb{z} = \pmb{\nu} -\pmb{\alpha}$ and consider another Lyapunov function candidate as given by
\begin{equation}
    \mathcal{V}_2 = \mathcal{V}_1 + \dfrac{1}{2}\pmb{z}^\top\pmb{Mz}. \label{eqn:V2_tv}
\end{equation}
The derivative of $\mathcal{V}_2$, given by \Cref{eqn:V2_tv}, with respect to time yields
\begin{align}\label{eqn:V2_dot_tv}
    \dot{\mathcal{V}}_2 &= \dot{\mathcal{V}}_1 + \pmb{z}^\top\pmb{M}\dot{\pmb{z}}.
\end{align}
Now, from \Cref{eqn:z_dot_ip}, we have $\dot{\pmb{z}}$ as 
\begin{align}
    \dot{\pmb{z}} = \dot{\pmb{\nu}} - \dot{\pmb{\alpha}}  &= \pmb{M}^{-1}\left[\pmb{\tau} - \pmb{C}(\pmb{\nu})\pmb{\nu} - \pmb{D}(\pmb{\nu})\pmb{\nu} \right] - \dot{\pmb{\alpha}}.  
    \label{eqn:z_dot_tv}
\end{align}
Choosing the control input, $\pmb{\tau}$, as
\begin{align}\label{eqn:tau_tv}
    \pmb{\tau} &= \pmb{C}(\pmb{\nu})\pmb{\nu} + \pmb{D}(\pmb{\nu})\pmb{\nu} + \pmb{M}\dot{\pmb{\alpha}} - \diag[k_{21}~~ k_{22}~~ k_{23}]\pmb{z}
    - \diag[\mu_{11}~~\mu_{12}~~\mu_{13}]\pmb{J}^\top \pmb{e}^{2p-1}, 
\end{align} 
and using \Cref{eqn:tau_tv,eqn:z_dot_tv,eqn:V2_dot_tv}, one can obtain
\begin{align}
    \dot{\mathcal{V}_2} = -\sum_{i=1}^{3}\dfrac{k_{1i} \varepsilon_i^{2p}}{1 - \varepsilon_i^{2p}} - \pmb{z}^\top \pmb{K}_2 \pmb{z} \leq 0.
\end{align}
By virtue of Lemma \ref{lem1}, the errors $e_{1i}(t)\forall i$ remain in the open set $e_{1i} \in (-k_{ai}(t), k_{bi}(t))~\forall\,\,t \geq 0$, provided that $e_{1i}(0) \in (-k_{ai}(0), k_{bi}(0))$ for $i \in \{1,2,3\}$.  Thus $-\underline{k}_{ci}(t)<\eta_{i}(t)<\bar{k}_{ci}(t)$. Lastly, we incorporate the bounds on the actuator into the design.

\subsection{Control under asymmetric dynamic state constraint with the actuator saturation}\label{sec:tv_ip}

In this subsection, apart from imposing dynamic motion constraints, we also consider the actuator bounds. To do so, we use the input saturation model given by \Cref{eqn:actuator} and consider an augmented, increased-order system:
\begin{subequations}
\begin{align}
    \dot{\pmb{\eta}}  &= \pmb{J}(\psi) \pmb{\nu},\\
    \dot{\pmb{\nu}} &= \pmb{M}^{-1}\left[\pmb{\tau} - \pmb{C}(\pmb{\nu})\pmb{\nu} - \pmb{D}(\pmb{\nu})\pmb{\nu}  + \pmb{b}\right], \\
    \dot{\pmb{\zeta}} &= \pmb{Q}(\pmb{I} - \pmb{G}_M) \pmb{\tau}_c - \pmb{\rho} \pmb{\zeta} + (\pmb{I}-\pmb{Q})(\pmb{I} - \pmb{G}_m) \pmb{\tau}_c, \quad \pmb{\zeta}(0) = \pmb{0}, \quad \pmb{\tau} = \pmb{\zeta}.
\end{align}
\end{subequations}
Now, consider another intermediate variable $\pmb{z}_2 = \pmb{\tau} - \pmb{\alpha}_2$. Using \Cref{eqn: dynamics}, previously defined variable 
 $\pmb{z} = \pmb{\nu} - \pmb{\alpha}$, and $\pmb{z}_2$, the time differentiation of $\pmb{z}$ results in 
\begin{align}
    \dot{\pmb{z}} = \dot{\pmb{\nu}} - \dot{\pmb{\alpha}}  &= \pmb{M}^{-1}\left[\pmb{\tau} - \pmb{C}(\pmb{\nu})\pmb{\nu} - \pmb{D}(\pmb{\nu})\pmb{\nu} \right] - \dot{\pmb{\alpha}}=\pmb{M}^{-1}\left[\pmb{z}_2 + \pmb{\alpha}_2 - \pmb{C}(\pmb{\nu})\pmb{\nu} - \pmb{D}(\pmb{\nu})\pmb{\nu} \right] - \dot{\pmb{\alpha}}.
    \label{eqn:z_dot_tv_ip}
\end{align}
To design the stabilizing function, $\pmb{\alpha}_2$, consider the Lyapunov function candidate given by \Cref{eqn:V2_tv}. On substituting the value of $\dot{\pmb{z}}$ from \Cref{eqn:z_dot_tv_ip} into \Cref{eqn:V2_dot_tv}, we obtain
\begin{align}
    \dot{\mathcal{V}}_2 &= \dot{\mathcal{V}}_1 + \pmb{z}^\top\pmb{M}\left(\pmb{M}^{-1}\left[\pmb{z}_2 + \pmb{\alpha}_2 - \pmb{C}(\pmb{\nu})\pmb{\nu} - \pmb{D}(\pmb{\nu})\pmb{\nu} \right] - \dot{\pmb{\alpha}}\right).
\end{align}
Using \Cref{eqn:V1dot_tv_ineq}, we can write $\dot{\mathcal{V}}_2$ as
\begin{align}
    \dot{\mathcal{V}}_2&\leq -\sum_{i=1}^{3}\dfrac{k_{1i} \varepsilon_i^{2p}}{1 - \varepsilon_i^{2p}} + \sum_{i=1}^{3} \left[  \mu_{1i}e_{1i}^{2p-1}J_i(\pmb{\eta})\pmb{z} \right]
    +\pmb{z}^\top\pmb{M}\left(\pmb{M}^{-1}\left[\pmb{z}_2 + \pmb{\alpha}_2 - \pmb{C}(\pmb{\nu})\pmb{\nu} - \pmb{D}(\pmb{\nu})\pmb{\nu} \right] - \dot{\pmb{\alpha}}\right)\label{eqn:V2_dot_tv_ip}
\end{align}
By choosing stabilizing function $\pmb{\alpha}_2$ as 
\begin{align}
    \pmb{\alpha}_2 &= \pmb{C}(\pmb{\nu})\pmb{\nu} + \pmb{D}(\pmb{\nu})\pmb{\nu} + \pmb{M}\dot{\pmb{\alpha}} - \diag[k_{21}~~ k_{22}~~ k_{23}]\pmb{z}
    - \diag[\mu_{11}~~\mu_{12}~~\mu_{13}]\pmb{J}^\top(\pmb{\eta}) \pmb{e}^{2p-1}, \label{eqn:alpha2_tv}
\end{align}
and using \Cref{eqn:alpha2_tv}, and \Cref{eqn:V2_dot_tv_ip}, we get
\begin{align}
    \dot{\mathcal{V}_2} \leq -\sum_{i=1}^{3}\dfrac{k_{1i} \varepsilon_i^{2p}}{1 - \varepsilon_i^{2p}} - \pmb{z}^\top \pmb{K}_2 \pmb{z} + \pmb{z}^\top \pmb{z}_2.
\end{align}
Next, to cancel out the cross-term ($ \pmb{z}^\top \pmb{z}_2$), consider another Lyapunov function candidate as given by
\begin{align}\label{eqn:V3_tv_ip}
    \mathcal{V}_3 = \mathcal{V}_2 + \dfrac{1}{2}\pmb{z}_2^\top \pmb{z}_2
\end{align}
On taking the time differentiation of \Cref{eqn:V3_tv_ip}, we get
\begin{align}
    \dot{\mathcal{V}}_3 = \dot{\mathcal{V}}_2 + \pmb{z}_2^\top \dot{\pmb{z}}_2 \label{eqn:V3_dot}
\end{align}
Differentiating $\pmb{z}_2$ with respect to time, one can obtain 
\begin{align}
    \dot{\pmb{z}}_2 = \dot{\pmb{\tau}} -\dot{\pmb{\alpha}}_2 
\end{align}
Substituting for $\dot{\pmb{\tau}}$ from \Cref{eqn:actuator}, we get 
\begin{align}
     \dot{\pmb{z}}_2 = \pmb{Q}(\pmb{I} - \pmb{G}_M) \pmb{\tau}_c - \pmb{\rho} \pmb{\zeta} + (\pmb{I}-\pmb{Q})(\pmb{I} - \pmb{G}_m) \pmb{\tau}_c -\dot{\pmb{\alpha}_2}. \label{eqn:z2_dot}
\end{align}
By choosing control input $\pmb{\tau}_c$ as 
\begin{align}
    \pmb{\tau}_c = \left[\pmb{Q}(\pmb{I} - \pmb{G}_M) + (\pmb{I}-\pmb{Q})(\pmb{I} - \pmb{G}_m) \right]^{-1}\left(\pmb{\rho \tau} + \dot{\pmb{\alpha}}_2\right) - \pmb{z}- \pmb{K}_3\pmb{z}_2 \label{eqn:tauc}
\end{align}
and using \Cref{eqn:tauc,eqn:z2_dot,eqn:V3_dot}, we obtain
\begin{align}
    \dot{\mathcal{V}}_3 \leq -\sum_{i=1}^{3}\dfrac{k_{1i} \varepsilon_i^{2p}}{1 - \varepsilon_i^{2p}} - \pmb{z}^\top \pmb{K}_2 \pmb{z} -\pmb{z}_2^\top \pmb{K}_3 \pmb{z}_2 \leq 0.
\end{align}
By virtue of Lemmas \ref{lem1} and \ref{lem2}, the errors, $e_{1i} \in (-k_{ai}(t), k_{bi}(t))~\forall\,\,t \geq 0$, provided that $e_{1i}(0) \in (-k_{ai}(0), k_{bi}(0))$ for $i \in \{1,2,3\}$.  Thus $-\underline{k}_{ci}(t)<\eta_{i}(t)<\bar{k}_{ci}(t)$. Additionally, $\tau_i \in (-\tau_{im},\tau_{iM})$ for $i \in \{u,v,r\}$ at all times, satisfying the actuator constraints.

\section{Simulation Results}\label{sec:simulation results}

In this section, we evaluate the performance of the proposed constrained motion control strategies. We used the CyberShip II model parameters \cite{skjetne2004modeling} for all numerical simulations. The CyberShip II is a $1:70$ scaled model of a supply ship. It is a fully actuated model and has a mass of 23.8 kg. The hydrodynamic derivatives and other model parameters are tabulated in \Cref{tab:paramstable}. 
\begin{table}[!ht]
\centering
\caption{Model parameters of the CyberShip II \cite{skjetne2004modeling}.}
\label{tab:paramstable}
\begin{tabular}{@{}lclclc@{}} 
\toprule
\textbf{Parameter} & \textbf{Value} & \textbf{Parameter} & \textbf{Value} & \textbf{Parameter} & \textbf{Value} \\ 
\midrule
$\text{m}$             & 23.800    & $X_{\dot{u}}$ & -2.0      & $X_u$         & -0.72253  \\
$Y_{v}$         & -2.0      & $I_z$         & 1.760     & $Y_{\dot{v}}$ & -10.0     \\
$X_{|u|u}$      & -1.32742  & $Y_{|v|v}$    & -36.47287 & $x_g$         & 0.046     \\
$Y_{\dot{r}}$   & -0.0      & $X_{uuu}$     & -5.86643  & $N_{v}$       & 0.03130   \\
$N_{\dot{v}}$   & -0.0      & $N_{\dot{r}}$ & -0.0      & $Y_{|r|v}$    & -0.805    \\
$N_{|r|v}$      & 0.130     & $Y_{r}$       & -7.250    & $N_{r}$       & -1.900    \\
$Y_{|v|r}$      & -0.845    & $N_{|v|r}$    & 0.080     & $N_{|v|v}$    & 3.95645   \\
$Y_{|r|r}$      & -3.450    & $N_{|r|r}$    & -0.750    &               &           \\
\bottomrule
\end{tabular}
\end{table}
We now demonstrate the performance of the motion control strategies by considering both simple and complex trajectories for various initial conditions. The equations for an elliptical trajectory considered are:
\begin{equation}
     x_d(t) = 4 \cos(0.05t) -4,\quad   y_d(t) = 6\sin(0.05t),\quad
    \psi_{d}(t) = 0.5\pi\cos(0.02t),\label{eqn:ellipse}
\end{equation}
and equations for an 8-shaped trajectory considered are:
\begin{equation}
     x_d(t) = 4(\cos(0.05t)-1),\quad  y_d(t) = 2.5\sin(0.1t), \quad \psi_{d}(t) = 0.5\pi\sin(0.02t).\label{eqn:eight}
\end{equation}
In \Cref{eqn:ellipse} and \Cref{eqn:eight}, the desired position coordinates are in meters, and the desired heading angle is in radians.
\subsection{ Tracking under static state constraints with input saturation}
In this subsection, we evaluate the performance of the proposed algorithm in \Cref{sec:st_ip}. This algorithm accounts for asymmetric constraints on USV states $\pmb{\eta}$, symmetric constraints on the body rates, $\pmb{\nu}$, and asymmetric actuator bounds.
The controller gain parameters are: $k_{11} = 1$, $k_{12} = 2$, $k_{13} = 3$, $\pmb{K}_2 = \diag[1~~1~~3]$, $\pmb{K}_3 = \diag[2~~3~~5]$, and $\pmb{\rho} = \diag [0.2~~0.2~~0.2]$. 
We test the efficacy of the algorithm for three initial configurations ($\pmb{\eta}_0=(x_0,y_0,\psi_0))$ denoted as $P_1:~(0.6\,\si{m},0.15 \,\si{m},105^\circ)$, $P_2:~(-0.5\,\si{m},0.20\,\si{m},76^{\circ})$, $P_3:~ (0.4\,\si{m},-0.2\,\si{m},80^\circ)$. 
In all cases, the USV moves at a surge speed of $0.1\, \si{m/s}$. The  state constraints imposed are as follows: $\pmb{k}_a = (0.8\,\si{m}, 0.4\,\si{m}, 15^\circ)$, $\pmb{k}_b = (1\,\si{m}, 0.5\,\si{m}, 20^\circ)$, $\pmb{k}_2 = (1\,\si{m/s},1\,\si{m/s},3\,\si{rad/s})$. Asymmetric control bounds considered are taken to be $\pmb{\tau}_{\text{max}} = [4~~4~~4]^\top$ and $\pmb{\tau}_{\text{min}} = [-3.5~-3.5~-3.5]^\top$.
\begin{figure*}[ht!]
    \centering
    \begin{subfigure}{0.5\linewidth}
    \centering
    \includegraphics[width=\linewidth]{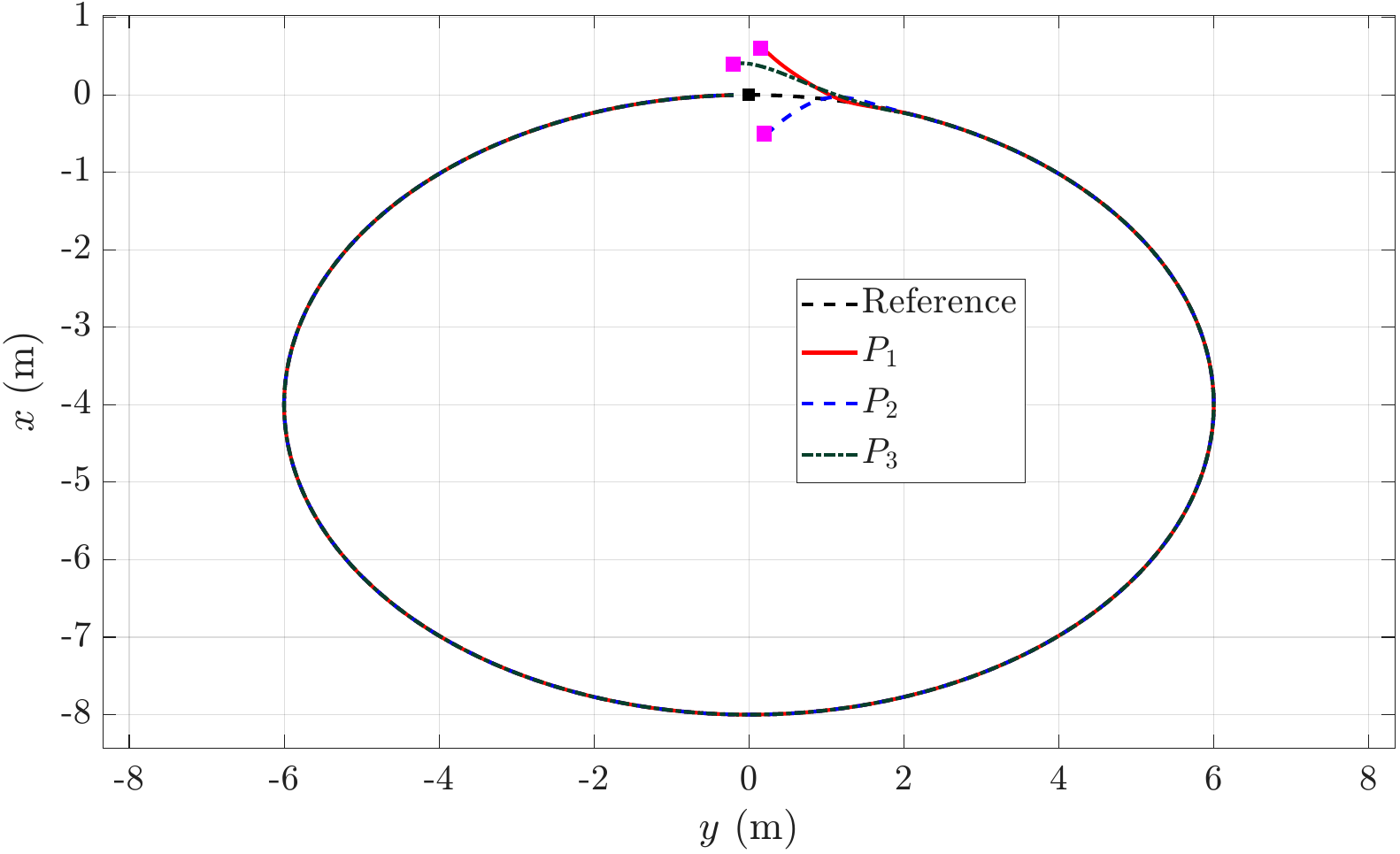}   
    \caption{Actual and desired trajectories.}
    \label{fig:asym_st_inp_sat_ellipse_path}
    \end{subfigure}%
    \begin{subfigure}{0.5\linewidth}
    \centering
     \includegraphics[width=\linewidth]{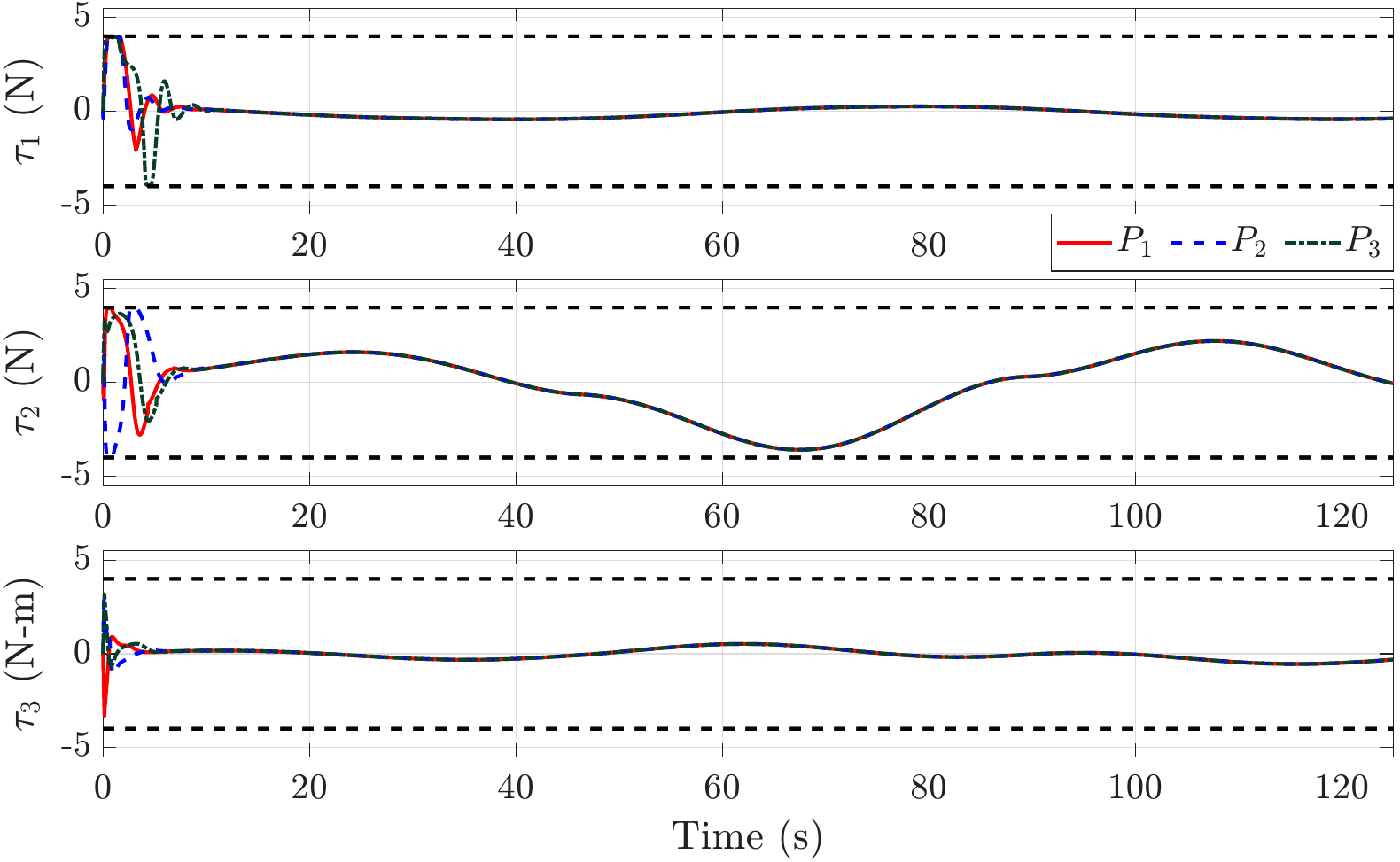}
    \caption{Control inputs.}
    \label{fig:asym_st_inp_sat_ellipse_tau}
    \end{subfigure}
    \begin{subfigure}{0.5\linewidth}
    \centering
    \includegraphics[width=\linewidth]{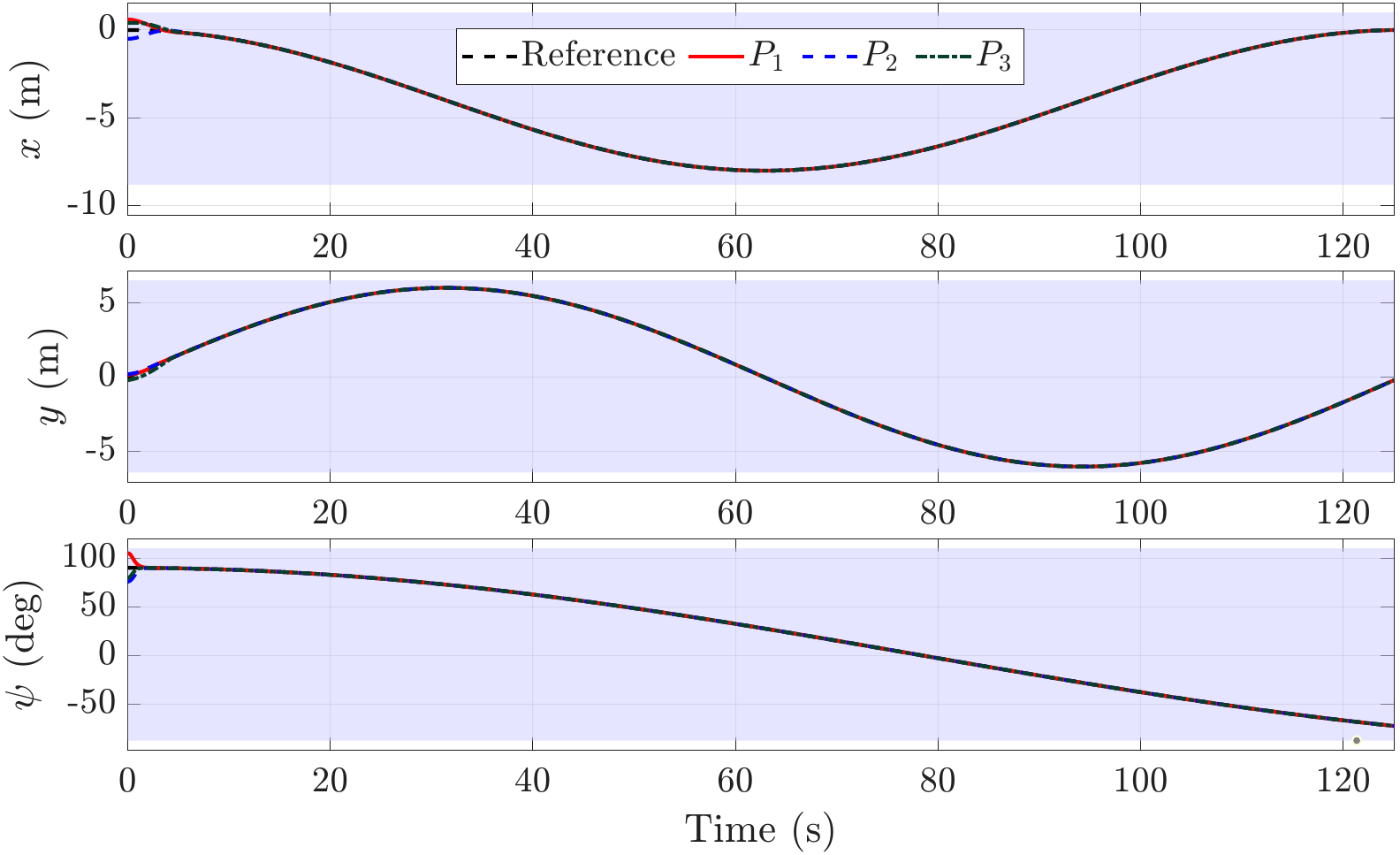}
    \caption{Position and heading of the vessel.}
    \label{fig:asym_st_inp_sat_ellipse_eta}
    \end{subfigure}%
    \begin{subfigure}{0.5\linewidth}
    \centering
    \includegraphics[width=\linewidth]{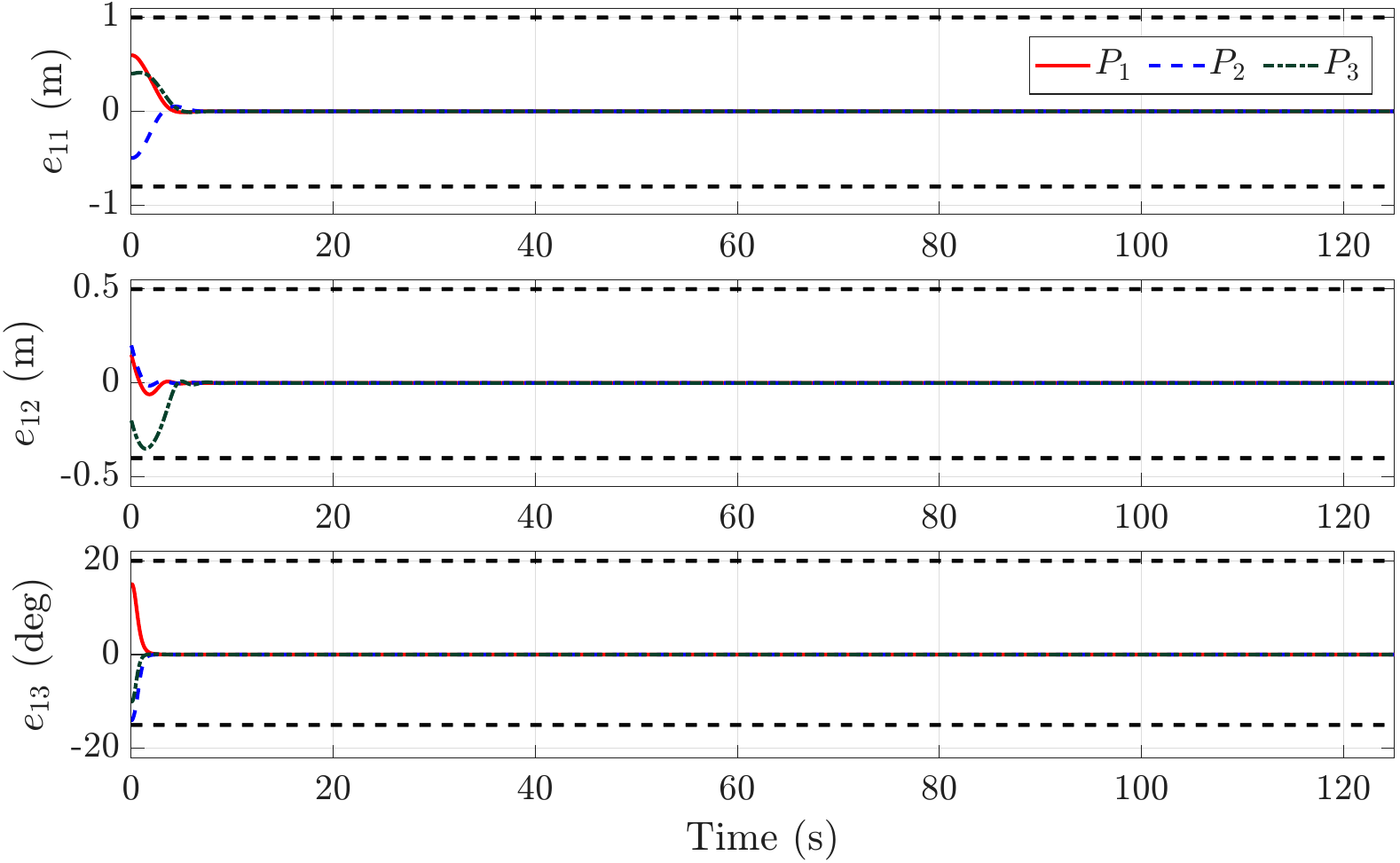}
    \caption{Errors in position and heading.}
    \label{fig:asym_st_inp_sat_ellipse_e1}
    \end{subfigure}
    \begin{subfigure}{0.5\linewidth}
    \centering
    \includegraphics[width=\linewidth]{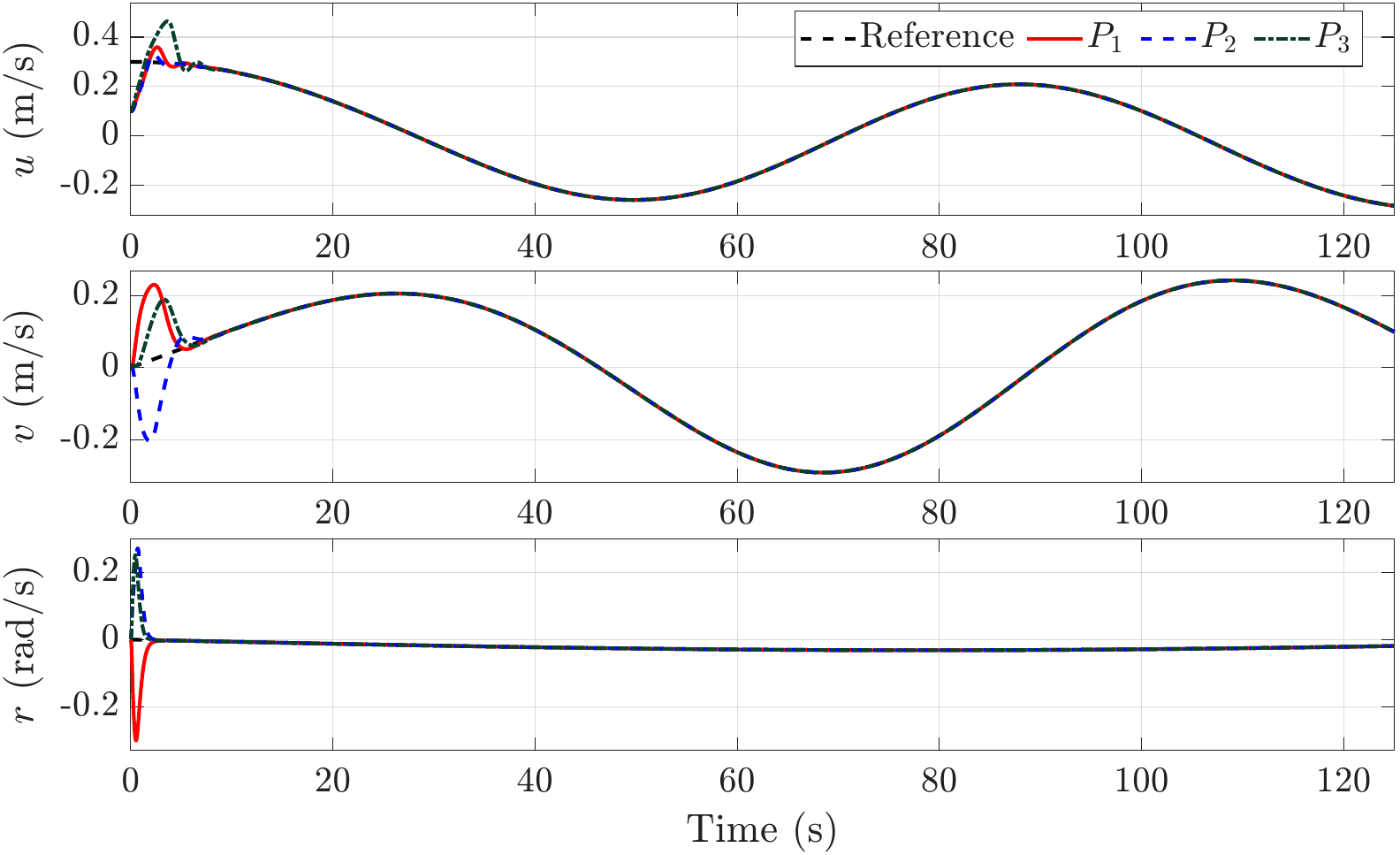}
    \caption{Surge, sway, and yaw rate.}
    \label{fig:asym_st_inp_sat_ellipse_nu}
    \end{subfigure}%
    \begin{subfigure}{0.5\linewidth}
    \centering
    \includegraphics[width=\linewidth]{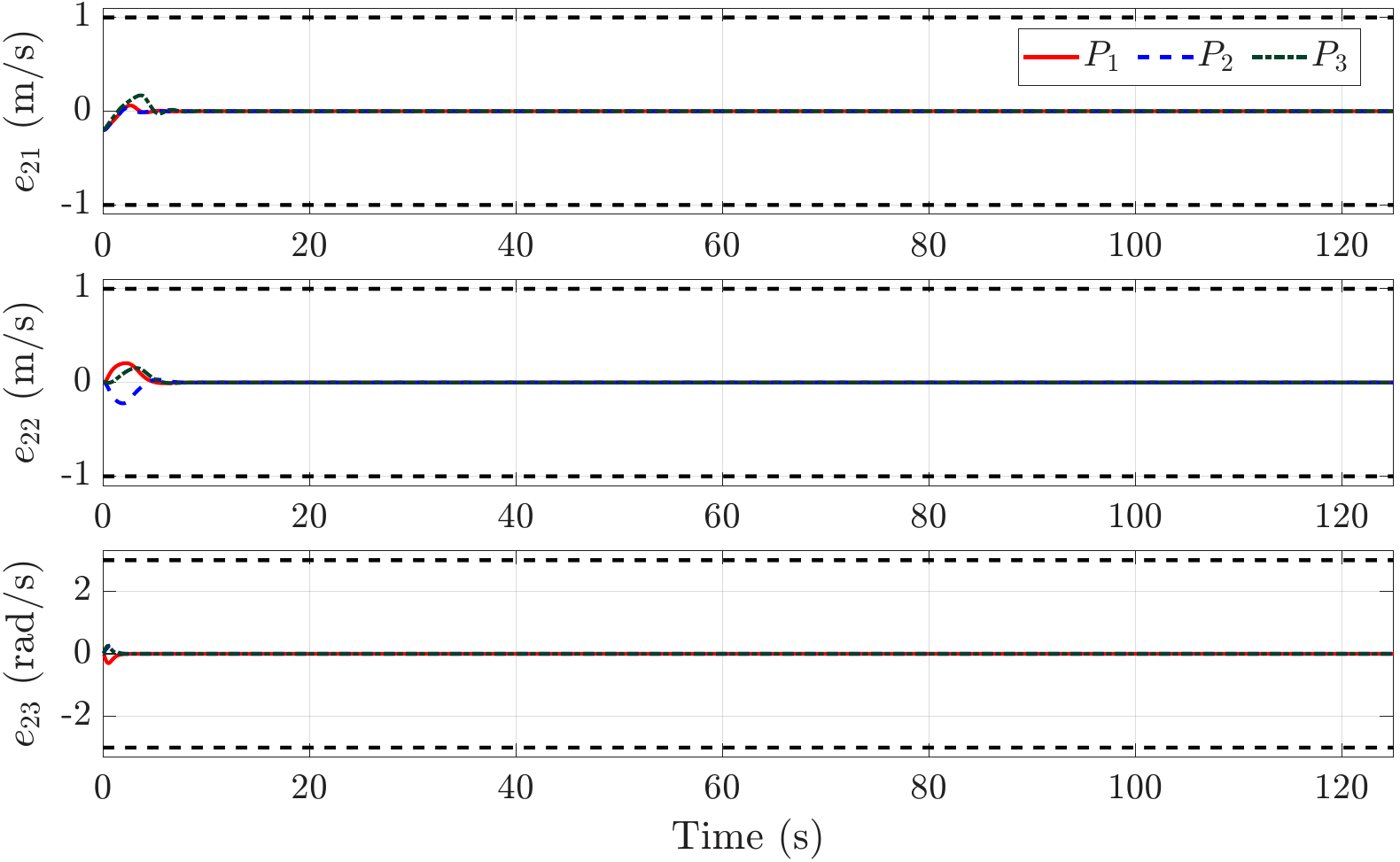}
    \caption{Error in velocities and yaw rate.}
    \label{fig:asym_st_inp_sat_ellipse_e2}
    \end{subfigure}
    \caption{Performance validation of proposed controller for a surface vessel with elliptical trajectory under static state constraints and input saturation.}
    \label{fig:asym_st_inp_sat_ellipse}
\end{figure*}
\cref{fig:asym_st_inp_sat_ellipse} demonstrates the tracking of an elliptical trajectory given by \Cref{eqn:ellipse}. 
As seen from \Cref{fig:asym_st_inp_sat_ellipse_path}, the USV tracks the desired trajectory starting from $P_1$, $P_2$, and $P_3$. 
It can be noted from \Cref{fig:asym_st_inp_sat_ellipse_e1,fig:asym_st_inp_sat_ellipse_e2}, that errors converge to zero and never violate the set bounds as indicated by black dashed horizontal lines. 
This can also be verified from \Cref{fig:asym_st_inp_sat_ellipse_eta,fig:asym_st_inp_sat_ellipse_nu}. Also, as guaranteed by the proposed algorithm, the control inputs also never cross the bounds (see \Cref{fig:asym_st_inp_sat_ellipse_tau}). Now, we illustrate the performance for an 8-shaped trajectory given by \Cref{eqn:eight}. 
\begin{figure*}[ht!]
    \centering
    \begin{subfigure}{0.5\linewidth}
    \centering
    \includegraphics[width=\linewidth]{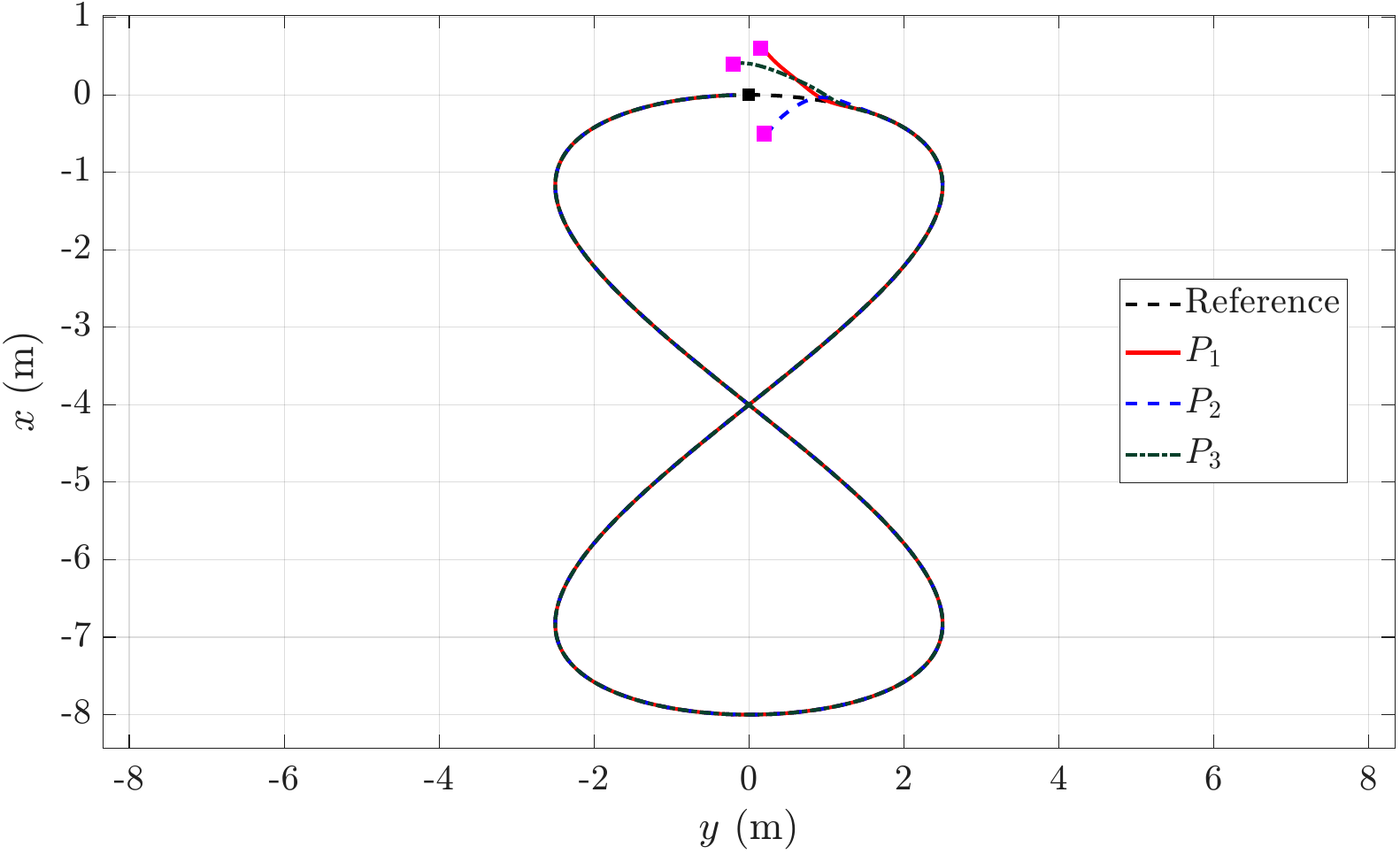}   
    \caption{Actual and desired trajectories.}
    \label{fig:asym_st_inp_sat_eight_path}
    \end{subfigure}%
    \begin{subfigure}{0.5\linewidth}
    \centering
     \includegraphics[width=\linewidth]{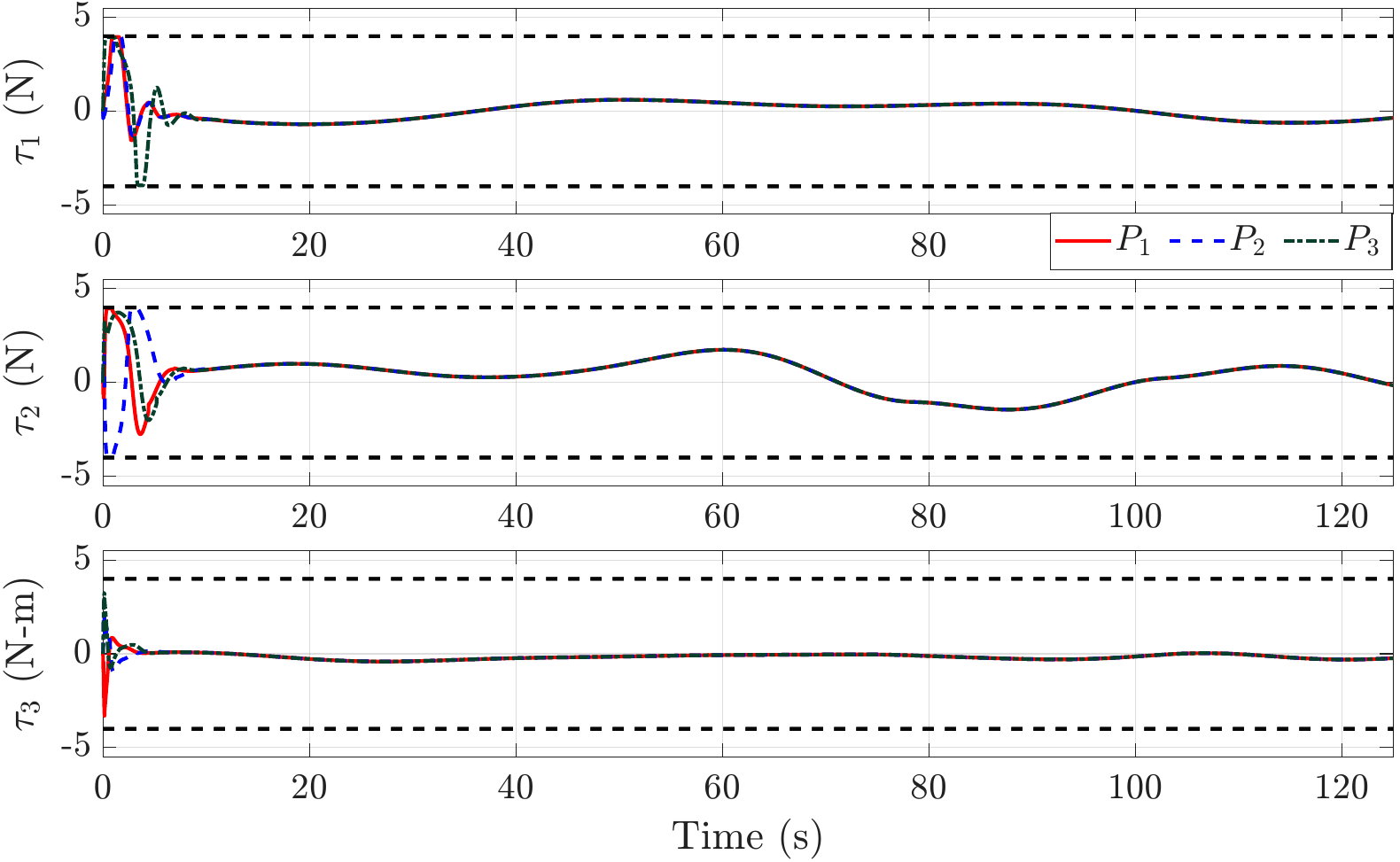}
    \caption{Control inputs.}
    \label{fig:asym_st_inp_sat_eight_tau}
    \end{subfigure}
    \begin{subfigure}{0.5\linewidth}
    \centering
    \includegraphics[width=\linewidth]{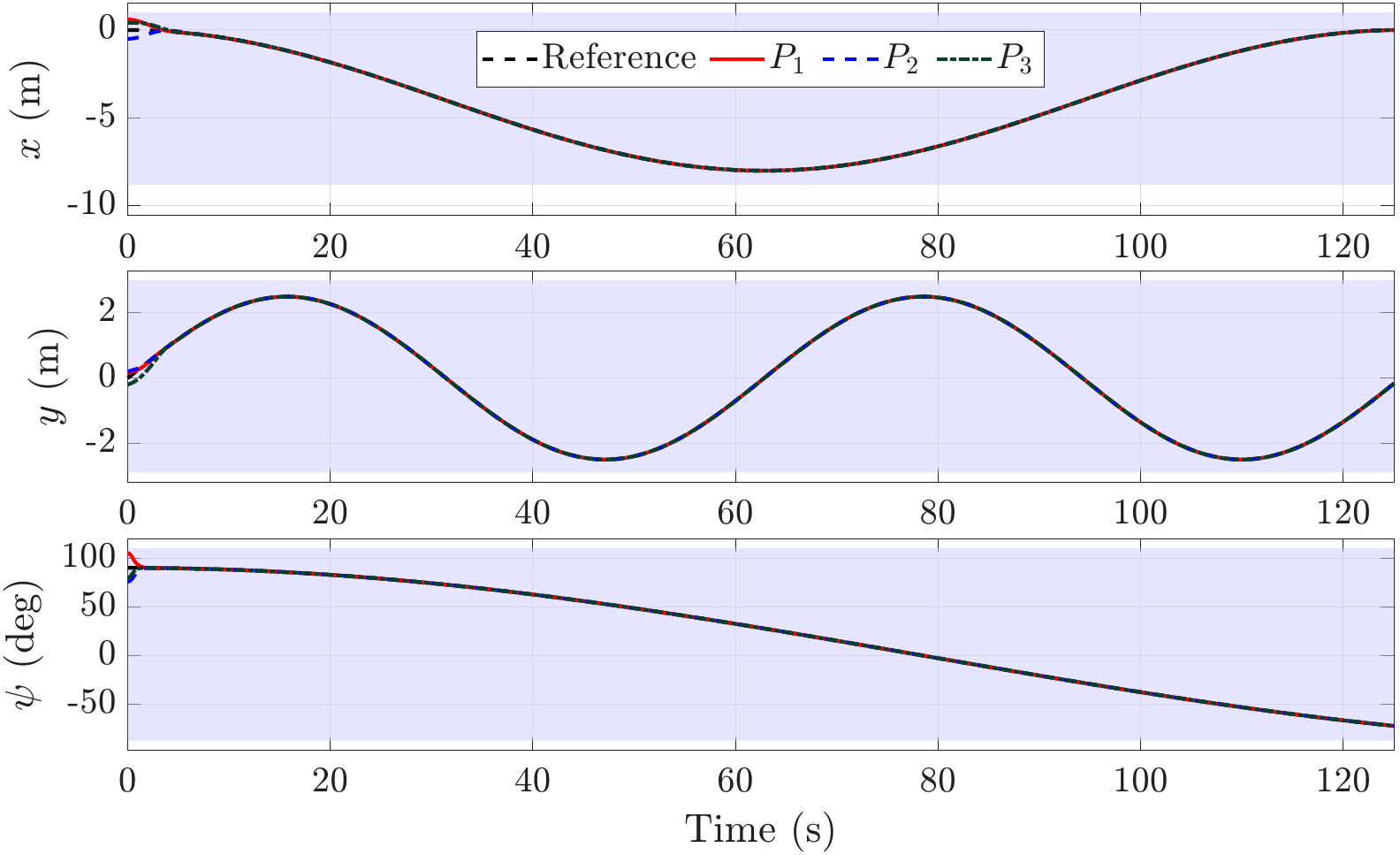}
    \caption{Position and heading of the vessel.}
    \label{fig:asym_st_inp_sat_eight_eta}
    \end{subfigure}%
    \begin{subfigure}{0.5\linewidth}
    \centering
    \includegraphics[width=\linewidth]{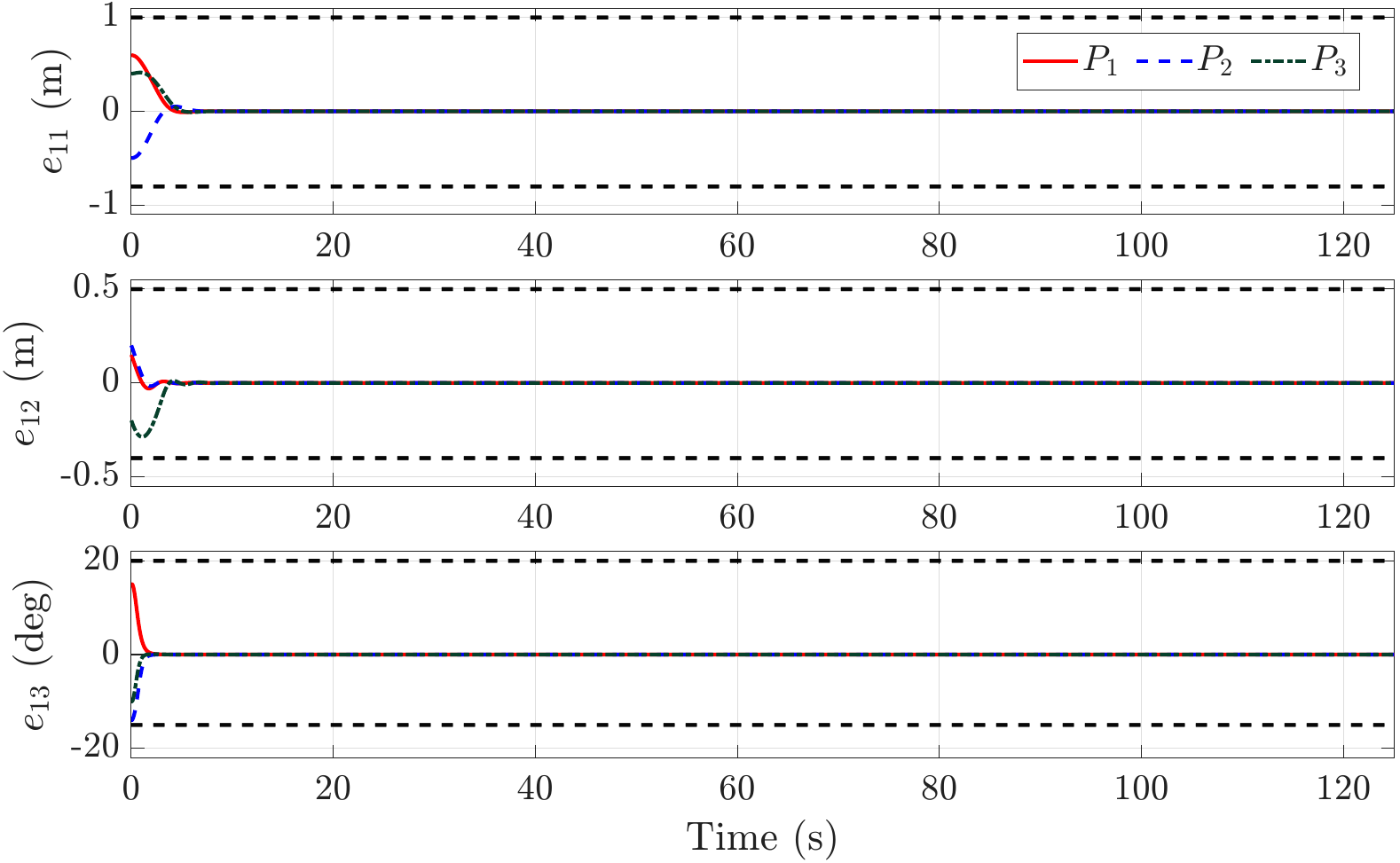}
    \caption{Errors in position and heading.}
    \label{fig:asym_st_inp_sat_eight_e1}
    \end{subfigure}
    \begin{subfigure}{0.5\linewidth}
    \centering
    \includegraphics[width=\linewidth]{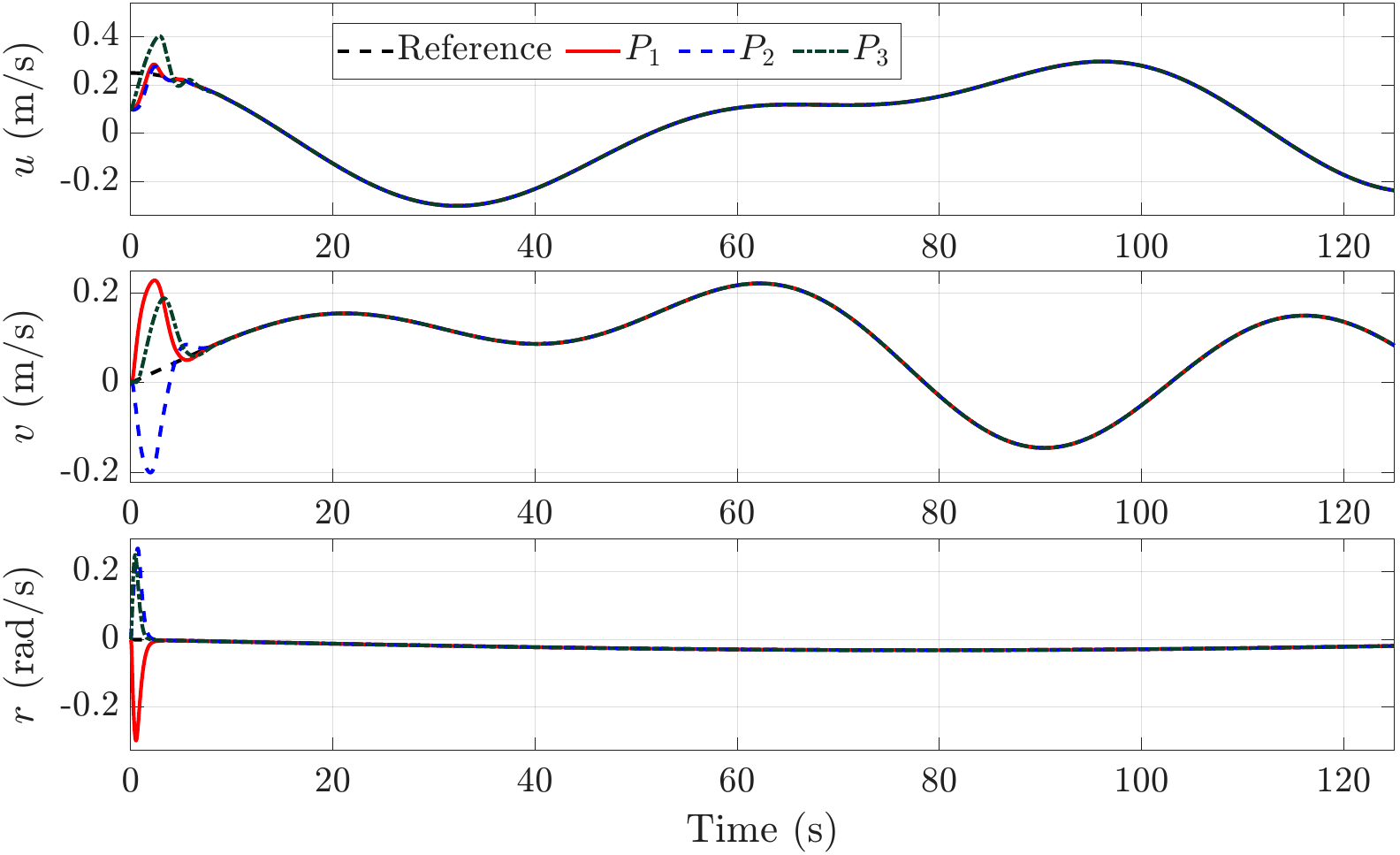}
    \caption{Surge, sway, and yaw rate.}
    \label{fig:asym_st_inp_sat_eight_nu}
    \end{subfigure}%
    \begin{subfigure}{0.5\linewidth}
    \centering
    \includegraphics[width=\linewidth]{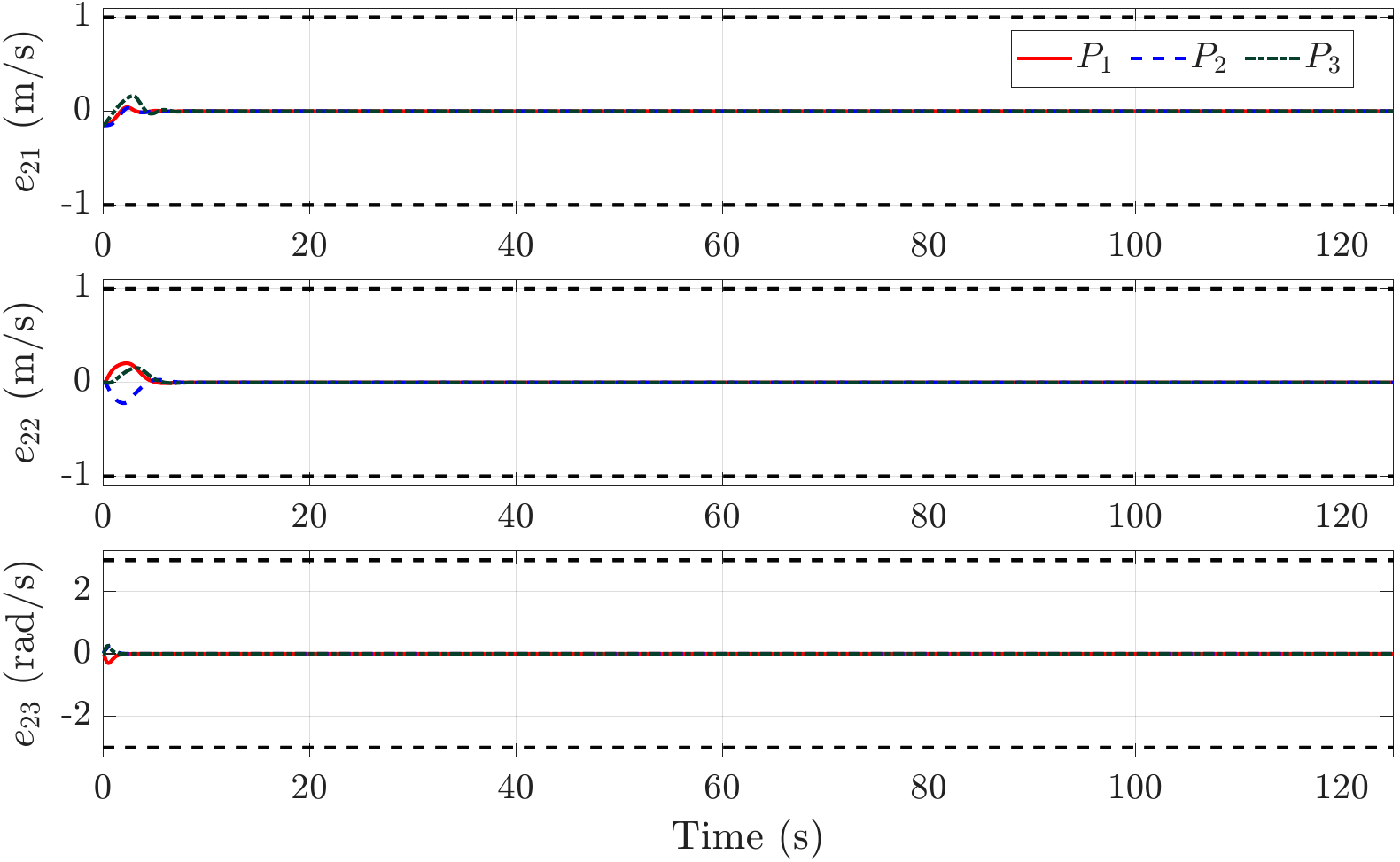}
    \caption{Error in velocities and yaw rate.}
    \label{fig:asym_st_inp_sat_eight_e2}
    \end{subfigure}
    \caption{Performance validation of proposed controller for a USV with an eight-shaped trajectory under static state constraints and input saturation.}
    \label{fig:asym_st_inp_sat_eight}
\end{figure*}
It can be seen from \Cref{fig:asym_st_inp_sat_eight_path} that the USV follows the path starting from different initial positions denoted with magenta squares. It is evident that the constraints on states (see
\Cref{fig:asym_st_inp_sat_eight_e1,fig:asym_st_inp_sat_eight_e2,fig:asym_st_inp_sat_eight_eta,fig:asym_st_inp_sat_eight_nu})
and actuator (see \Cref{fig:asym_st_inp_sat_eight_tau}) are always satisfied with errors converging to zero. 
From \Cref{fig:asym_st_inp_sat_eight_tau}, we can see that during the transients, the control inputs applied are different, but once the USV starts tracking the desired trajectory, the control inputs for all the cases coincide. 

\subsection{Tracking under dynamic state constraints}
In this subsection, we evaluate the performance of the proposed algorithm in \Cref{sec:tv}. The dynamic constraints case is a generalized version of the static constraints case. However, we are validating it separately because, in this case, we have imposed dynamic constraints only on the position and heading, not on body rates. This is motivated by a real-life example of a USV navigating through a curvy river leading to a time-varying boundary case. The controller gain parameters are systematically chosen as: $k_{11} = 0.1$, $k_{12} = 0.4$, $k_{13} = 0.3$, $\pmb{K}_2 = \diag[1~~1~~3]$, $a_{1}=0.1$, $a_2 = 0.1$, $a_3=0.1$. Note that all real-world actuators have bounded control availability, and since the control bounds are not explicitly accounted for in this design, we impose them using a saturation block. 
The control input saturation bounds in surge, sway, and yaw are taken to be $\pmb{\tau}_{\text{max}}=[4~~4~~4]^\top$ and $\pmb{\tau}_{\text{min}}=[-4~-4~-4]^\top$. When the demanded control effort exceeds these bounds, the applied control effort saturates at the bounds. 
As in the previous case, we validate the proposed strategy against two trajectories: an elliptical and an 8-shaped trajectory.

\subsubsection{Elliptical trajectory}
The elliptical trajectory is given by \Cref{eqn:ellipse}, and the boundaries are dictated by
\begin{subequations}\label{eqn:kc_ellipse}
\begin{align}
    \bar{\pmb{k}_{c1}}(t) &= \begin{bmatrix}
        \sin(0.05t + \pi/4) &  6 + 2\cos(0.05t- \pi/6) & 0.8 + 0.4\pi\cos(0.02t- \pi/10)
    \end{bmatrix} \\
    \underline{\pmb{k}}_{c1}(t)&= \begin{bmatrix}
        -8+\sin(0.05t + \pi/4) &  -6 +\sin(0.05t) & -0.5 + 0.6\pi\cos(0.02t)
    \end{bmatrix}.
\end{align}
\end{subequations}
We test the efficacy of the algorithm for three initial configurations ($\pmb{\eta}_0=(x_0,y_0,\psi_0))$ denoted as $P_1:~(0.6\,\si{m},-1 \,\si{m},110^\circ)$, $P_2:~(-1\,\si{m},2\,\si{m},100^{\circ})$, $P_3:~ (-2\,\si{m},-2\,\si{m},80^\circ)$. In all the cases, the USV starts with a surge speed of $0.1\, \si{m/s}$. 
\begin{figure*}[ht!]
    \centering
    \begin{subfigure}{0.5\linewidth}
    \centering
    \includegraphics[width=\linewidth]{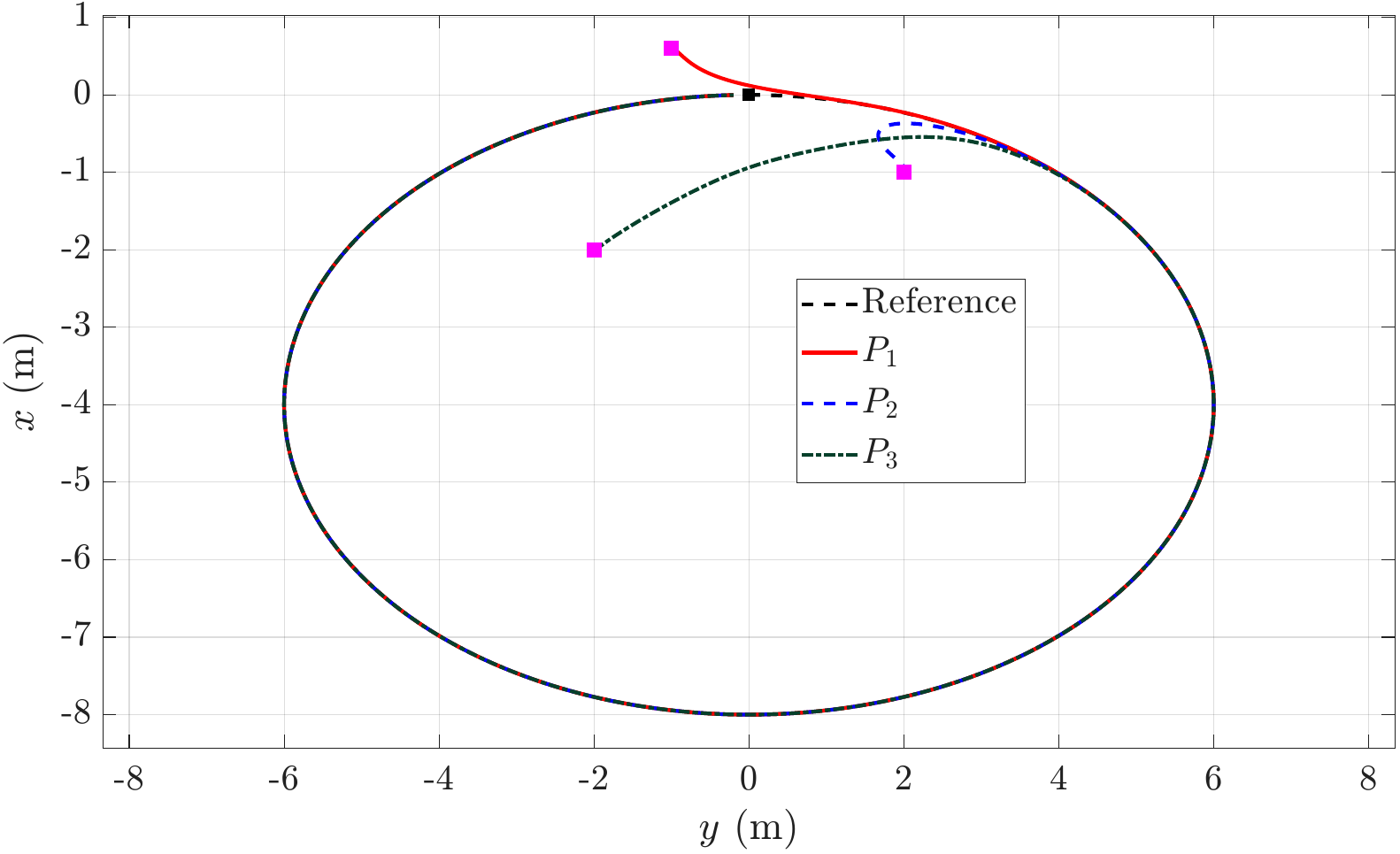}   
    \caption{Actual and desired trajectories.}
    \label{fig:asym_tv_ellipse_path}
    \end{subfigure}%
    \begin{subfigure}{0.5\linewidth}
    \centering
     \includegraphics[width=\linewidth]{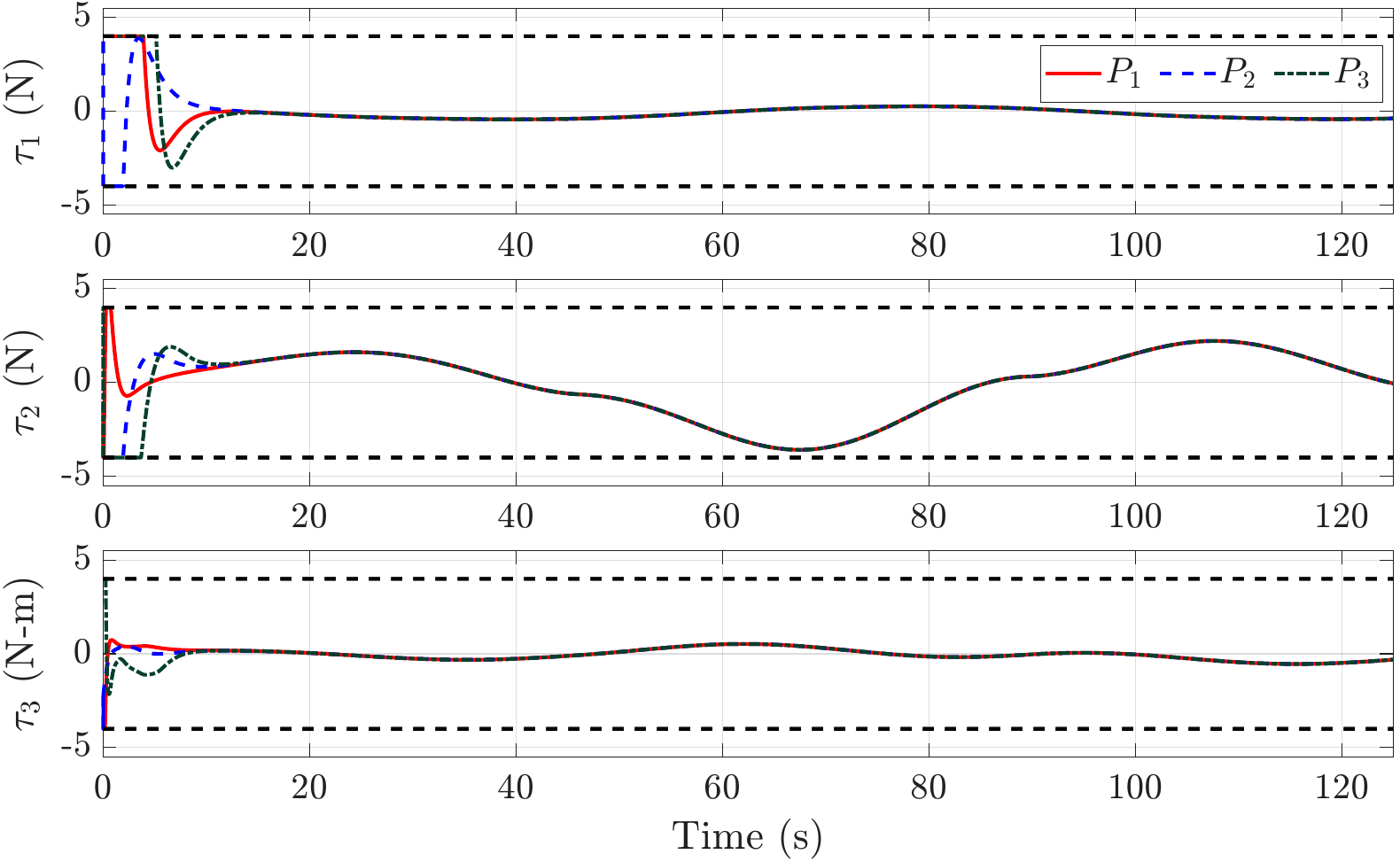}
    \caption{Control inputs.}
    \label{fig:asym_tv_ellipse_tau}
    \end{subfigure}
    \begin{subfigure}{0.5\linewidth}
    \centering
    \includegraphics[width=\linewidth]{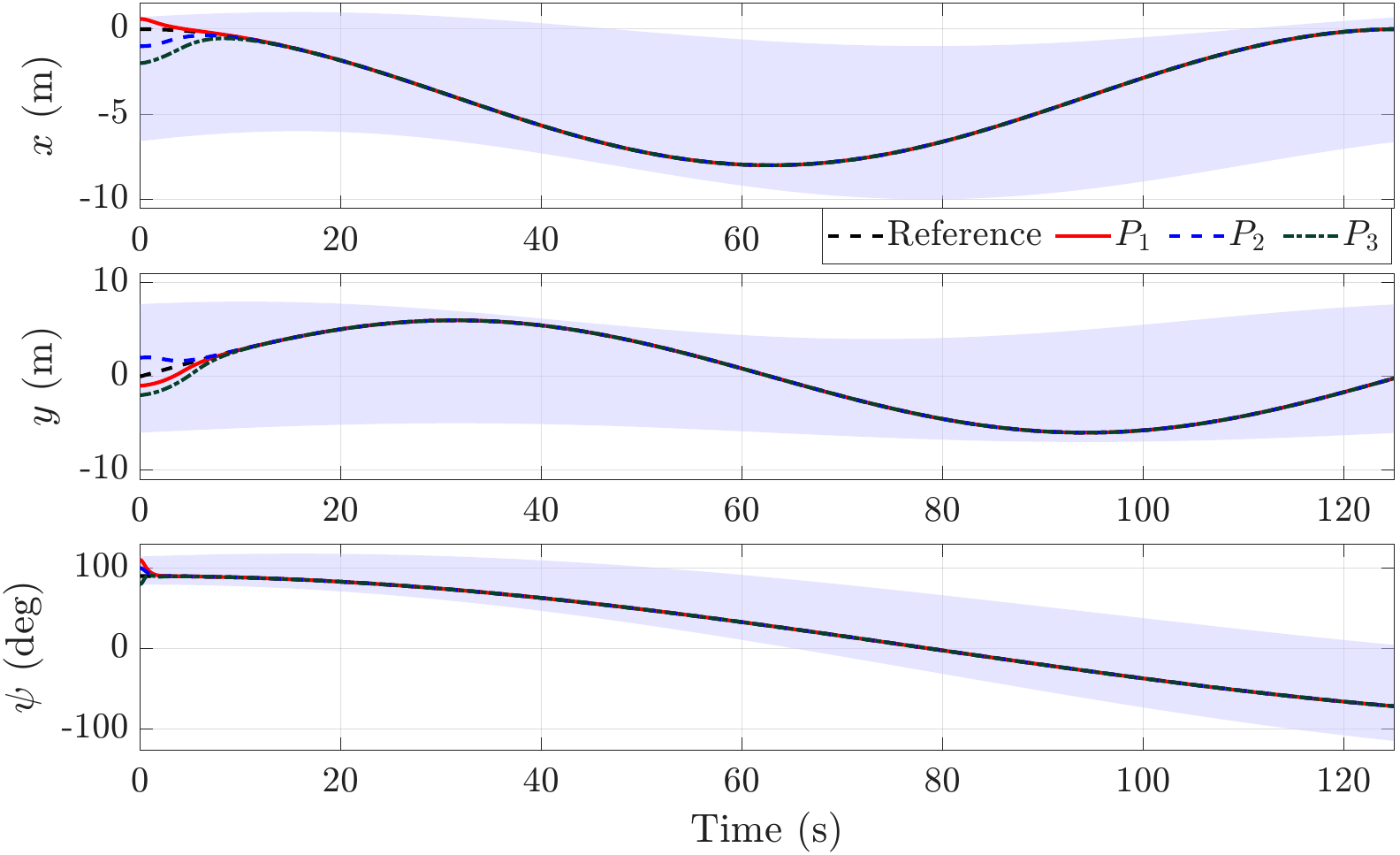}
    \caption{Position and heading of the vessel.}
    \label{fig:asym_tv_ellipse_eta}
    \end{subfigure}%
    \begin{subfigure}{0.5\linewidth}
    \centering
    \includegraphics[width=\linewidth]{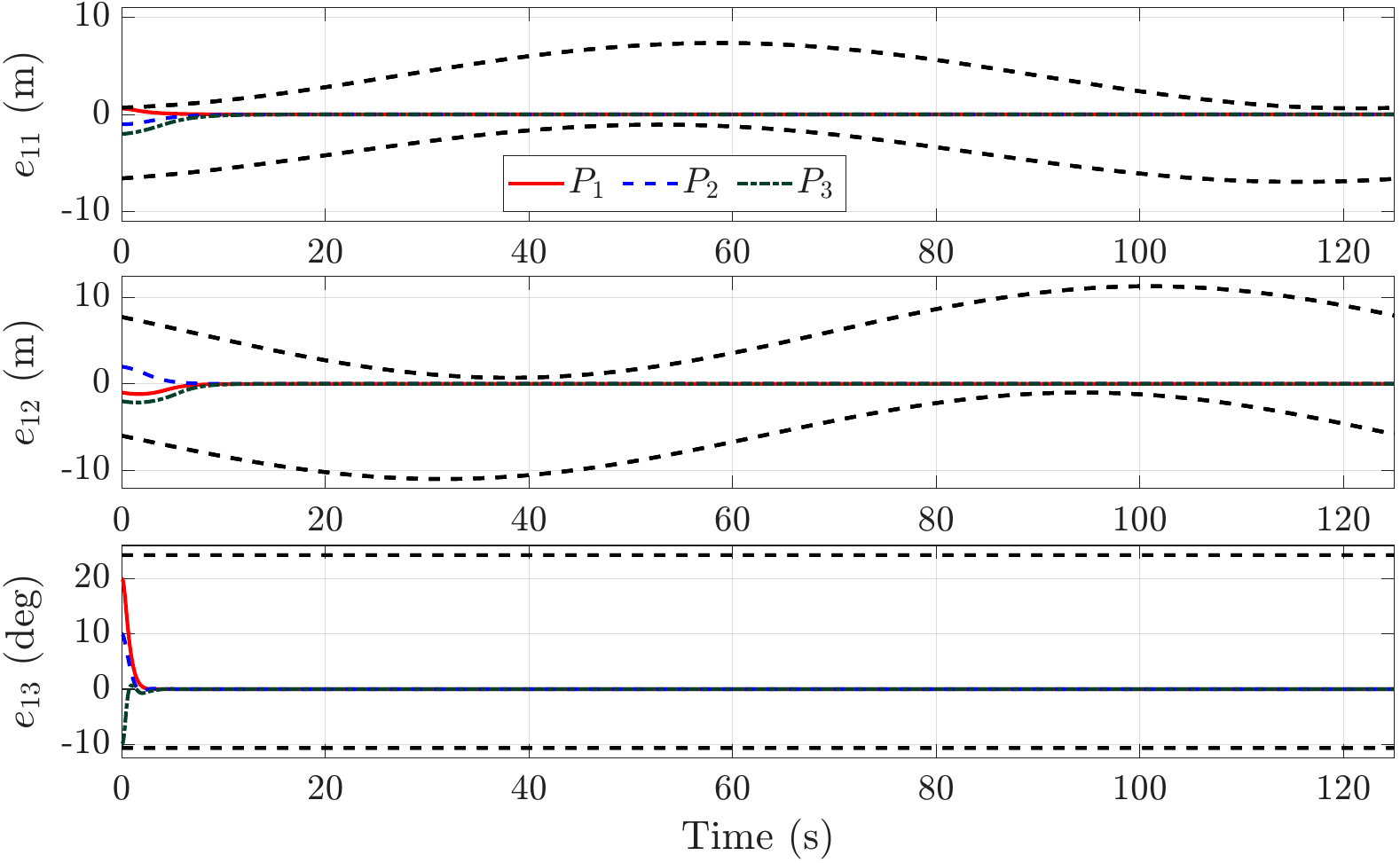}
    \caption{Errors in position and heading.}
    \label{fig:asym_tv_ellipse_e1}
    \end{subfigure}
    \begin{subfigure}{0.5\linewidth}
    \centering
    \includegraphics[width=\linewidth]{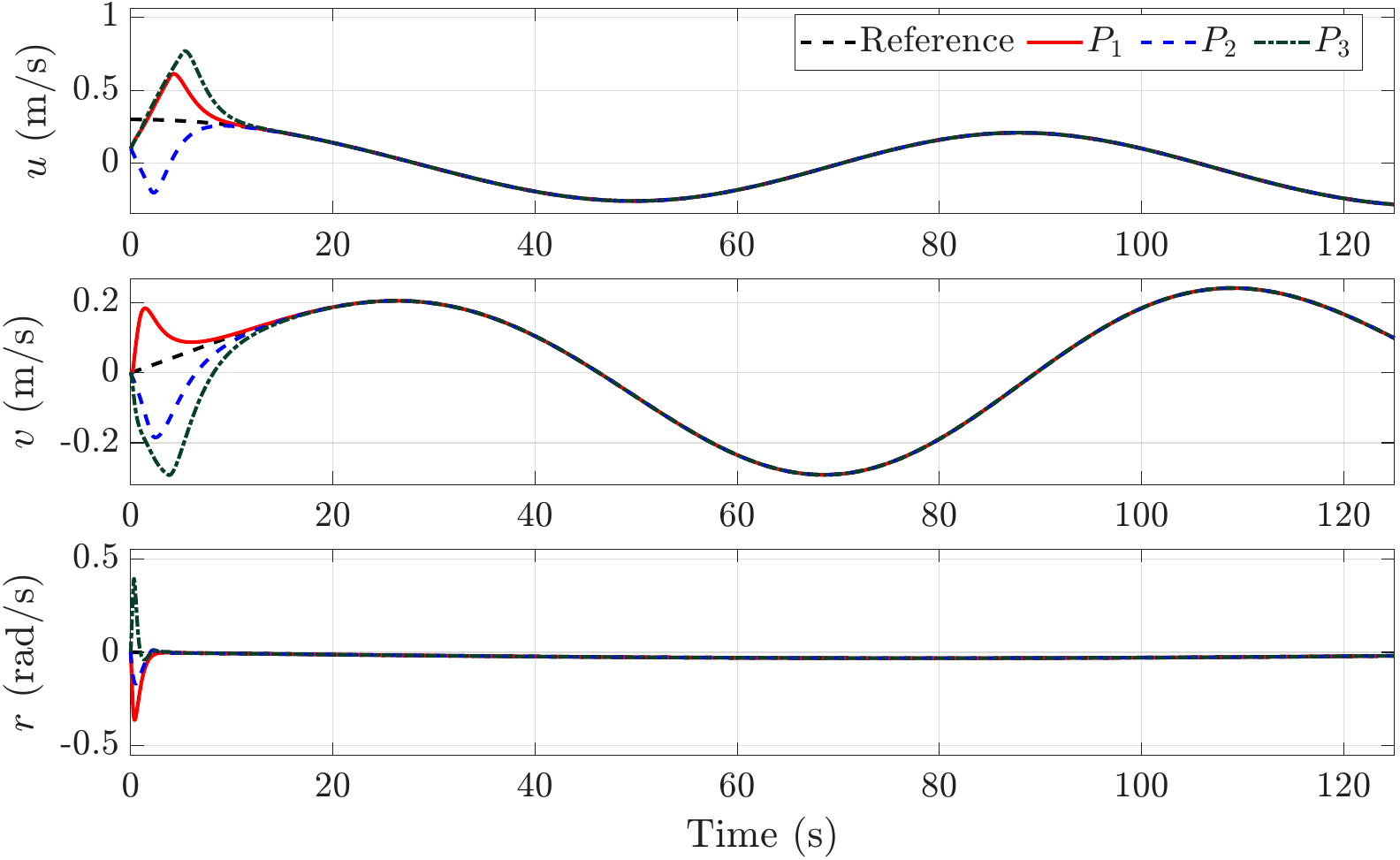}
    \caption{Surge, sway, and yaw rate.}
    \label{fig:asym_tv_ellipse_nu}
    \end{subfigure}%
    \begin{subfigure}{0.5\linewidth}
    \centering
    \includegraphics[width=\linewidth]{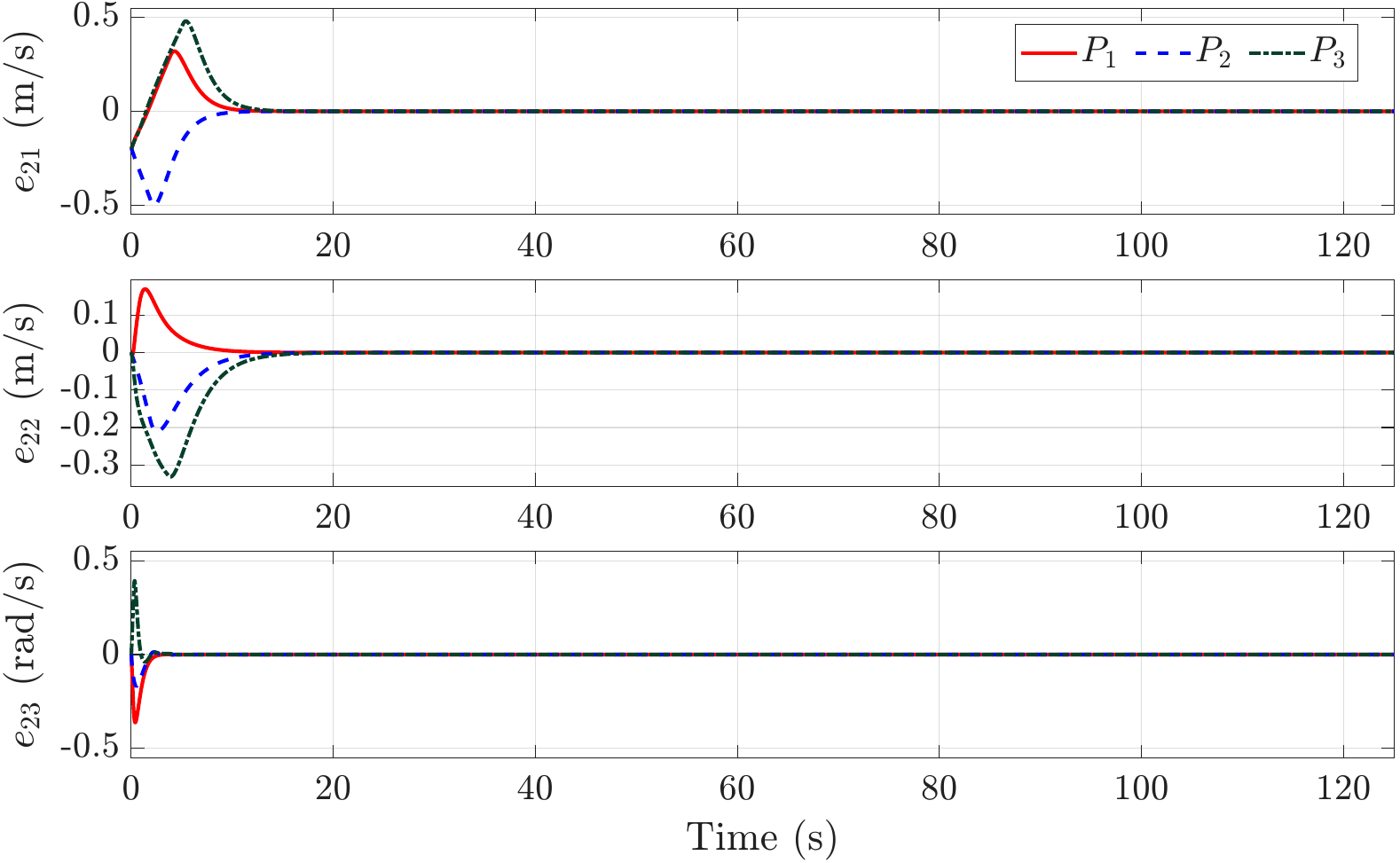}
    \caption{Error in velocities and yaw rate.}
    \label{fig:asym_tv_ellipse_e2}
    \end{subfigure}
    \caption{Performance validation of proposed controller for a USV with elliptical trajectory under asymmetric dynamic constraints on position and heading of the USV.}
    \label{fig:asym_tv_ellipse}
\end{figure*}
\cref{fig:asym_tv_ellipse} depicts the performance of the USV navigating under time-varying constraints on the position and heading of the USV. The USV follows the desired trajectory precisely (see  \Cref{fig:asym_tv_ellipse_path,fig:asym_tv_ellipse_eta,fig:asym_tv_ellipse_nu}) with errors remaining within specified bounds (see \Cref{fig:asym_tv_ellipse_e1}) and converging to zero (see \Cref{fig:asym_tv_ellipse_e1,fig:asym_tv_ellipse_e2}). The black dashed lines in \Cref{fig:asym_tv_ellipse_e1} show the bounds on the tracking error. Furthermore, the control input profiles are depicted in \Cref{fig:asym_tv_ellipse_tau}. It can be observed that at the beginning of the maneuver, the control demands saturate and later reduce to small values as the error approaches zero.

\subsubsection{Eight-shaped trajectory}
The eight-shaped trajectory is given by \Cref{eqn:eight}, and the dynamic boundary is given by 
\begin{subequations}\label{eqn:kc_eight}
\begin{align}
    \bar{\pmb{k}}_{c1} &= \begin{bmatrix}
        1+ \sin(0.05t) & 3 + .5\cos(0.1t) & 0.6 + 0.4\pi\cos(0.02t)
    \end{bmatrix}^\top \\
    \underline{\pmb{k}} _{c1} &= \begin{bmatrix}
        -9 + 2\cos(0.05t) & -3 + \sin(0.1t) & -0.5 + 0.6\pi\cos(0.02t)
    \end{bmatrix}
\end{align}
\end{subequations}
We test the proposed algorithm for three different initial configurations ($\pmb{\eta}_0=(x_0,y_0,\psi_0))$ denoted as $P_1:~(-1\,\si{m},-1 \,\si{m},105^\circ)$, $P_2:~(0.99\,\si{m},1\,\si{m},95^{\circ})$, $P_3:~ (0.99\,\si{m},-2\,\si{m},80^\circ)$.
In all the cases, the USV starts with a surge speed of $0.1\, \si{m/s}$. 
\begin{figure*}[h!]
    \centering
    \begin{subfigure}{0.5\linewidth}
    \centering
    \includegraphics[width=\linewidth]{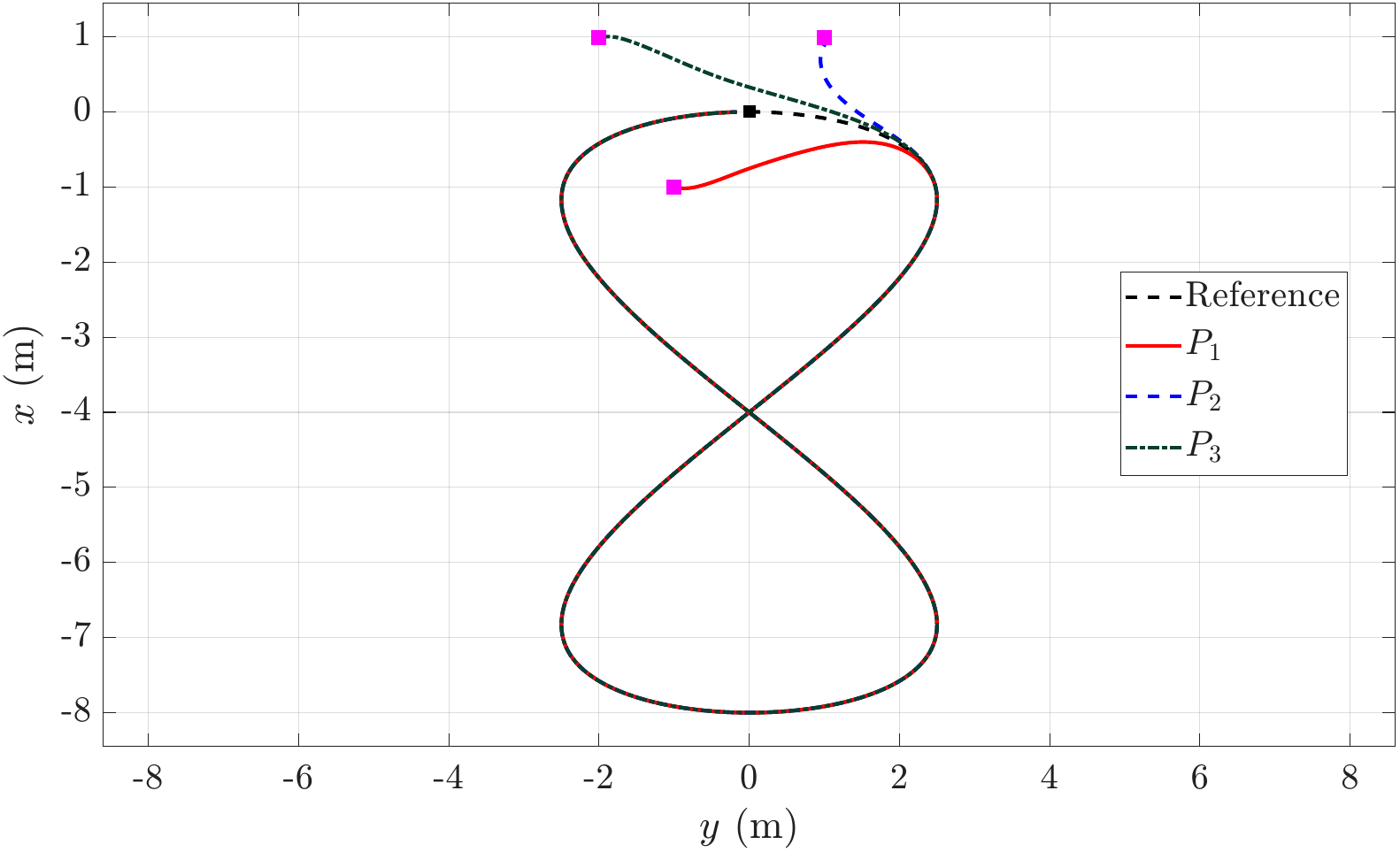}   
    \caption{Actual and desired trajectories.}
    \label{fig:asym_tv_eight_path}
    \end{subfigure}%
    \begin{subfigure}{0.5\linewidth}
    \centering
     \includegraphics[width=\linewidth]{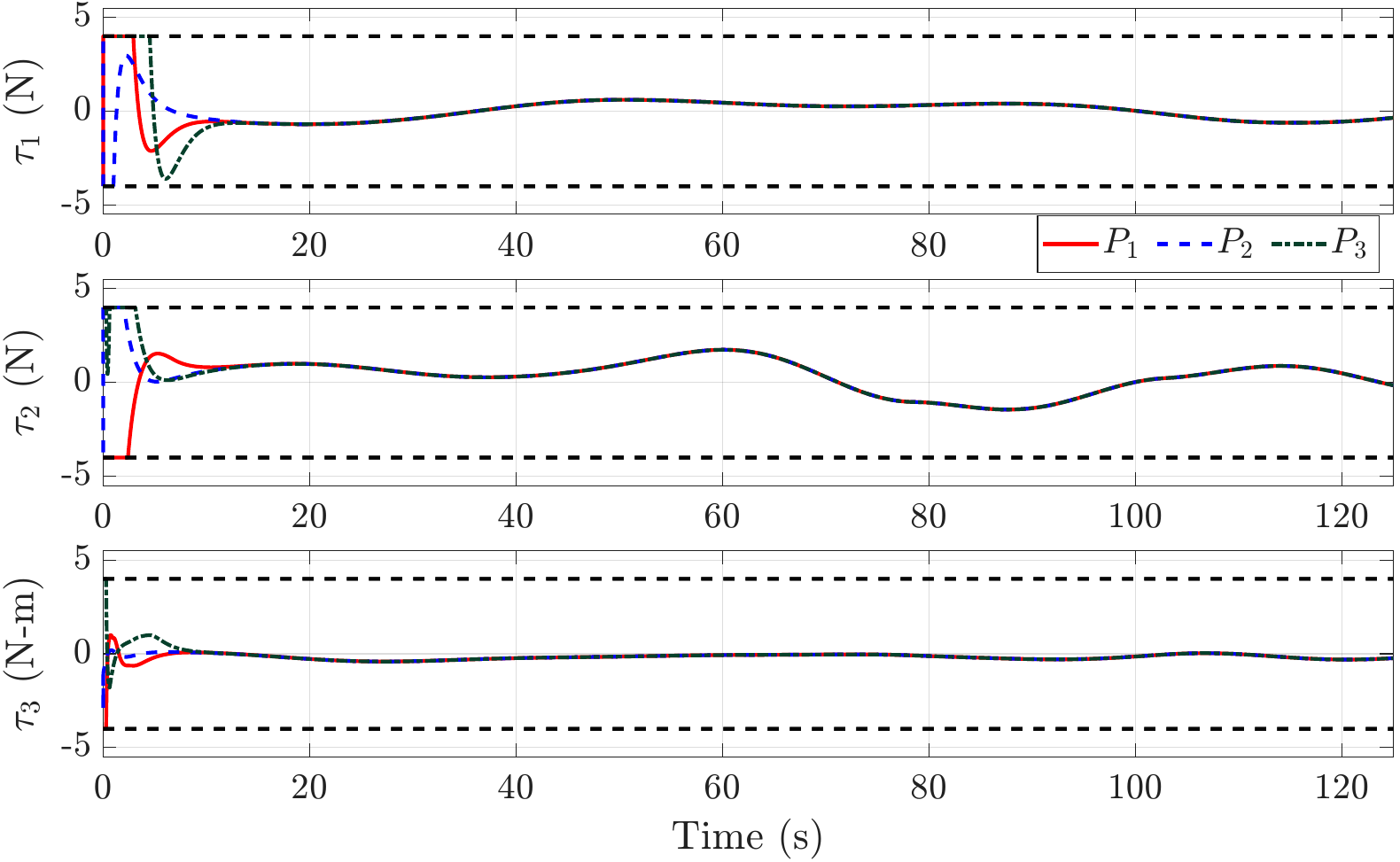}
    \caption{Control inputs.}
    \label{fig:asym_tv_eight_tau}
    \end{subfigure}
    \begin{subfigure}{0.5\linewidth}
    \centering
    \includegraphics[width=\linewidth]{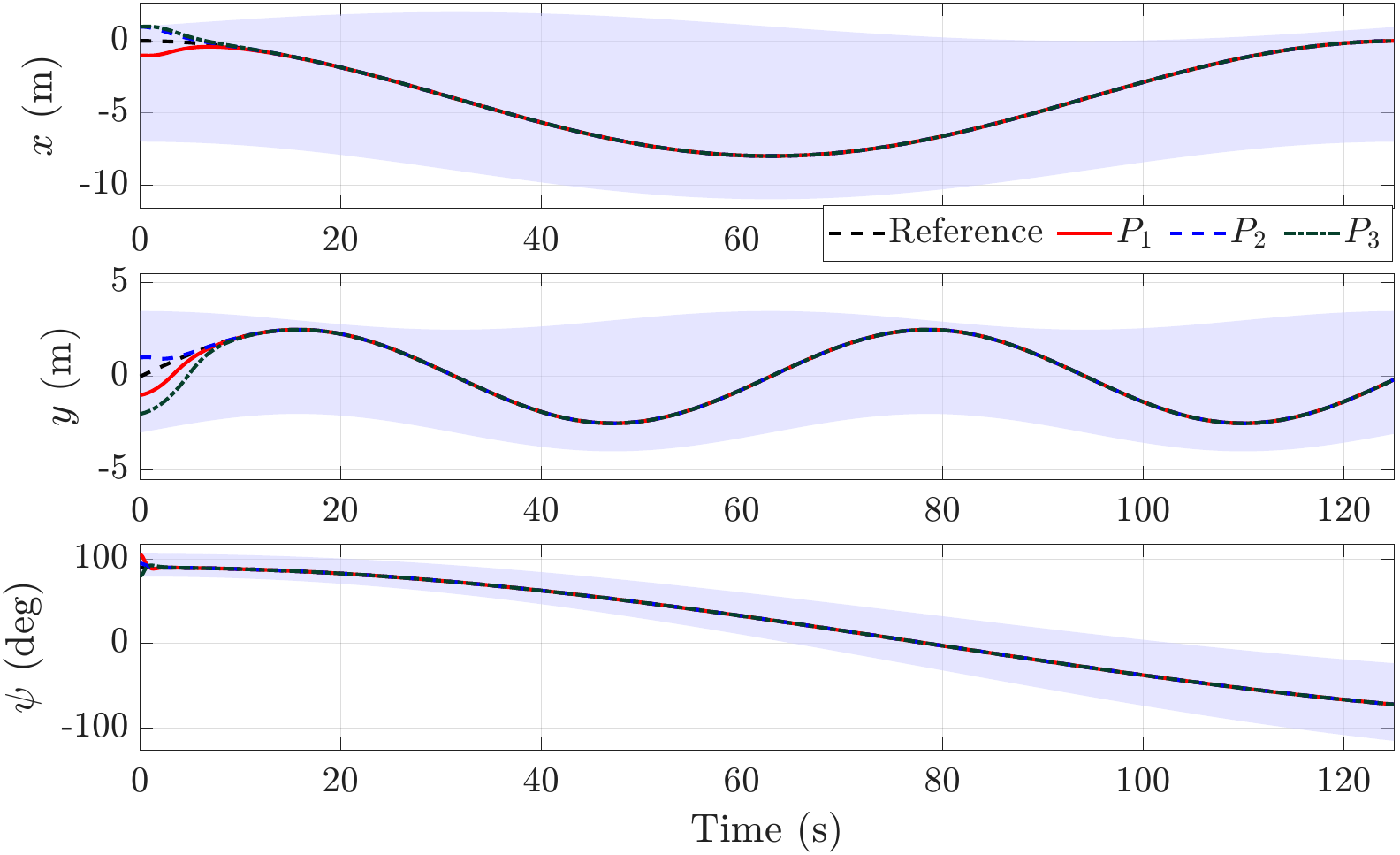}
    \caption{Position and heading of the vessel.}
    \label{fig:asym_tv_eight_eta}
    \end{subfigure}%
    \begin{subfigure}{0.5\linewidth}
    \centering
    \includegraphics[width=\linewidth]{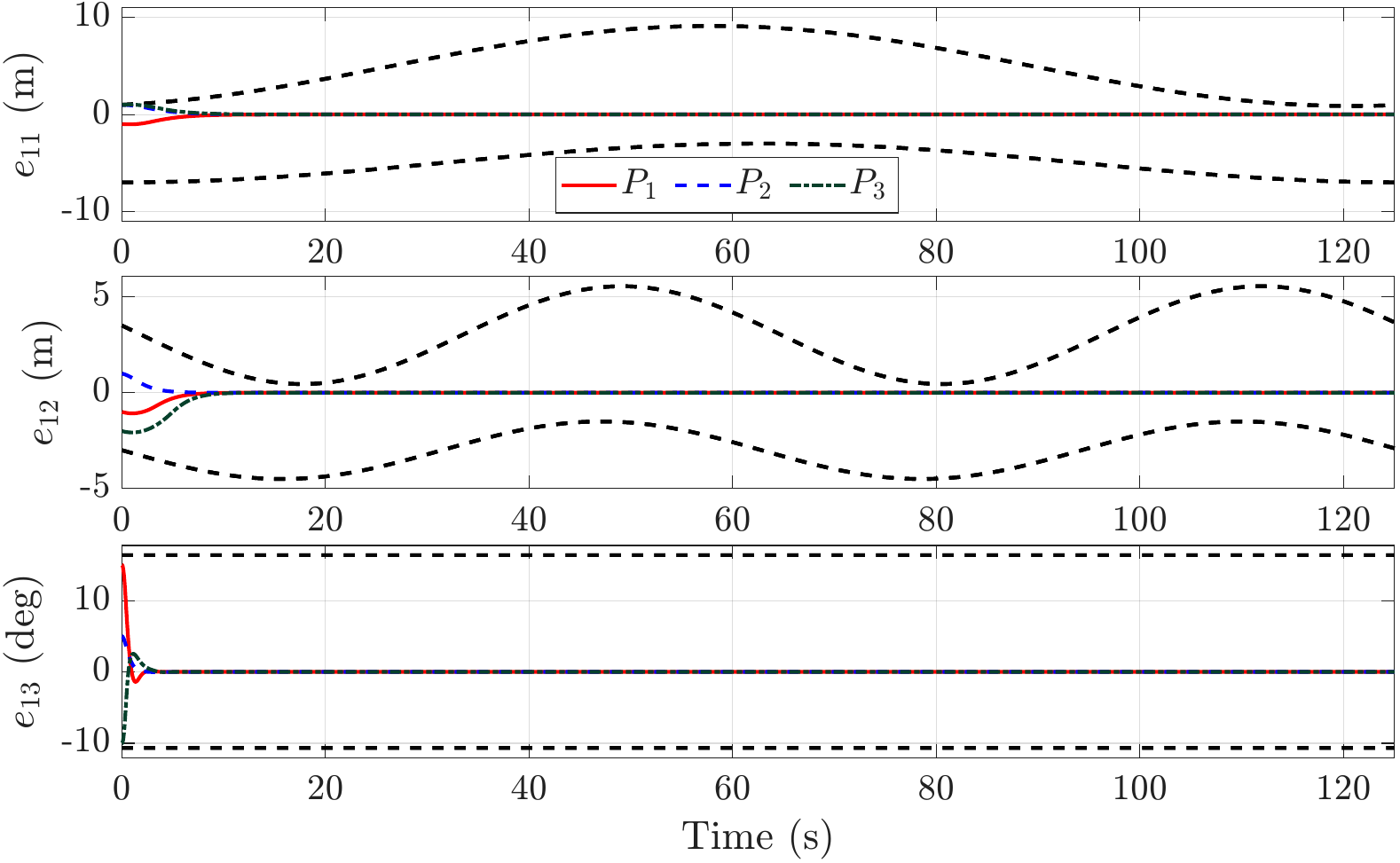}
    \caption{Errors in position and heading.}
    \label{fig:asym_tv_eight_e1}
    \end{subfigure}
    \begin{subfigure}{0.5\linewidth}
    \centering
    \includegraphics[width=\linewidth]{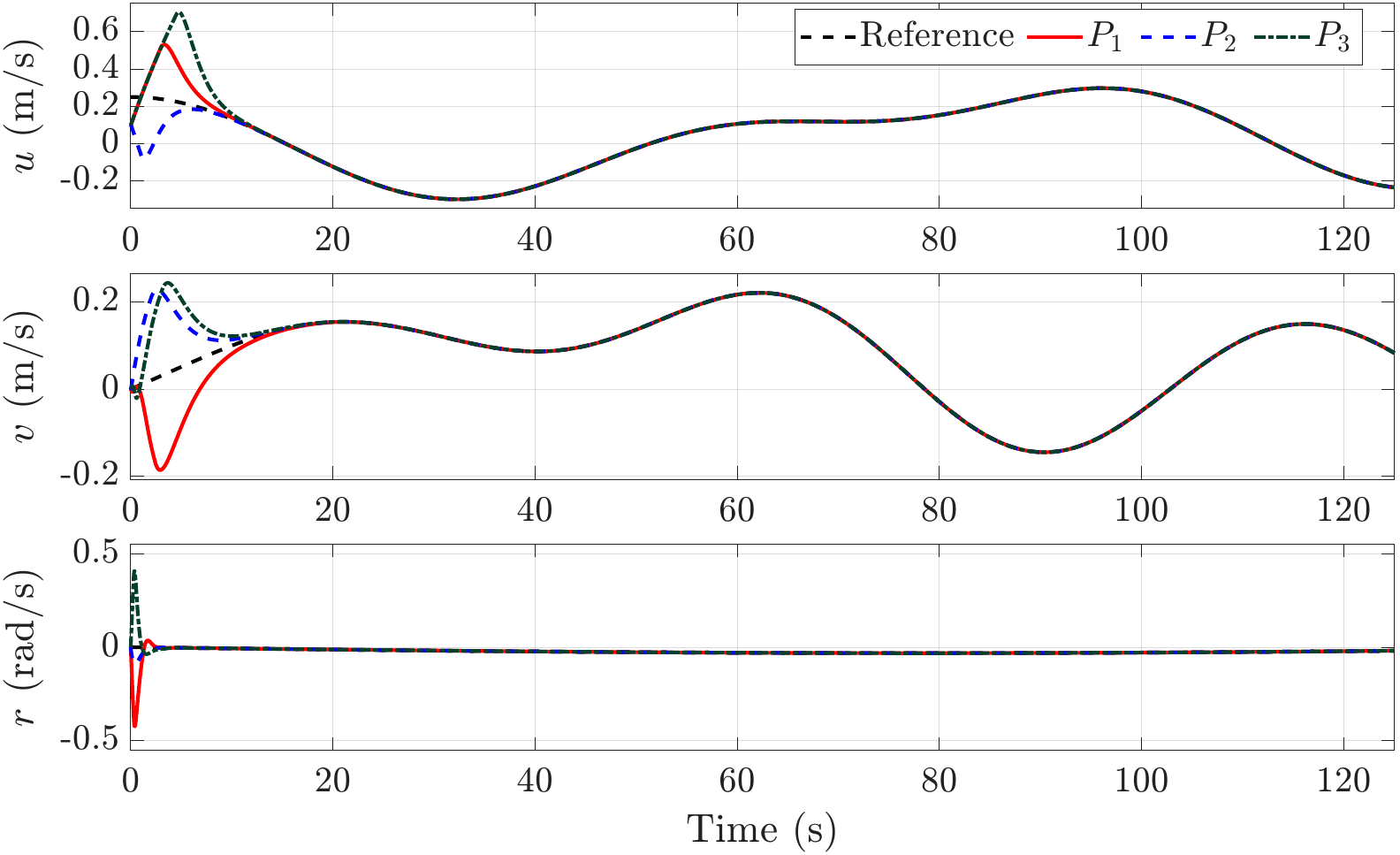}
    \caption{Surge, sway, and yaw rate.}
    \label{fig:asym_tv_eight_nu}
    \end{subfigure}%
    \begin{subfigure}{0.5\linewidth}
    \centering
    \includegraphics[width=\linewidth]{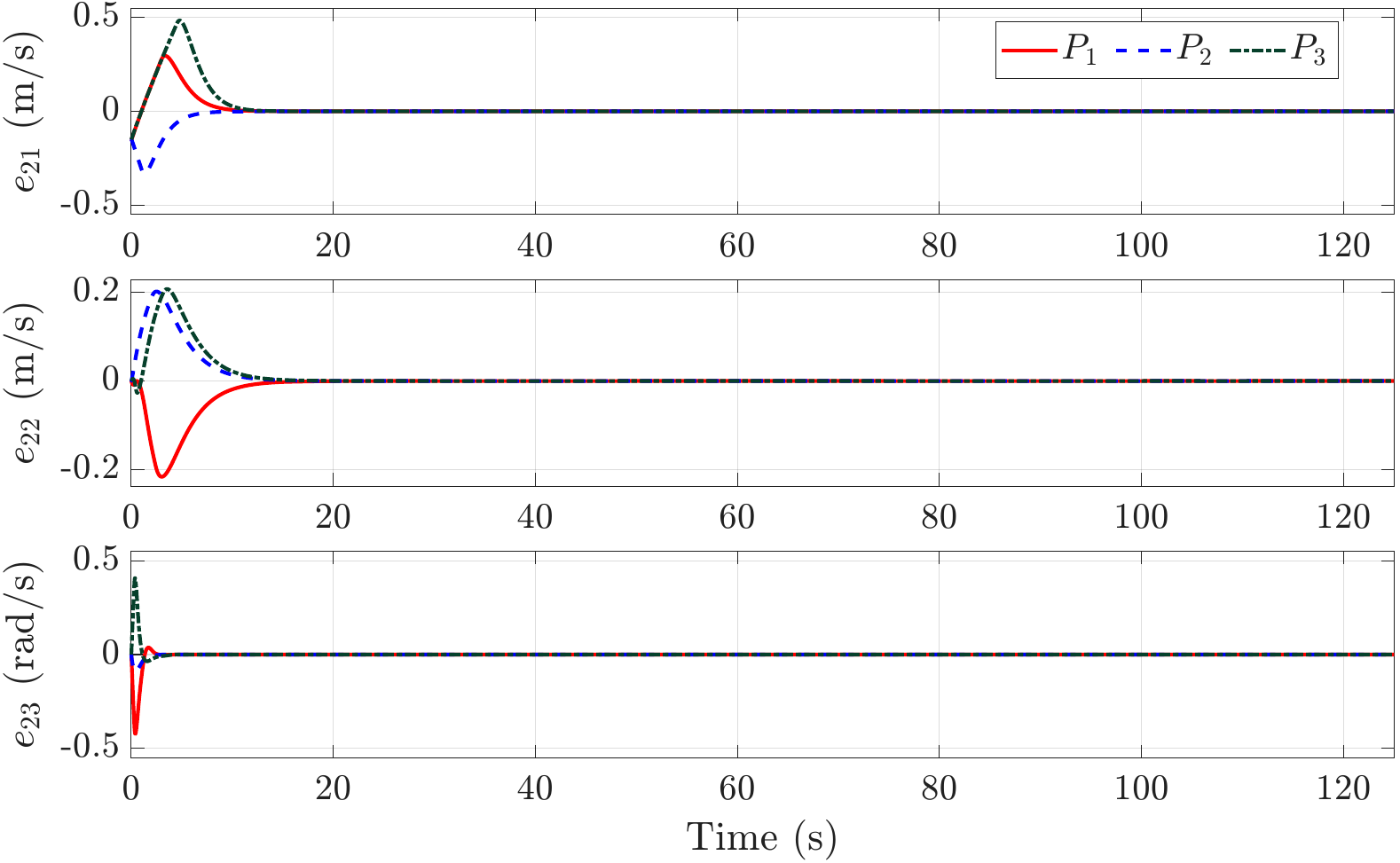}
    \caption{Error in velocities and yaw rate.}
    \label{fig:asym_tv_eight_e2}
    \end{subfigure}
    \caption{Performance validation of proposed controller for a USV following an eight-shaped trajectory under asymmetric dynamic constraints on position and heading of the USV.}
    \label{fig:asym_tv_eight}
\end{figure*}
As evident from \Cref{fig:asym_tv_eight}, the USV can successfully pursue an 8-shaped trajectory while satisfying all the dynamic state constraints, using the proposed controllers. Note that in the case of $P_2$ and $P_3$, the USV is placed close to the boundary (see \Cref{fig:asym_tv_eight_eta}). This shows the proposed strategy working even in cases when the USV is closer to the boundary. Furthermore, one more observation that can be made from \Cref{fig:asym_tv_ellipse_tau,fig:asym_tv_eight_tau} is on the presence of sharp corners in the control profiles due to saturating the controls at the bounds. Recall that this is one of the motivations to explicitly account for the actuator bounds in the controller design.

\subsection{Tracking under dynamic state constraints with input saturation}
In this subsection, we evaluate the performance of the proposed algorithm in \Cref{sec:tv_ip}. In addition to imposing time-varying constraints on the USV's position and heading, this strategy commands the control inputs by explicitly accounting for the control bounds. The asymmetric control bounds considered are $\pmb{\tau}_{\text{max}} = [4~~4~~4]^\top$ and $\pmb{\tau}_{\text{min}} = [-3.5~-3.5~-3.5]^\top$. 
The controller gain parameters are systematically chosen as: $k_{11} = 0.1$, $k_{12} = 0.4$, $k_{13} = 0.3$, $\pmb{K}_2 = \diag[1~~1~~3]$, $\pmb{K}_3 = \diag[2~~3~~5]$, $\pmb{\rho} = \diag [0.2~~0.2~~0.2]$, $a_{1}=0.1$, $a_2 = 0.1$, $a_3=0.1$.  
We evaluate the performance of the proposed scheme for both elliptical and 8-shaped trajectories. 
The dynamic state constraint for the elliptical trajectory is given by \Cref{eqn:kc_ellipse} and for the 8-shaped trajectory is given by \Cref{eqn:kc_eight}.
The three different initial configurations of the USV ($\pmb{\eta}_0=(x_0,y_0,\psi_0))$ under consideration for elliptical trajectory are denoted as $P_1:~(0.6\,\si{m},-1 \,\si{m},110^\circ)$, $P_2:~(-1\,\si{m},2\,\si{m},100^{\circ})$, $P_3:~ (-2\,\si{m},-2\,\si{m},80^\circ)$. 
In all cases, the USV is moving with a surge speed of $0.1\, \si{m/s}$.
\begin{figure*}[h!]
    \centering
    \begin{subfigure}{0.5\linewidth}
    \centering
    \includegraphics[width=\linewidth]{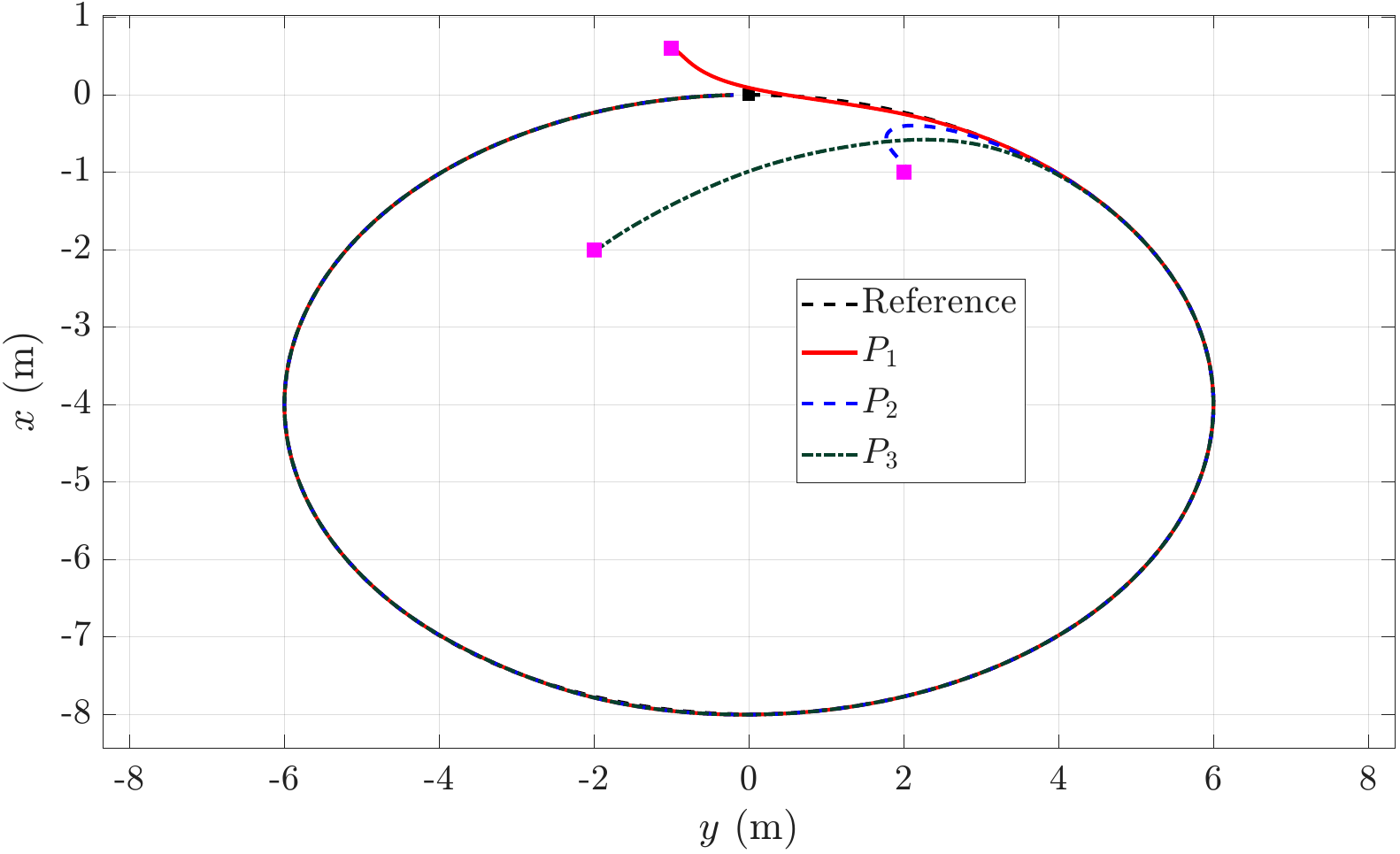}   
    \caption{Actual and desired trajectories.}
    \label{fig:asym_tv_inp_sat_ellipse_path}
    \end{subfigure}%
    \begin{subfigure}{0.5\linewidth}
    \centering
     \includegraphics[width=\linewidth]{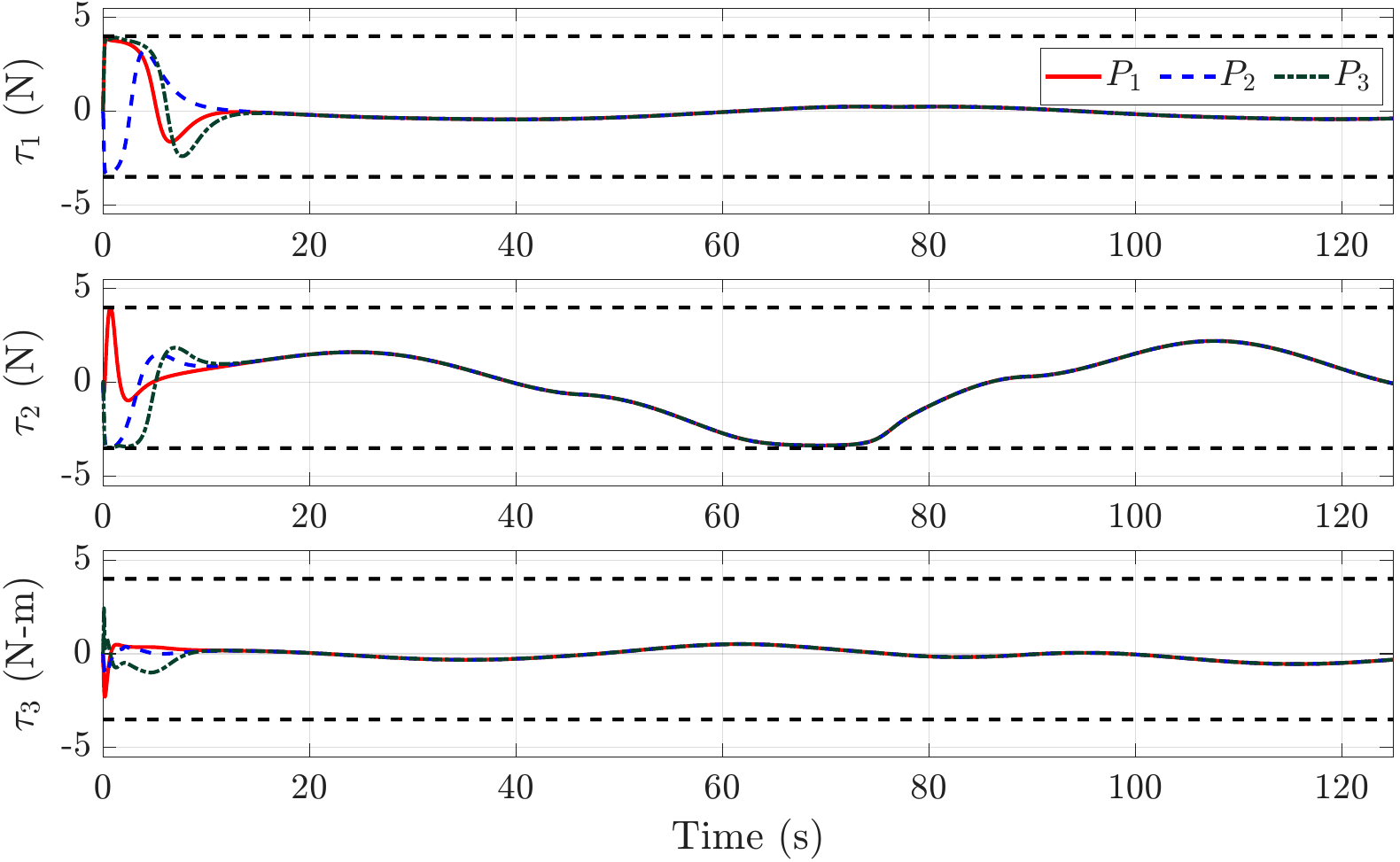}
    \caption{Control inputs.}
    \label{fig:asym_tv_inp_sat_ellipse_tau}
    \end{subfigure}
    \begin{subfigure}{0.5\linewidth}
    \centering
    \includegraphics[width=\linewidth]{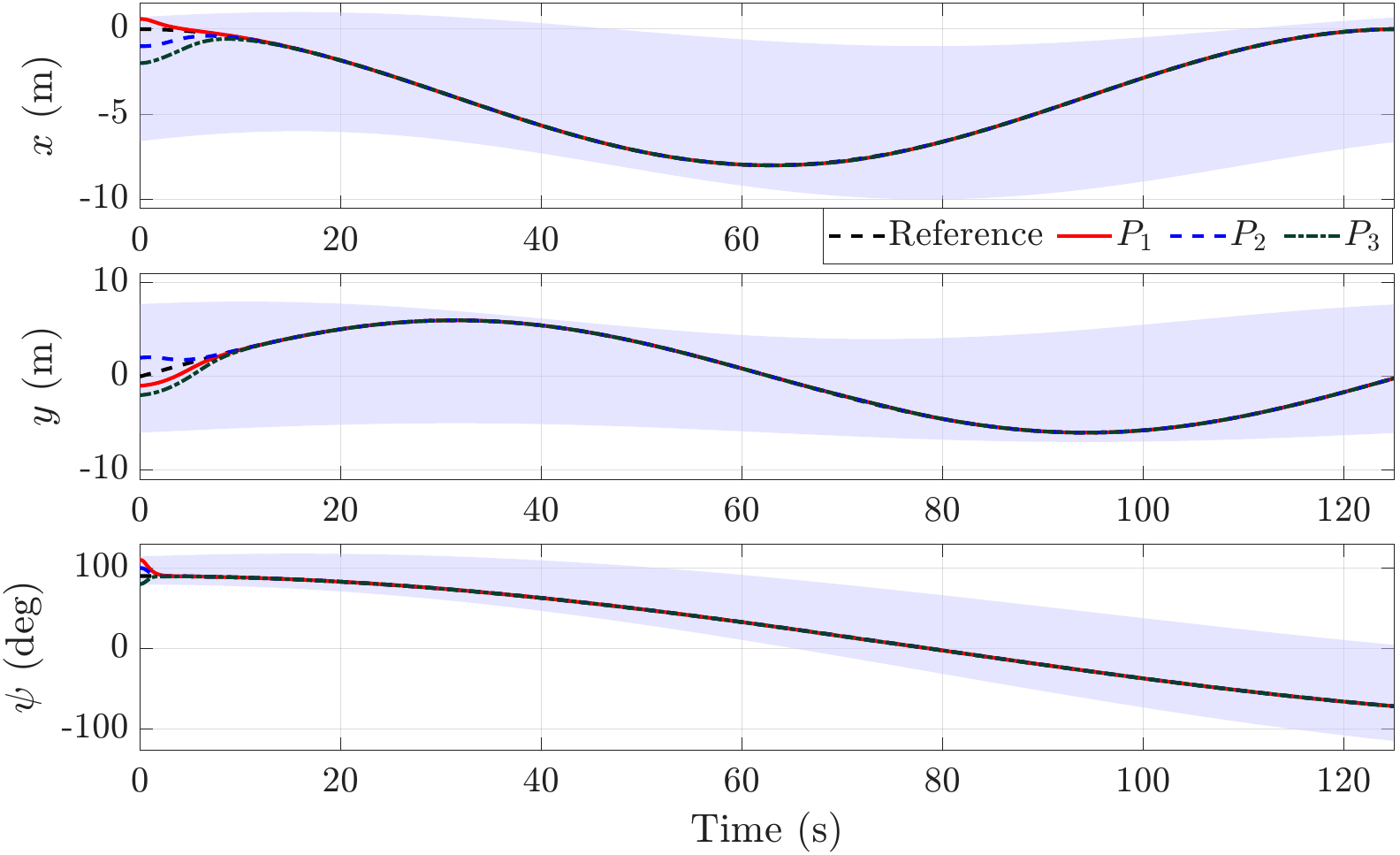}
    \caption{Position and heading of the vessel.}
    \label{fig:asym_tv_inp_sat_ellipse_eta}
    \end{subfigure}%
    \begin{subfigure}{0.5\linewidth}
    \centering
    \includegraphics[width=\linewidth]{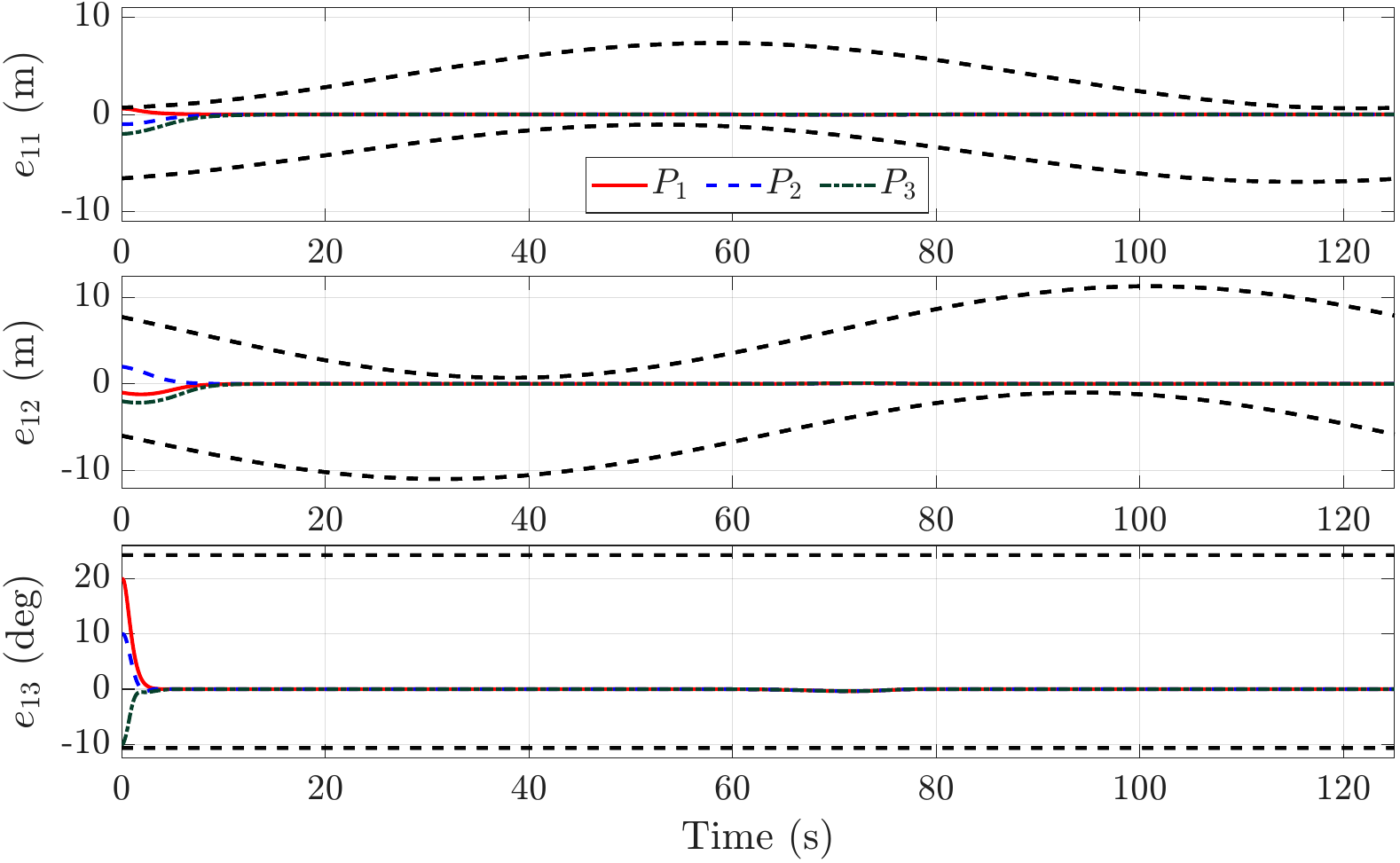}
    \caption{Errors in position and heading.}
    \label{fig:asym_tv_inp_sat_ellipse_e1}
    \end{subfigure}
    \begin{subfigure}{0.5\linewidth}
    \centering
    \includegraphics[width=\linewidth]{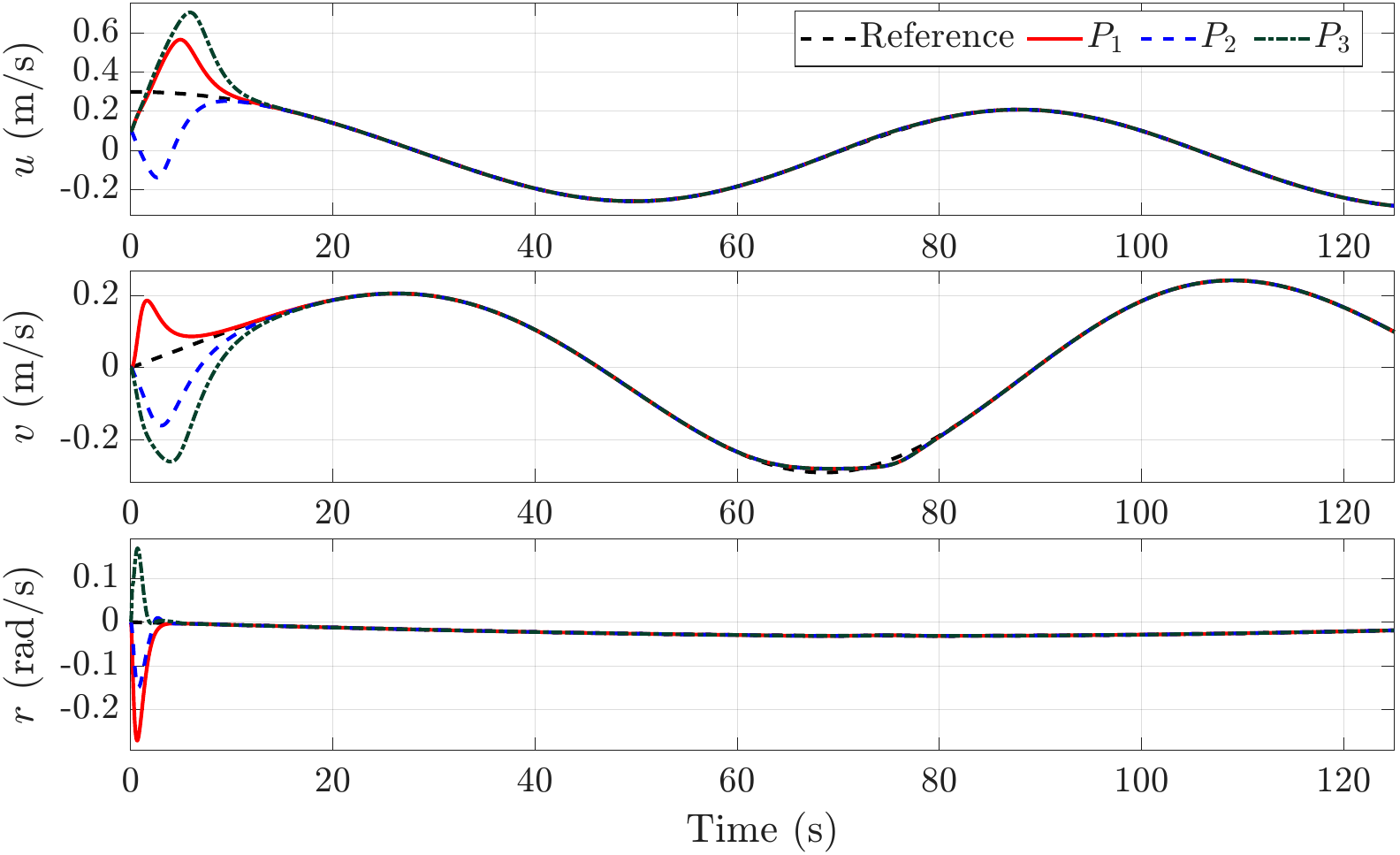}
    \caption{Surge, sway, and yaw rate.}
    \label{fig:asym_tv_inp_sat_ellipse_nu}
    \end{subfigure}%
    \begin{subfigure}{0.5\linewidth}
    \centering
    \includegraphics[width=\linewidth]{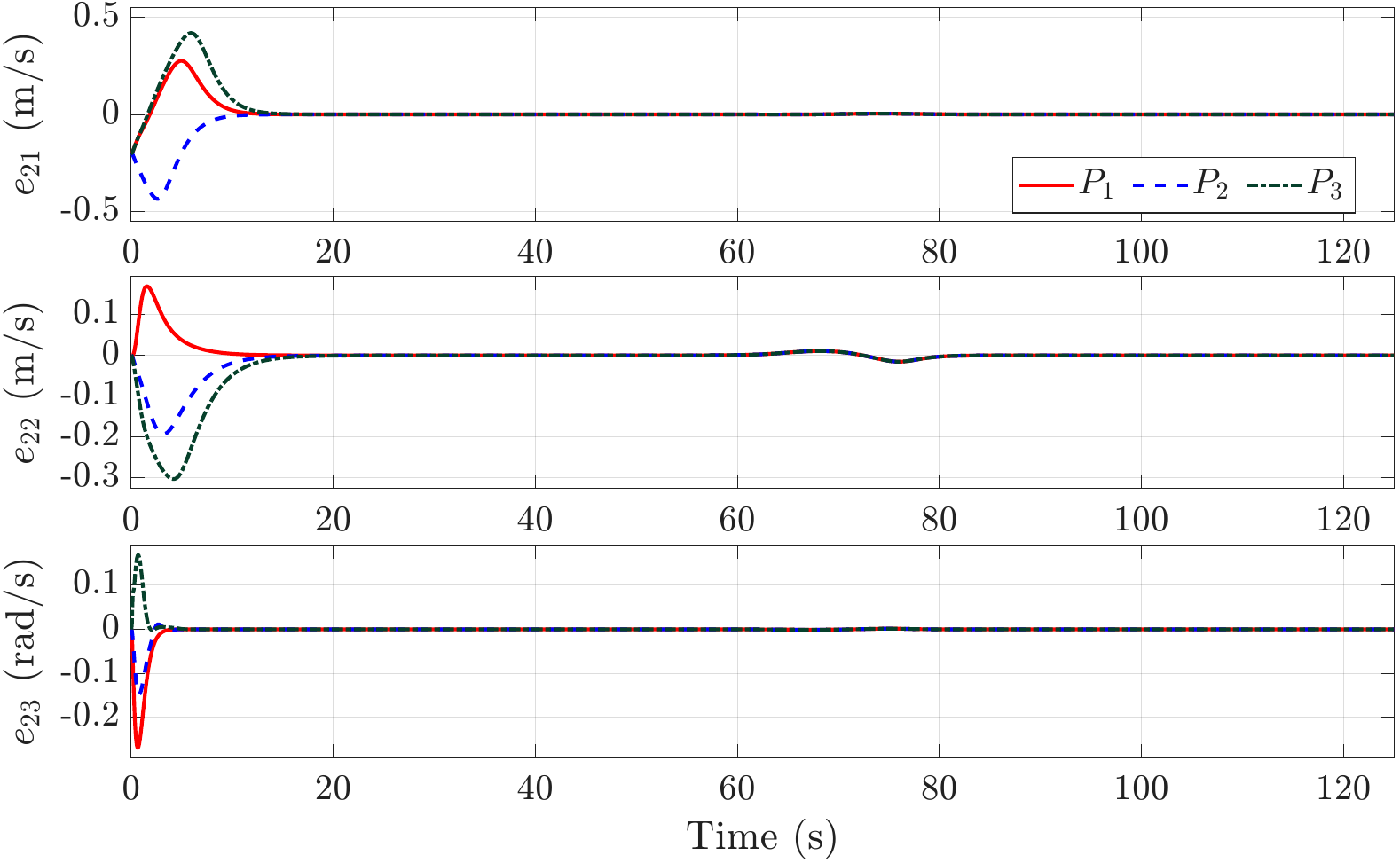}
    \caption{Error in velocities and yaw rate.}
    \label{fig:asym_tv_inp_sat_ellipse_e2}
    \end{subfigure}
    \caption{Performance validation of proposed controller for a USV following an elliptical trajectory under asymmetric dynamic constraints and input saturation.}
    \label{fig:asym_tv_inp_sat_ellipse}
\end{figure*}
\cref{fig:asym_tv_inp_sat_ellipse} illustrates the USV following an elliptical trajectory without violating state constraints (see \Cref{fig:asym_tv_inp_sat_ellipse_e1} and actuator bounds (see \Cref{fig:asym_tv_inp_sat_ellipse_tau}). 
The three different initial configurations of the USV for an 8-shaped trajectory are denoted as $P_1:~(-1\,\si{m},-1 \,\si{m},105^\circ)$, $P_2:~(0.99\,\si{m},1\,\si{m},95^{\circ})$, $P_3:~ (0.99\,\si{m},-2\,\si{m},80^\circ)$. 
\begin{figure*}[h!]
    \centering
    \begin{subfigure}{0.5\linewidth}
    \centering
    \includegraphics[width=\linewidth]{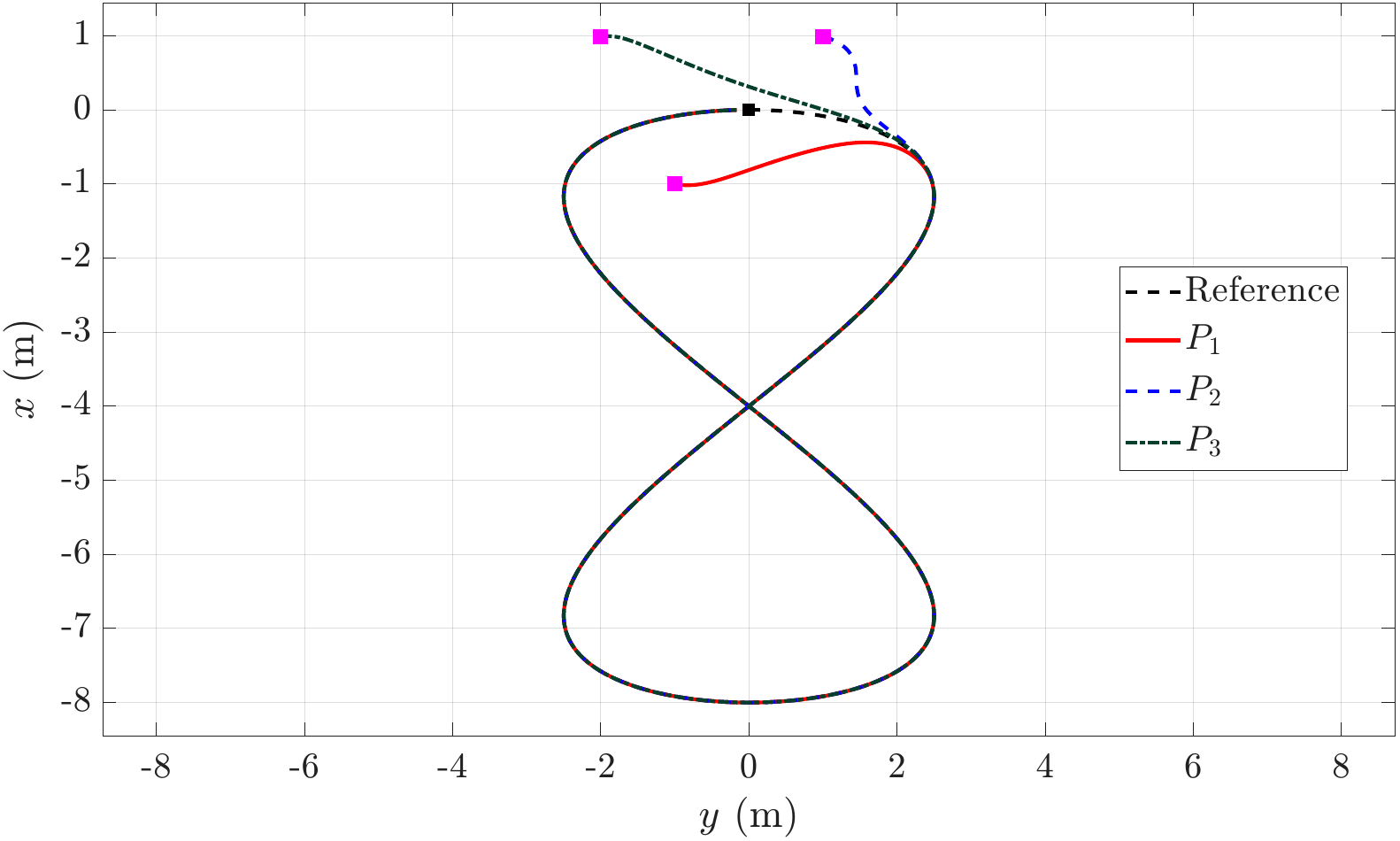}   
    \caption{Actual and desired trajectories.}
    \label{fig:asym_tv_inp_sat_eight_path}
    \end{subfigure}%
    \begin{subfigure}{0.5\linewidth}
    \centering
     \includegraphics[width=\linewidth]{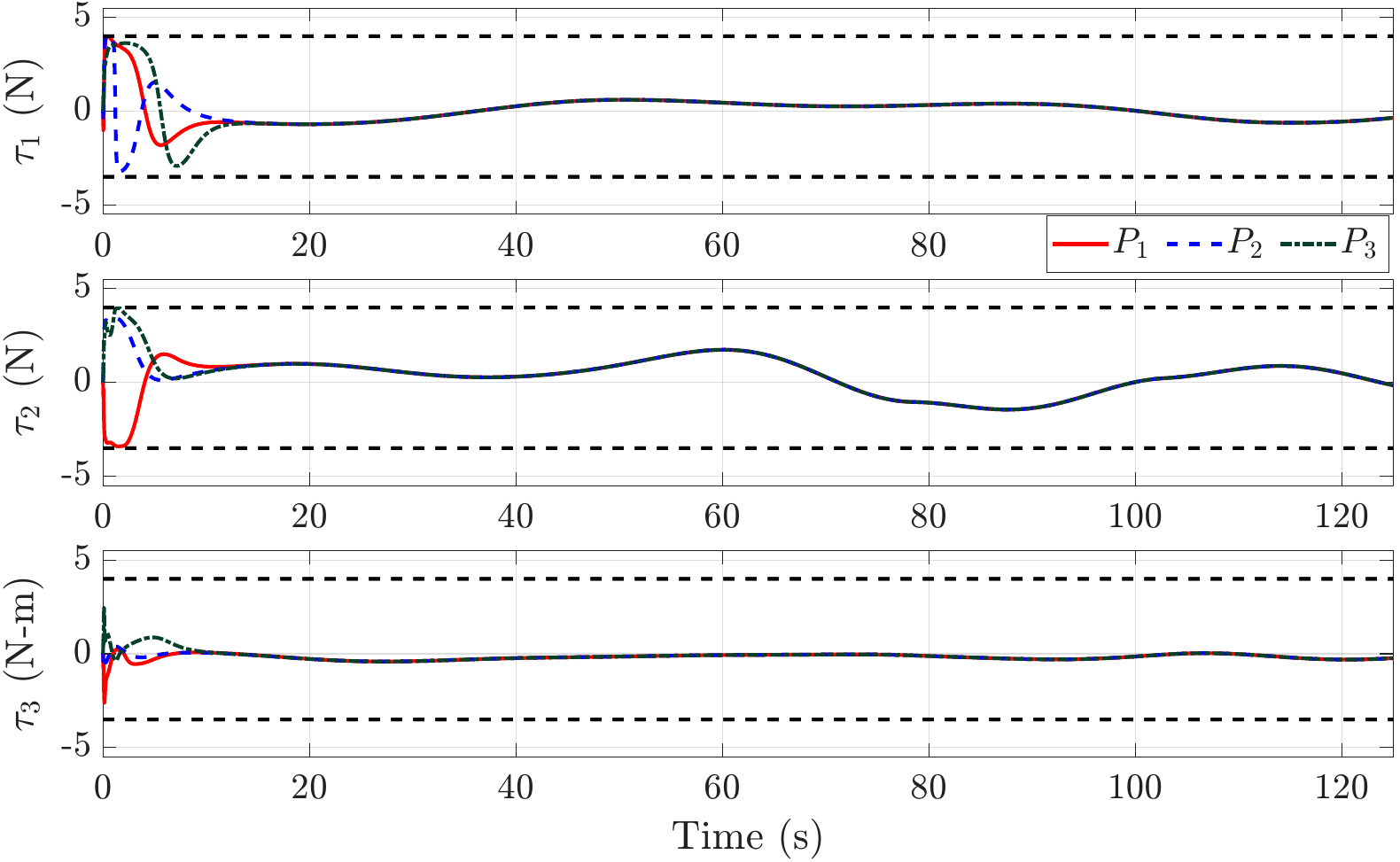}
    \caption{Control inputs.}
    \label{fig:asym_tv_inp_sat_eight_tau}
    \end{subfigure}
    \begin{subfigure}{0.5\linewidth}
    \centering
    \includegraphics[width=\linewidth]{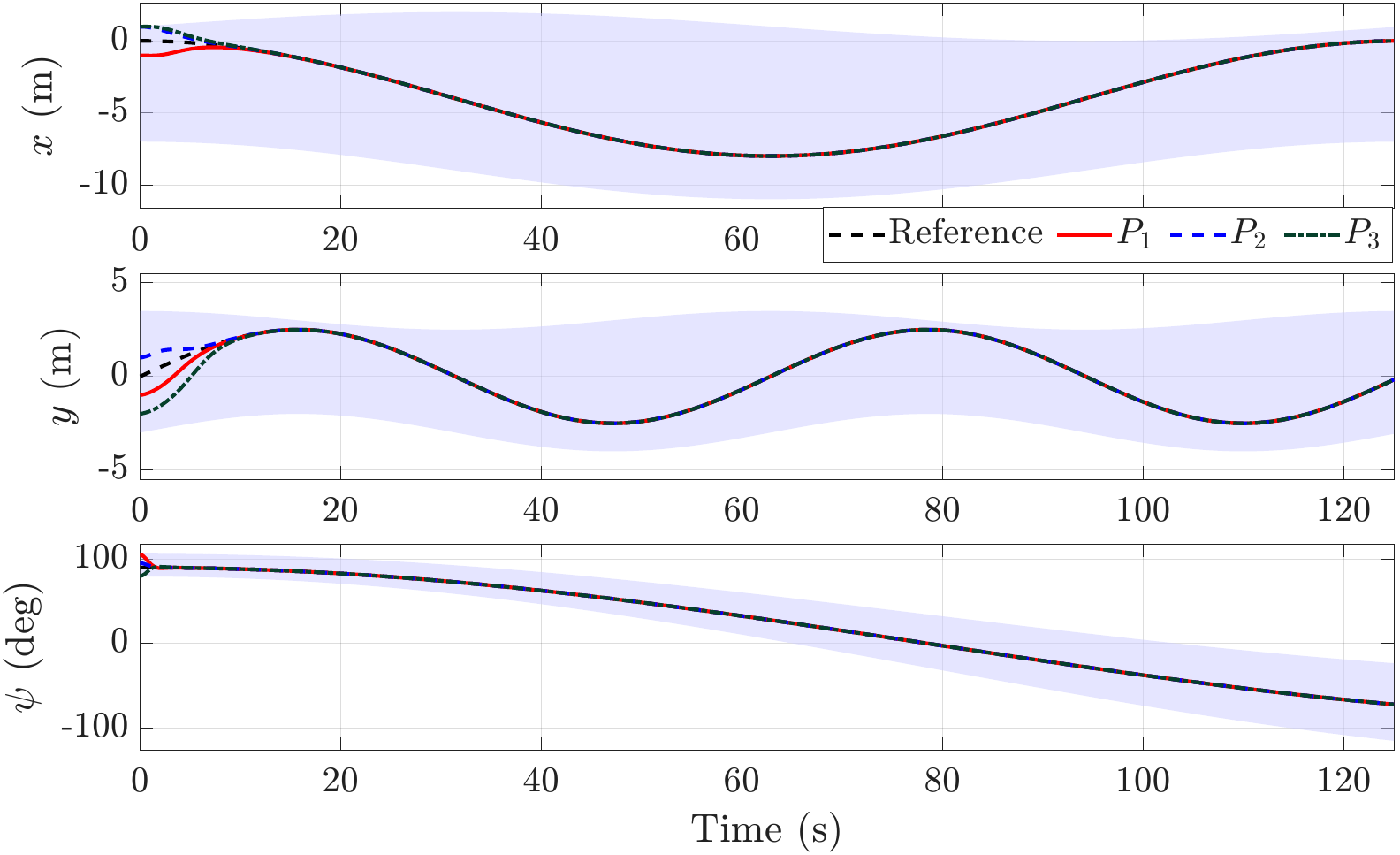}
    \caption{Position and heading of the vessel.}
    \label{fig:asym_tv_inp_sat_eight_eta}
    \end{subfigure}%
    \begin{subfigure}{0.5\linewidth}
    \centering
    \includegraphics[width=\linewidth]{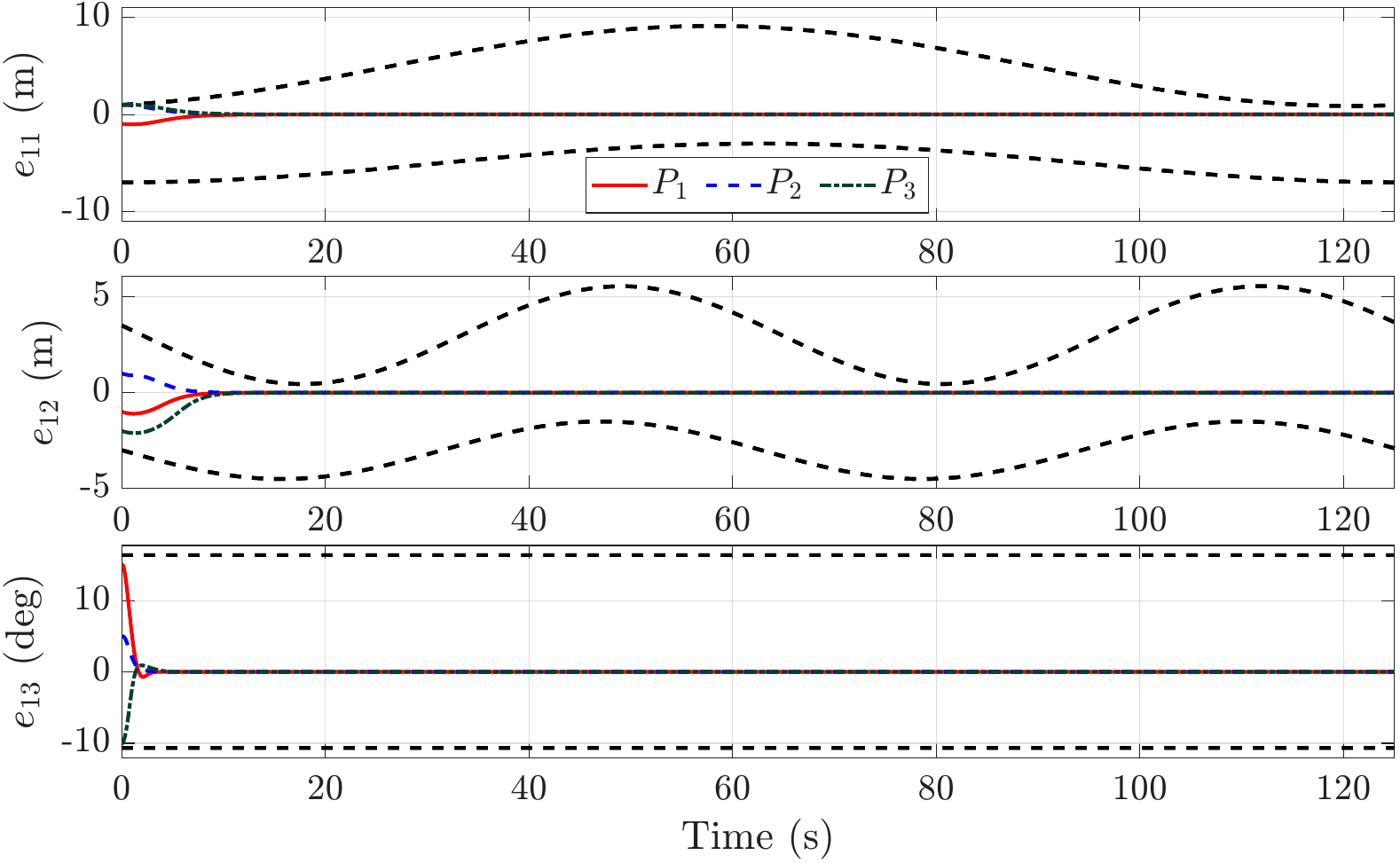}
    \caption{Errors in position and heading.}
    \label{fig:asym_tv_inp_sat_eight_e1}
    \end{subfigure}
    \begin{subfigure}{0.5\linewidth}
    \centering
    \includegraphics[width=\linewidth]{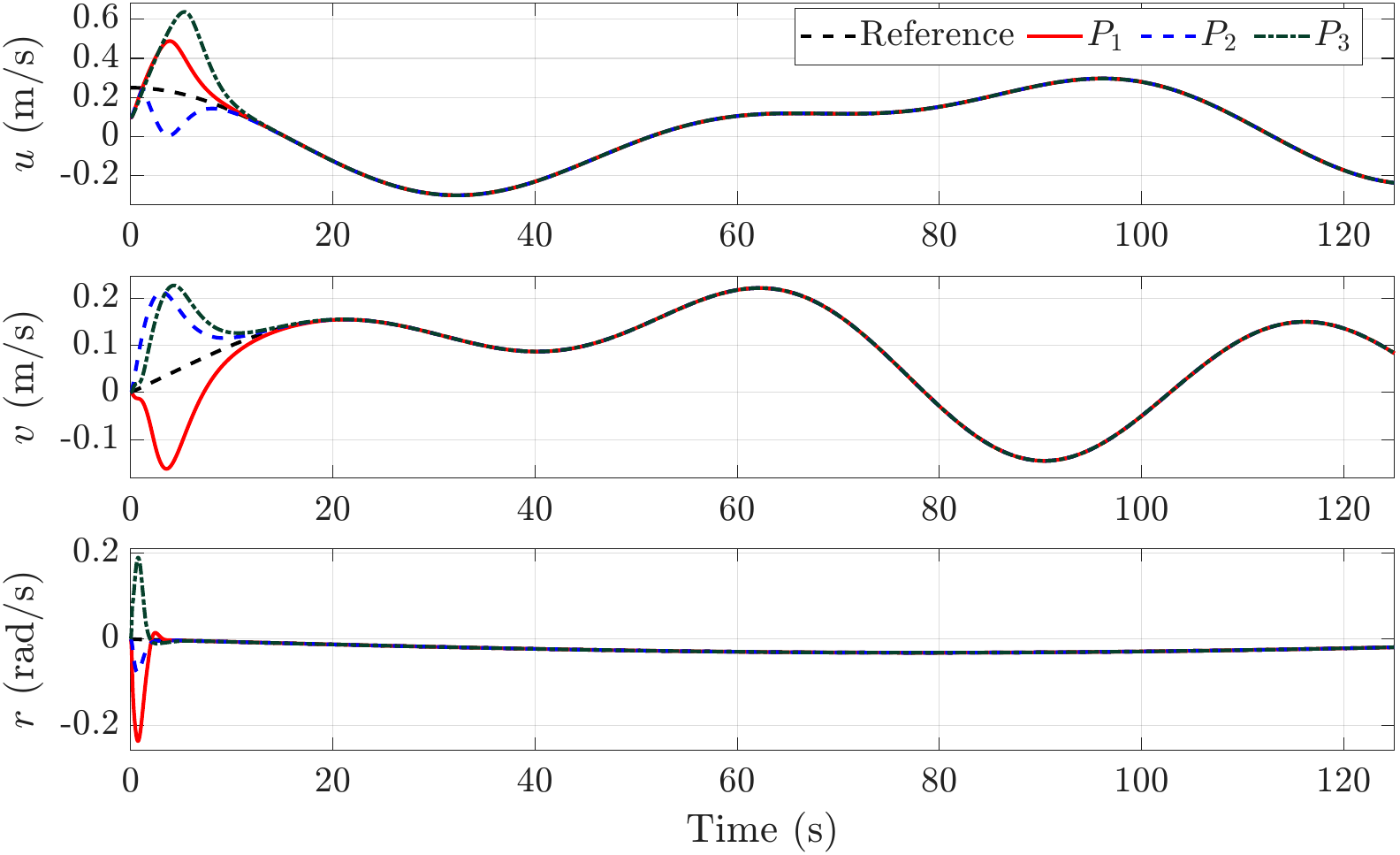}
    \caption{Surge, sway, and yaw rate.}
    \label{fig:asym_tv_inp_sat_eight_nu}
    \end{subfigure}%
    \begin{subfigure}{0.5\linewidth}
    \centering
    \includegraphics[width=\linewidth]{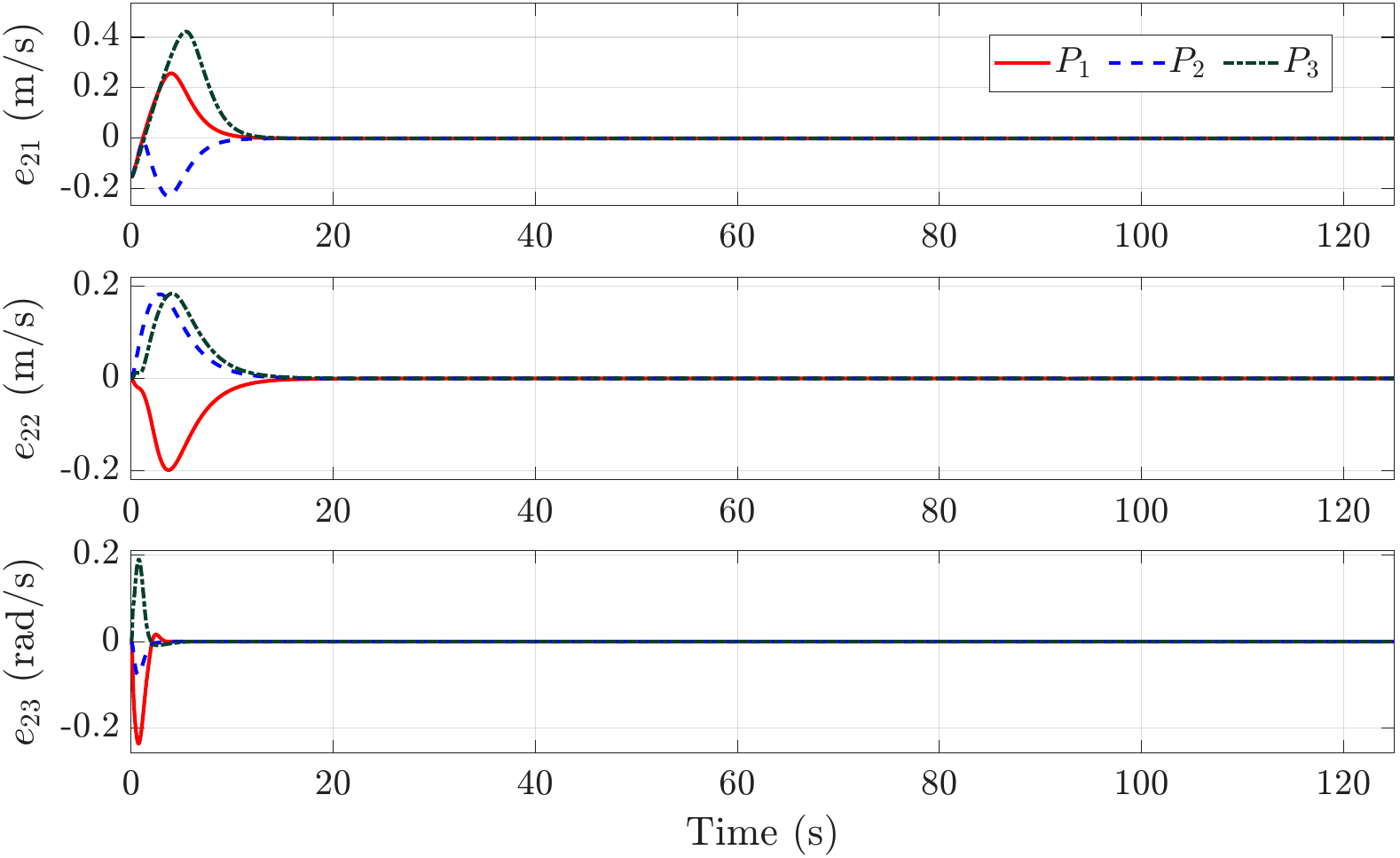}
    \caption{Error in velocities and yaw rate.}
    \label{fig:asym_tv_inp_sat_eight_e2}
    \end{subfigure}
    \caption{Performance validation of proposed controller for a USV following an eight-shaped trajectory under asymmetric dynamic constraints and input saturation.}
    \label{fig:asym_tv_inp_sat_eight}
\end{figure*}
In order to validate the proposed scheme, we consider an 8-shaped trajectory under the dynamic constraints given by \Cref{eqn:kc_eight}.  As evidenced by \Cref{fig:asym_tv_inp_sat_eight}, the USV can steer itself on the desired trajectory without violating the imposed constraints on state and the input. The tracking error converges to zero as illustrated in \Cref{fig:asym_tv_inp_sat_eight_e1}.

\section{Conclusions}\label{sec:conclusion}
This work proposed nonlinear motion control strategies for a USV following a prespecified trajectory under static and dynamic state constraints. We employed a barrier Lyapunov function to address both static and time-varying state constraints. Furthermore, the algorithm incorporates asymmetric actuator bounds into the design by augmenting a smooth input saturation model. The derived control strategies were mathematically proven to be stable. This is unlike the ad hoc approach to restricting the control input, which loses the guarantee of system stability. We performed numerous simulations against several initial conditions for elliptical and 8-shaped trajectories. The proposed scheme enables the USVs to follow the trajectory effectively while respecting state constraints (whether static, such as in a pond, or dynamic, such as in a river) without exceeding the available control authority. Addressing these critical issues enables the proposed algorithm to successfully perform missions in a constrained space while satisfying the physical limitations of the actuators. Subsequent research will focus on enhancing the algorithm's robustness against parameter uncertainties and environmental disturbances.
\bibliography{references.bib}
\end{document}